{}
{}

\documentclass{article}

\usepackage[utf8]{inputenc}
\usepackage{lmodern}
\usepackage[dvipsnames]{xcolor}
\usepackage{graphicx,caption}
\usepackage{caption}
\usepackage{subcaption}
\usepackage{cite}

\usepackage{comment}
\usepackage{microtype}

\usepackage{tikz}
\usepackage{pgfplots}
\pgfplotsset{compat=1.18}
\usepackage{wrapfig}
\usepackage{capt-of}

\usepackage{geometry}
\usepackage{afterpage}
\usepackage{multirow}
\usepackage{tabularx}

\usepackage{bm}
\usepackage{mathtools,slashed}
\usepackage{amsmath, amssymb}

\numberwithin{equation}{section}
\usepackage{mathrsfs}
\usepackage{physics} 
\usepackage{bigints}

\usepackage{url}
\usepackage[bottom,hang,flushmargin]{footmisc}
\usepackage{fancyhdr}
\usepackage{nicefrac}

\usepackage{subfiles}
\usepackage{enumitem}\setlist[enumerate]{font=\bfseries}

\usepackage{titlesec}
\usepackage[toc,page]{appendix}
\usepackage[breaklinks]{hyperref}
\usepackage{cleveref}

\titleformat*{\section}{\LARGE\bfseries}
\titleformat*{\subsection}{\Large\bfseries}
\titleformat*{\subsubsection}{\large\bfseries}

\newcommand{\ha}[1]{\widehat{#1}}

\newcommand{\comma}{\text{,\space\space\space\space\space}}

\newcommand{\ol}[1]{\overline{#1}}

\newcommand{\ra}{\rightarrow}

\newcommand{\hg}{\mathfrak{g}}

\newcommand{\ad}{\text{ad}}

\newcommand{\slNC}{\mf{sl}_{\C}(N)}
\newcommand{\SLNR}{\text{SL}_{\R}(N)}
\newcommand{\slNR}{\mf{sl}_{\R}(N)}
\newcommand{\ft}{\mathfrak{t}}
\newcommand{\C}{\mathbb{C}}
\newcommand{\Z}{\mathbb{Z}}
\newcommand{\R}{\mathbb{R}}

\newcommand{\mc}[1]{\mathcal{#1}}
\newcommand{\mf}[1]{\mathfrak{#1}}

\newcommand{\Ad}{\text{Ad}}

\newcommand{\Lc}{\mathcal{L}}
\newcommand{\Rc}{\mathcal{R}}
\newcommand{\vp}{\varphi}
\newcommand{\p}{\partial}
\newcommand{\CP}{\mathbb{CP}^1}
\newcommand{\Ph}{\widehat{\mathbb{P}}}
\newcommand{\Zh}{\widehat{\mathbb{Z}}}
\newcommand{\g}{\mathfrak{g}}
\newcommand{\Ac}{\mathcal{A}}
\newcommand{\ri}{{\rm i}}
\newcommand{\s}{\sigma}
\newcommand{\Ab}{\mathbb{A}}
\newcommand{\gh}{\widehat{g}}
\newcommand{\zt}{\widetilde{z}}
\newcommand{\vpt}{\widetilde{\varphi}}
\newcommand{\Dc}{\mathcal{D}}
\newcommand{\Id}{\mathbb{I}}
\newcommand{\kf}{\mathfrak{k}}
\newcommand{\Tc}{\mathcal{T}}
\newcommand{\tf}{\mathfrak{t}}
\newcommand{\res}{\operatorname{res}}
\newcommand{\m}{\bm{m}}
\newcommand{\e}[2]{e^{#1}_{\phantom{#1}#2}}

\DeclareFontEncoding{LS1}{}{}
\DeclareFontSubstitution{LS1}{stix}{m}{n}
\DeclareSymbolFont{stixsymbols}{LS1}{stixscr}{m}{n}
\SetSymbolFont{stixsymbols}{bold}{LS1}{stixscr}{b}{n}
\DeclareMathSymbol{\kay}{\mathalpha}{stixsymbols}{"6B}

\makeatletter
\let\@keywords\@empty
\let\@subject\@empty
\providecommand{\keywords}[1]{\gdef\@keywords{#1}}
\providecommand{\subject}[1]{\gdef\@subject{#1}}
\def\thetitle{\@title}
\def\theauthor{\@author}
\def\thesubject{\@subject}
\def\thedate{\@date}
\def\thekeywords{\@keywords}
\makeatother
\AtBeginDocument{
\hypersetup{pdftitle={\thetitle}}
\hypersetup{pdfauthor={\theauthor}}
\hypersetup{pdfsubject={\thesubject}}
\hypersetup{pdfkeywords={\thekeywords}}}

\title{Geometry of the spectral parameter and renormalisation of integrable sigma-models}

\author{Sylvain Lacroix and Anders Wallberg}

\newcommand{\tightoverset}[2]{%
  \mathop{#2}\limits^{\vbox to -.5ex{\kern-0.75ex\hbox{$#1$}\vss}}}

\begin{document}

\begin{titlepage}

\begin{flushright}
{\ttfamily CERN-TH-2024-008}
\end{flushright}

\begin{center}

\vspace*{2cm}

\begingroup\Large\bfseries
Geometry of the spectral parameter and \\ renormalisation of integrable $\bm\s$-models
\par\endgroup

\vspace{1.cm}

\begingroup
Sylvain Lacroix$^{a,}$\footnote{E-mail:~sylvain.lacroix@eth-its.ethz.ch} and
Anders Wallberg$^{b,c,d,}$\footnote{E-mail:~anders.heide.wallberg@cern.ch}
\endgroup

\vspace{1cm}

\begingroup
$^a$\it Institute for Theoretical Studies, ETH Z\"urich, \\ Clausiusstrasse 47, 8092 Z\"urich, Switzerland\\~\\

$^b$\it  Department of Theoretical Physics, CERN,\\

1211 Meyrin, Switzerland\\~\\

$^c$\it  Laboratory for Theoretical Fundamental Physics, EPFL, \\

Rte de la Sorge, 1015 Lausanne, Switzerland\\~\\

$^d$\it  Institut f\"ur Theoretische Physik, ETH Z\"urich,\\
Wolfgang-Pauli-Strasse 27, 8093 Z\"urich, Switzerland\\
\endgroup

\end{center}

\vspace{1cm}

\begin{abstract}
\noindent
In the past few years, the unifying frameworks of 4-dimensional Chern-Simons theory and affine Gaudin models have allowed for the systematic construction of a large family of integrable $\s$-models. These models depend on the data of a Riemann surface $C$ (here of genus 0 or 1) and of a meromorphic 1-form $\omega$ on $C$, which encodes the geometry of their spectral parameter and the analytic structure of their Lax connection. The main subject of this paper is the renormalisation of these theories and in particular two conjectures describing their 1-loop RG-flow in terms of the 1-form $\omega$. These conjectures were put forward in \cite{delduc_rg_2021} and \cite{derryberry_lax_2021} and were proven in a variety of cases. After extending the proposal of \cite{delduc_rg_2021} to the elliptic setup (with $C$ of genus 1), we establish the equivalence of these two conjectures and discuss some of their applications. Moreover, we check their veracity on an explicit example, namely an integrable elliptic deformation of the Principal Chiral Model on $\SLNR$.

\end{abstract}
\end{titlepage}

\tableofcontents

\thispagestyle{empty}
\clearpage

\setcounter{page}{1}
\setcounter{footnote}{0} 

\section{Introduction}\label{sec:Intro}

Integrable $\s$-models form an important class of two-dimensional field theories, whose highly-symmetric nature allows for the development of various exact methods in the computation of their physical observables. They find applications in high-energy physics, string theory and condensed matter and as toy models for deepening our understanding of field theories. The construction and study of these models have been a growing topic of research for now more than 40 years. More recently, two unifying approaches of this domain have been developed, based on the formalisms of \textit{affine Gaudin models}~\cite{levin_hitchin_2003,feigin_quantization_2009,vicedo_integrable_2019,delduc_assembling_2019} and the \textit{semi-holomorphic 4-dimensional Chern-Simons theory}~\cite{costello_gauge_2019}. These frameworks, deeply related to one another~\cite{vicedo_4d_2021, levin_2d_2022}, have been shown to encompass most known classical integrable $\s$-models and also allowed the systematic construction of many new examples (see for instance the reviews~\cite{lacroix_4-dimensional_2022,Lacroix:2023gig} and the references therein).

Beyond the organising principles that they offer at the classical level, a natural question is whether these frameworks also provide new interesting insights on the quantum properties of integrable $\s$-models, for instance their renormalisation. It is a long-standing conjecture (see \textit{e.g.}~\cite{Fateev:1992tk,Fateev:1996ea,Lukyanov:2012zt}) that the large amount of symmetries characterising these models also ensures their renormalisability, at least at 1-loop. In that case, the continuous parameters entering the action start depending on the renormalisation energy scale $\mu$. This dependence is described by the Renormalisation Group (RG) flow, in the form of a differential equation with respect to $\mu$. In the case of integrable $\s$-models coming from the 4D Chern-Simons theory or affine Gaudin models, two conjectures have been put forward in~\cite{delduc_rg_2021,derryberry_lax_2021} (and partially proven in~\cite{hassler_rg_2021,Hassler:2023xwn}), proposing an explicit and universal formulation of the 1-loop RG-flow. The main goal of this paper is to review and, when needed, extend these conjectures, show that they are equivalent and check their veracity on a recently constructed novel example~\cite{Lacroix:2023qlz} (namely an elliptic integrable deformation of the Principal Chiral Model on $\SLNR$).\\

Let us now describe the main topics and results of the paper in a bit more detail. The classical integrability of $\s$-models is most often established using the notion of a \textit{Lax connection}. This is a connection on the 2-dimensional worldsheet of the model, whose light-cone components are denoted $\Lc_\pm(z)$, valued in a complexified Lie algebra $\g^\C$ and depending meromorphically on an auxiliary complex variable $z$ called the \textit{spectral parameter}. The Lax-connection is built from the fields of the $\s$-model in such a way that the equations of motion of these fields are equivalent to the flatness of $\Lc_\pm(z)$, for all values of $z$. This property ensures the existence of an infinite number of conserved quantities, which is a key feature of integrable field theories.

The formalisms of affine Gaudin models and 4-dimensional Chern-Simons theory allow for the systematic construction of a large class of $\s$-models whose equations of motion automatically admit a Lax representation. These integrable $\s$-models are built from certain defining ingredients. The first one is simply the choice of the Lie algebra $\g$ in which the Lax connection $\Lc_\pm(z)$ is valued (or more precisely its complexification). The next three ingredients $(C,\omega,\Zh^\pm)$ form what we will call the \textit{geometric data}: they will play a crucial role in this paper and are deeply related to the analytic structure of $\Lc_\pm(z)$ as a function of the spectral parameter $z$. The first element in this data is the choice of a compact Riemann surface $C$, which prescribes the space in which $z$ is valued. In this article, this surface will either be the Riemann sphere $\CP = \C \sqcup \lbrace \infty \rbrace$ (of genus 0), in which case the Lax connection $\Lc_\pm(z)$ is a rational function of $z$; or a torus (of genus 1), in which case $\Lc_\pm(z)$ is an elliptic function of $z$. The corresponding integrable $\s$-models will be respectively called rational and elliptic, according to this distinction.

The second component of the geometric data is a meromorphic 1-form $\omega = \vp(z)\,\dd z$ on $C$, which we will call the \textit{twist 1-form} and which we suppose has en even number $2M$ of zeroes, all simple. Finally, the last component is a splitting of these zeroes into two equal-size subsets $\Zh^\pm = \lbrace \ha{z}^{\,\pm}_1, \dots, \ha{z}^{\,\pm}_M \rbrace$. The points $\Zh^\pm \subset C$ exactly correspond to the poles of the light-cone component $\Lc_\pm(z)$ of the Lax connection and are thus quite natural characteristics of the integrable structure of the model. To summarise the results of the formalisms, the choice of the geometric data $(C,\omega,\Zh^\pm)$ completely fixes the form of $\Lc_\pm(z)$ as a meromorphic function of $z$. As an illustration, let us consider the rational setup $C=\CP$: in this case, the Lax connection is given by
\begin{equation}\label{eq:LaxIntro}
    \Lc_\pm(z) = \sum_{r=1}^M \frac{U_{\pm,r}}{z-\ha{z}^{\,\pm}_r} + U_{\pm,0}\,,
\end{equation}
for some Lie algebra-valued currents $U_{\pm,r}$ $(r=0,\dots,M)$. When $C$ is a torus, a similar expression holds, with the simple fractions $1/(z-\ha{z}^{\,\pm}_r)$ replaced by appropriately chosen elliptic functions (see section \ref{sec:IntSigma} of the main text for details).

Besides its zeroes, whose interpretation we just described, the poles of the twist 1-form $\omega$ also play a natural role in the construction of the integrable theory: the fundamental fields of the  $\s$-model are canonically associated with the poles of $\omega$. The last steps in the construction are then to relate these fundamental fields with the currents $U_{\pm,r}$ appearing in the Lax connection \eqref{eq:LaxIntro} and to specify the action of the $\s$-model in terms of these fields, in such a way that the induced equations of motion can be recast as the flatness of this Lax connection. The concrete way these various points are implemented will essentially not play any role in this article and will thus be only briefly glimpsed upon in the main text.\\

All of the above discussion was at the classical level: we now turn to quantum aspects. For a general $\s$-model, it is well known that the 1-loop renormalisation takes the form of a generalised Ricci flow of the metric and B-field~\cite{Ecker:1972bm,Honerkamp:1971sh,Friedan:1980jf,Curtright:1984dz}. For the integrable models introduced above, the metric and the B-field depend on a finite number of continuous parameters, essentially encoded in the geometric data $(C,\omega,\Zh^\pm)$\footnote{In addition to the ones encoded in $(C,\omega,\Zh^\pm)$, the models can also depend on some additional parameters which appear in the construction of the fundamental fields and their relation to the Lax connection. These aspects were skipped in the summary of the classical above and they will not play any role in our discussion of the 1-loop renormalisation. Essentially, the corresponding additional parameters completely decouple from the RG-flow and stay independent of the renormalisation scale $\mu$. Thus, we will focus here on the geometric data $(C,\omega,\Zh^\pm)$.}, e.g. the positions of the poles and zeroes of $\omega$ and the periods of the torus $C$ in the elliptic case. The 1-loop renormalisability of these models is then equivalent to saying that the Ricci flow can be reabsorbed as a running of these parameters with respect to the renormalisation scale $\mu$. It was conjectured that this is always the case and two explicit formulations of the 1-loop RG-flow of the geometric data $(C,\omega,\Zh^\pm)$ were proposed in~\cite{delduc_rg_2021,derryberry_lax_2021}.

The proposal of~\cite{delduc_rg_2021} directly describes the flow of the twist 1-form $\omega=\vp(z)\,\dd z$ and takes the form
\begin{equation}\label{eq:dPsiIntro}
    \frac{\dd\;}{\dd \tf} \omega = \p_z\Psi(z)\,\dd z\,,
\end{equation}
where $\tf = \frac{1}{4\pi}\log(\mu)$ and $\Psi(z)$ is a specific meromorphic function of the spectral parameter, which is explicitly built from the geometric data $(C,\omega,\Zh^\pm)$ -- see section \ref{sec:GeoRG} of the main text for details. We will refer to this formula as the \textit{$\p\Psi$-conjecture}. It was first put forward in~\cite{delduc_rg_2021} for certain rational models (\textit{i.e.} with $C=\CP$), whose twist 1-form $\omega$ has a double pole at infinity: in that case, it was recently proven in the works~\cite{hassler_rg_2021,Hassler:2023xwn}. The extension to an arbitrary pole structure at infinity was proposed in~\cite{Kotousov:2022azm} and checked in various examples, although a proof of this general case is not available at the moment. Its generalisation to elliptic models (for which $C$ is a torus) has not been studied yet in the literature and will be the first new result of the present paper. We conjecture that in the elliptic case, the flow takes the exact same form \eqref{eq:dPsiIntro} as in the rational setup, but with a different definition of the function $\Psi(z)$, now involving certain Weierstrass quasi-elliptic functions.

The main interest of the $\p\Psi$-conjecture \eqref{eq:dPsiIntro} is that it gives an explicit and direct formula for the RG-flow of the twist 1-form $\omega$. Moreover, as we will explain in the main text, its structure makes it easy to extract the corresponding RG-flow of various natural parameters encoded in $\omega$, for example its poles and zeroes. However, its main disadvantage is that this RG-flow generally takes a rather complicated form as a system of coupled non-linear differential equations. In particular, this formulation does not make it apparent how to solve the renormalization flow.

This is in contrast with the second conjecture that we will discuss in this paper, which was proposed by Costello and reported in the work~\cite{derryberry_lax_2021} of Derryberry. It is formulated in terms of the periods of $\omega$, \textit{i.e.} its integrals over well-chosen paths in $C$. Explicitly, it reads
\begin{equation}
		\frac{\dd}{\dd \mathfrak{t}}\Bigg[\oint\omega\Bigg]=0\,, \qquad
		\frac{\dd}{\dd \mathfrak{t}}\Bigg[\int_{\hat{z}^\pm_r}^{\hat{z}^\pm_s}\omega\Bigg]=0\,, \qquad
		\frac{\dd}{\dd \mathfrak{t}}\Bigg[\int_{\hat{z}^+_r}^{\hat{z}^-_s}\omega\Bigg]=2\hbar\,c_{\hg}\,, \label{eq:RGperiodsIntro}
  \end{equation}
with $c_{\g}$ the dual Coxeter number of the Lie algebra $\g$ and where we recall that the points $\ha{z}^{\,\pm}_r$ are the zeroes of $\omega$, which are split into the two subsets $\Zh^\pm = \lbrace \ha{z}^{\,\pm}_r \rbrace_{r=1}^M$. We will call this formula the \textit{period-conjecture}. The paths of integration considered in the first equation of \eqref{eq:RGperiodsIntro} are closed contours in $C$: the corresponding integrals are called the \textit{absolute periods} of $\omega$ and are then RG-invariants according to the conjecture. Typical examples of such absolute periods are the residues of $\omega$ at its poles, obtained by taking the contours as small circles encircling these points. These residues form all the absolute periods in the rational case, while they have to be supplemented with two additional quantities in the elliptic case, namely the integrals of $\omega$ over the A- and B-cycles of the torus. The second and third equations of the flow \eqref{eq:RGperiodsIntro} concern the so-called \textit{relative periods}, which are integrals of $\omega$ between two of its zeroes. The conjecture states that such a relative period is an RG-invariant if the zeroes belong to the same subset $\Zh^+$ or $\Zh^-$, while it grows linearly with the RG-parameter $\tf$ when the zeroes belong to different subsets. The data of the absolute and relative periods is essentially equivalent to that of $(C,\omega)$: the formula \eqref{eq:RGperiodsIntro} thus completely encodes the RG-flow of the twist 1-form, albeit in a more implicit way than the $\p\Psi$-conjecture \eqref{eq:dPsiIntro}. The main interest of the period-formulation is that, contrarily to the $\p\Psi$-conjecture and quite remarkably, the resulting RG-flow is trivially solved. In particular, this shows that the periods of $\omega$ are quite natural objects to discuss some of the quantum properties of integrable $\s$-models. The second main result of the present paper is the proof that the two conjectures \eqref{eq:dPsiIntro} and \eqref{eq:RGperiodsIntro} are equivalent to one another. This was suggested in~\cite{derryberry_lax_2021} for a specific class of models and will be shown in full generality here, relying on some of the special properties of the function $\Psi(z)$.

Finally, to see an explicit example of these conjectures in action, we will prove the 1-loop renormalisability of a specific elliptic integrable $\s$-model introduced recently in~\cite{Lacroix:2023qlz}, which corresponds to the simplest choice of twist 1-form $\omega$ on the torus, with one double pole and two simple zeroes, and which takes the form of a deformation of the Principal Chiral Model on $\SLNR$. In particular, we will derive the explicit 1-loop RG-flow of this model using Ricci-techniques and will use this result to check the validity of the $\p\Psi$- and period-conjectures for this example (hence providing their first test in the elliptic setup).\\

The plan of the paper is as follows. In section \ref{sec:IntSigma} we will review the class of integrable $\s$-models arising from 4D Chern-Simons theory and affine Gaudin models, although we will not explain their origin explicitly. We will put a special emphasis on the geometric data $(C,\omega,\Zh^\pm)$ appearing in their definition. The 1-loop RG-flow of this data will be the subject of section \ref{sec:GeoRG}: in particular, we will discuss in detail the $\p\Psi$- and period-conjectures and will prove their equivalence. Section \ref{sec:DPCM} will be devoted to the explicit 1-loop RG-flow of the elliptic integrable deformed Principal Chiral Model and the check of the conjectures in this example. Finally, we will conclude and discuss various perspectives in section \ref{sec:Conc}. Some technical computations as well as various reviews and reminders will be gathered in appendices \ref{app:Belavin} to \ref{sec:RGAppendix}.

\section{Integrable \texorpdfstring{$\bm\sigma$}{sigma}-models with twist 1-form}
\label{sec:IntSigma}

In this section, we review the properties of a large family of classical integrable $\sigma$-models which will be the main subjects of interest of this article. These theories are naturally described in terms of some geometrical data related to their spectral parameter, including the choice of what we will call the twist 1-form, which will play a crucial role throughout the paper. We will be interested in two variants of these theories, which we will refer to as \textit{rational} and \textit{elliptic} models, respectively. We will treat these two cases in parallel. In addition to summarising the main characteristics of these models, we will sketch the interpretation of the twist 1-form in their Hamiltonian formulation and will briefly explain their origin from semi-holomorphic 4D Chern-Simons (4D-CS) theory in the accompanying appendix \ref{app:4dCS}. We start by reviewing some basic facts about integrable $\s$-models.

\subsection[The basics of integrable \texorpdfstring{$\sigma$}{sigma}-models]{The basics of integrable \texorpdfstring{$\bm\sigma$}{sigma}-models}
\label{subsec:Basics}

\paragraph{Worldsheet and target space.} A $\s$-model is a 2-dimensional field theory defined on a space-time manifold $\Sigma$, which we call the \textit{worldsheet} and which we describe in terms of two coordinates $(t,x)$. We choose $\Sigma$ to be either the plane, in which case the spatial coordinate $x$ is on the real line $\R$, or a cylinder, in which case $x$ is subject to the periodicity condition $x \sim x+2\pi$. In this article, we will consider only relativistic $\s$-models, for which $\Sigma$ is equipped with a flat Lorentzian metric. In particular, it will be convenient to work with light-cone coordinates and their derivatives:
\begin{equation}
    x^\pm = t \pm x \qquad \text{ and } \qquad \p_\pm = \frac{1}{2}(\p_t \pm \p_x)\,.  
\end{equation}
The fundamental dynamical degree of freedom of the $\sigma$-model is a field
\begin{equation}
    \phi : \Sigma \longrightarrow \Tc\,,
\end{equation}
valued in a (pseudo)-Riemannian manifold $\Tc$, which we call the \textit{target space}. This space is equipped with a metric $G$, \textit{i.e.} a symmetric non-degenerate 2-tensor on $\Tc$, and a B-field $B$, \textit{i.e.} a skew-symmetric 2-form on $\Tc$.  We will denote by $d=\dim\Tc$ its dimension. The data $(\Tc,G,B)$, which defines the geometry of the target space, completely characterises the $\s$-model.

\paragraph{Action and equations of motion.} To describe this model more explicitly, let us consider a set of (local) coordinates on the target space $\Tc$. In this chart, the $\Tc$-valued field $\phi(x^+,x^-)$ can equivalently be seen as a collection of $d$ scalar fields $\lbrace \phi^{\,i}(x^+,x^-) \rbrace_{i=1}^d$, while the metric and B-field can be seen as symmetric and skew-symmetric tensors $G_{ij}(\phi)=G_{ji}(\phi)$ and $B_{ij}(\phi)=-B_{ji}(\phi)$ respectively. The $\s$-model is then defined by the action
\begin{equation}\label{eq:action}
    S[\phi] = \iint_\Sigma \dd x^+\,\dd x^- \,\bigl( G_{ij}(\phi) + B_{ij}(\phi) \bigr)\,\p_+\phi^{\,i}\,\p_-\phi^{\,j}\,.
\end{equation}
The dynamic of the theory is governed by the equations of motion derived from varying this action. These equations take the form
\begin{equation}\label{eq:EoMSigma}
    \p_+\p_- \phi^{\,i} + (\Gamma^{i}_{\phantom{i}jk} - H^{i}_{\phantom{i}jk} \bigr)\, \p_+ \phi^{\,j}\,\p_-\phi^{k} = 0\,,
\end{equation}
where $\Gamma^{i}_{\phantom{i}jk}$ are the Christoffel symbols of the metric $G_{ij}$ and $H=\dd B$ is the torsion 3-form associated with the B-field.

\paragraph{Lax connection and $\bm\Rc$-matrix.} In this paper, we will focus on \textit{integrable} $\sigma$-models, which correspond to very particular choices of target space $(\Tc,G,B)$, for which the dynamic \eqref{eq:EoMSigma} admits an infinite number of symmetries / conserved quantities. This property of integrability is often defined through the existence of a so-called \textit{Lax connection}. The latter is formed by two light-cone components $\Lc_\pm(z\,;x^+,x^-)$, which are matrices valued in a complexified Lie algebra $\g^\C$, built from the fields $\phi^{\,i}(x^+,x^-)$ and depending meromorphically on an auxiliary complex parameter $z$, which we call the \textit{spectral parameter}. The definition of this connection in terms of the fields $\phi^{\,i}(x^+,x^-)$ should be such that the equations of motion \eqref{eq:EoMSigma} are equivalent to the flatness condition
\begin{equation}\label{eq:flatness}
    \p_+ \Lc_-(z) - \p_- \Lc_+(z) + \bigl[ \Lc_+(z), \Lc_-(z) \bigr] = 0\,, \qquad \forall\,z\,.
\end{equation}
This property ensures the existence of an infinite number of conserved charges, built from the monodromy of the Lax matrix $\Lc_x(z) = \Lc_+(z) - \Lc_-(z)$. Indeed, the trace of this monodromy is conserved along the time evolution $\p_t=\p_+ + \p_-$ due to the flatness condition \eqref{eq:flatness} on $\Lc(z)$. Moreover, since the latter holds for all values of the spectral parameter $z$, one can extract an infinite number of conserved charges from this construction, for instance by considering a power series expansion in $z$. This illustrates the crucial role played by the spectral parameter in the integrable structure of the theory.

The integrability of the model also requires these conserved quantities to be pairwise Poisson-commuting in the Hamiltonian formulation. This question is ultimately related to the Poisson algebra obeyed by the various components of two Lax matrices $\Lc_x(z_1)$ and $\Lc_x(z_2)$, evaluated at different values $z_1,z_2$ of the spectral parameter. In~\cite{maillet_kac-moody_1985,maillet_new_1986}, Maillet proposed a sufficient condition on this Poisson algebra which ensures the Poisson-commutation of the monodromy charges and thus that the model is integrable in the Hamiltonian sense. For brevity, we will not exhibit the explicit form of this Maillet bracket here and refer for instance to~\cite[Section 2.1]{Lacroix:2023qlz} for a detailed review. The main new ingredient appearing in this bracket is the so-called \textit{$\Rc$-matrix}, which is an element $\Rc(z_1,z_2)$ of the tensor product $\g^\C \otimes \g^\C$, depending meromorphically on the two spectral parameters $z_1,z_2$ and satisfying the classical Yang-Baxter equation. This $\Rc$-matrix depends on the model under consideration and characterises its Hamiltonian integrable structure.

\paragraph{Form of the Lax connection.} Let us now discuss in more detail the form of the Lax connection $\Lc_\pm(z)$. We introduce a basis $\lbrace T_\alpha \rbrace$ of the Lie algebra $\g^\C$. As mentioned above, $\Lc_\pm(z)$ is valued in this algebra and is built from the fields $\phi^{\,i}(x^+,x^-)$. In all the known examples of integrable $\s$-models, it is more precisely linear in the derivatives $\p_\pm\phi^{\,i}$ of these fields and take the form
\begin{equation}
    \Lc_\pm(z) = \Lambda_{(\pm),\,i\,}^{\,\alpha}(z,\phi)\,\p_\pm\phi^{\,i}\;T_\alpha\,,
\end{equation}
where a summation over the repeated indices $i,\alpha$ is implicitly assumed. Here, the coefficients $\Lambda_{(\pm),\,i\,}^{\,\alpha}(z,\phi)$ are meromorphic functions of the spectral parameter $z$ which also depend on the fields $\phi^{\,i}$, but not their derivatives. This dependence can generally be quite complicated and reflects the non-linearity of the $\s$-model dynamics \eqref{eq:EoMSigma}, as encoded in the flatness of $\Lc_\pm(z)$.\\

Of particular interest for our discussion will be the analytic structure of $\Lc_\pm(z)$ -- or equivalently $\Lambda_{(\pm),\,i\,}^{\,\alpha}(z,\phi)$ -- as a function of the spectral parameter. For instance, we will consider two main classes of integrable $\s$-models, depending on the nature of this $z$-dependence: the \textit{rational} ones, for which $\Lc_\pm(z)$ is a rational function of $z$, and the \textit{elliptic} ones, for which $\Lc_\pm(z)$ is an elliptic function of $z$ (which can be seen as a doubly-periodic function on the complex plane).

Another key characteristic of the analytic structure of $\Lc_\pm(z)$ is the data of its poles and of their multiplicities: in this article, we will focus on the case of first-order poles for simplicity and will denote by $\Zh^\pm = \lbrace \ha{z}^{\,\pm}_1,\dots,\ha{z}^{\,\pm}_M \rbrace$ their positions\footnote{In the rational case, there are by definition finitely many such poles. In the elliptic one, there is technically an infinite 2d-lattice of poles due to the periodicity properties of $\Lc_\pm(z)$: schematically, $\Zh^\pm$ is then defined as a set of independent representatives in this lattice and is therefore finite (we refer to subsection \ref{subsec:SigmaModel} for a more precise statement, where we will see that we need to take into account a finer property of quasi-periodicity of $\Lc_\pm(z)$). Note that we expect $\Zh^+$ and $\Zh^-$ to be of the same size, which we call $M$, due to the symmetric treatment of the two light-cone directions in relativistic $\s$-models.}. In the models that we will consider, the poles $\Zh^+$ of $\Lc_+(z)$ will always be pairwise distinct from the ones $\Zh^-$ of $\Lc_-(z)$\footnote{This can be justified heuristically as follows: The existence of a common pole in $\Lc_+(z)$ and $\Lc_-(z)$ would generally create a second order singularity in the last term of the flatness equation \eqref{eq:flatness}, which cannot be cancelled by the first two terms using only the equations of motion \eqref{eq:EoMSigma}.}.

\subsection{Towards a general construction}

In the previous subsection, we reviewed the basics of integrable $\s$-models. We note that the condition for integrability, namely the existence of a Lax connection $\Lc_\pm(z)$ satisfying a Maillet bracket, is very constraining and thus requires an extremely fine-tuned choice of target space $(\Tc, G,B)$. In other words, integrable $\s$-models are very rare occurrences in the space of all $\s$-models. It is thus natural to ask the following questions:\vspace{-2pt}
\begin{itemize}\setlength\itemsep{2pt}
    \item Is there a systematic way to find such integrable $\s$-models?
    \item What are the natural parameters entering the definition of their target spaces $(\Tc,G,B)$?\vspace{-2pt}
\end{itemize}
In recent years, our understanding of these questions have improved greatly by realising the integrable $\s$-models using two formalisms, called the \textit{affine Gaudin models} (AGMs)~\cite{levin_hitchin_2003,feigin_quantization_2009,vicedo_integrable_2019,delduc_assembling_2019} and the \textit{4-dimensional Chern-Simons (4D-CS) theory}~\cite{costello_gauge_2019}, which are deeply related to one another~\cite{vicedo_4d_2021, levin_2d_2022}. The rest of this section is devoted to reviewing the integrable $\s$-models obtained through these unifying frameworks.\\

As a motivation, let us first come back to the second question asked above, namely how to describe the parameters that define an integrable $\s$-model.  As we saw earlier, integrability is related to the existence of a Lax connection $\mc{L}(z)$. We will initially focus on the dependence of this connection on the spectral parameter $z$ and the geometric structure underlying it. For instance, recall that we distinguished the classes of rational and elliptic models, depending on the nature of $\Lc_\pm(z)$ as a function of $z$. This information (including the value of the periods of $\Lc_\pm(z)$ in the elliptic case) is part of the geometric data defining the model and will formally be encoded in the choice of a Riemann surface $C$, of genus 0 or 1. Other natural parameters are the poles $\Zh^\pm$ of the Lax connection $\Lc_\pm(z)$: by construction, they appear in the equations of motion of the theory and thus should enter the choice of the target space in some way. Geometrically, they form two sets $\Zh^\pm$ of marked points in the Riemann surface $C$. As we will see in the rest of this section, these points appear as part of a richer geometric structure in the AGMs/4d-CS framework, which we will call the \textit{twist 1-form} $\omega$. Namely, $\omega$ will be a meromorphic 1-form on $C$ with zeroes exactly at the points $\Zh^\pm$. The other parameters encoded in $\omega$, for instance its poles, will also play an important role in the definition of the model and its integrable structure. The goal of subsection \ref{subsec:GeoTwist} is to define precisely and in detail this geometric data, which will then serve as the starting point for the rest of the construction. Moreover, this will introduce the main notions and terminologies needed to formulate the results of section \ref{sec:GeoRG}.\\

Beyond the geometry of the spectral parameter, the $\s$-model will also depend on some algebraic data. Most notably, the Lax-connection takes its values in (the complexification of) a certain Lie algebra $\hg$: the choice of this algebra then has to appear in the construction of the $\s$-model in some way. As we will see later in subsection \ref{subsec:SigmaModel}, this choice, together with other algebraic structures, will also be part of the defining data of the model and will play an important role in its description. 

\subsection{Geometry of the spectral parameter and twist 1-form}
\label{subsec:GeoTwist}

\paragraph{The Riemann surface $\bm{C}$.} The starting point of our construction is the choice of a compact \textit{Riemann surface} $C$ of genus $g=0$ or $g=1$.
\begin{itemize}
    \item For the rational models, we take this surface to be the Riemann sphere:
    \begin{equation}
       \text{Rational (}g=0\text{):}   \qquad C=\CP\,.\vspace{-4pt}
    \end{equation}
    \item For the elliptic models, we take this surface to be a complex torus with half-periods $\bm\ell=(\ell_1,\ell_2)$:
    \begin{equation}
        \text{Elliptic (}g=1\text{):} \qquad  C=\C/\Gamma \qquad \text{ with }\qquad \Gamma=2\Z\,\ell_1 \oplus 2\Z\,\ell_2
    \end{equation}
    the 2-dimensional period lattice.
\end{itemize}
In what follows, we will describe $C$ using a local complex coordinate $z$. In the rational case, the Riemann sphere $C=\CP$ corresponds to the complex $z$-plane to which we add the point $z=\infty$. In the elliptic case, the torus $C=\C/\Gamma$ corresponds to the complex $z$-plane quotiented by the identification $z \sim z+2n_1\ell_1+2n_2\ell_2$, $n_1,n_2\in\Z$. As the notation suggests, we will later identify this complex variable $z$ with the \textit{spectral parameter} of the underlying integrable $\sigma$-model.

\paragraph{The twist 1-form $\bm\omega$.} The next ingredient needed for the construction is a meromorphic 1-form $\omega$ on the Riemann surface $C$, which we will call the \textit{twist 1-form}. We will write it in terms of the local coordinate $z$ as
\begin{equation}
    \omega = \vp(z)\,\dd z\,.
\end{equation}
The object $\vp(z)$ appearing in this expression is then often referred to as the twist function\footnote{This is the standard terminology used in the literature, which takes its origins in the Hamiltonian formulation of these integrable models (cf. the brief review in subsection \ref{subsec:SigmaModel}). We stress however that geometrically, $\omega=\vp(z)\,\dd z$ behaves as a 1-form on $C$ under a change of the coordinate $z$ (see later in this subsection). This is the reason why we introduced the terminology ``twist 1-form'' for $\omega$, to recall the geometric nature of this object.}:
\begin{itemize}
    \item For $C=\CP$, it is a rational function of $z$.
    \item For $C=\C/\Gamma$, it is an elliptic function of $z$, \textit{i.e.} it satisfies the double-periodicity condition\vspace{-3pt}
    \begin{equation}\label{eq:PhiEll}
        \vp(z+2\ell_i) = \vp(z)\,.
    \end{equation}
\end{itemize}
We will denote by 
\begin{equation}
    \Ph = \lbrace \ha{p}_1,\dots, \ha{p}_{n} \rbrace
\end{equation}
the set of poles of $\omega$ on $C$ and by $\m=(m_1,\dots,m_n)\in\Z_{\geq 1}^n$ their multiplicities. Moreover, we will suppose that $\omega$ only has simple zeroes and that there are an even number $2M$ of them. We will separate these zeroes into two subsets\vspace{-2pt}
\begin{equation}\label{eq:zeroes}
    \Zh^\pm = \lbrace \ha{z}^{\,\pm}_1,\dots, \ha{z}^{\,\pm}_{M} \rbrace \vspace{-2pt}
\end{equation}
of equal size $M$. Note that the Riemann-Hurwitz formula implies\vspace{-2pt}
\begin{equation}
    2M = \sum_{r=1}^{n} m_r + 2(g-1)\,.
\end{equation}
In agreement with our previous notations, the points $\Zh^\pm$ will later be identified with the poles of the Lax connection of the underlying $\s$-model. As a warning, let us note that from now on, we will mostly refer to these points as zeroes of $\omega$, rather than poles of the Lax connection since $\omega$ will be the more fundamental object throughout the article. In particular, they should not be confused with the points $\Ph$, which are the poles of $\omega$ (but are neither poles nor zeros of the Lax connection).

\paragraph{Reality conditions.} To ensure that the integrable $\sigma$-model associated with the data $(C,\omega)$ is real, we will have to impose certain reality conditions on these objects. For instance, in the elliptic case, we will ask that the half-periods of the torus $\C/\Gamma$ satisfy\vspace{-2pt}
\begin{equation}
    \ell_1 \in \R_{> 0} \qquad \text{ and } \qquad \ell_2 \in \ri\, \R_{> 0} \,.\vspace{-2pt}
\end{equation}
Moreover, we will suppose in both cases that the zeroes $\ha{z}^{\,\pm}_r$ of $\omega$ are all real and that the poles $\ha{p}_r$ are either real or come in pairs of complex conjugate points. Together with an appropriate reality condition on its overall factor, these requirements translate to the following property of $\vp(z)$:\vspace{-2pt}
\begin{equation}
    \ol{\vp(z)} = \vp(\ol{z})\,.
\end{equation}

\paragraph{Change of coordinate and moduli space of abelian differentials.} In the previous paragraphs, we described $(C,\omega)$ using a complex coordinate $z$ and writing $\omega$ as $\vp(z)\,\dd z$. Let us now consider a change of this coordinate $z \mapsto \zt=h(z)$, where $h$ is a biholomorphism (\textit{i.e.} a holomorphic invertible map whose inverse is also holomorphic). Since $\omega$ is a 1-form, it can be expressed in terms of the new coordinate $\zt$ as
\begin{equation}\label{eq:ChangeOmega}
    \omega = \vpt(\zt)\,\dd \zt = \vp\bigl( h^{-1}(\zt) \bigr)\, \p_{\tilde{z}} \bigl(h^{-1}(\zt) \bigr)\, \dd \zt\,.
\end{equation}
The integrable model that we will introduce in the next subsections can be constructed using either of the coordinates $z$ or $\zt$ and will in the end be independent of this choice: the change $z\mapsto \zt$ will simply correspond to a redefinition of the spectral parameter of this theory, but will not affect its target space or its integrable structure. These will instead depend only on the intrinsic geometric structure of $(C,\omega)$: in the mathematical terminology, this corresponds to a point in the so-called \textit{moduli space of abelian differentials}\footnote{Since $\omega$ is allowed to have poles and non-zero residues, it is more precisely an abelian differential of the 3rd kind.}. To describe this space in more explicit terms, we will need to distinguish between the rational and the elliptic case.\\

We start with the rational case, for which $C=\CP$ is formed by the complex $z$-plane, to which we add a point at infinity. The allowed changes of coordinate are then the M\"obius transformations
\begin{equation}\label{eq:Mobius}
    z \longmapsto \zt = h(z) = \frac{az+b}{cz+d}\,,
\end{equation}
where $(a,b,c,d)$ are real\footnote{Here, we impose that $a,b,c,d$ are real in order to preserve the reality conditions discussed earlier.} 
 parameters (with $ad-bc \neq 0$ and considered up to rescalings $\lambda(a,b,c,d)$), forming the group $\text{PSL}_\R(2)$. 
For a given choice of pole multiplicities $\m=(m_1,\dots,m_n)$, one way of parameterising the moduli $(C,\omega)$ is then through the choice of a global proportionality factor and the positions of the poles and zeroes of $\omega$ with respect to the coordinate $z$, quotiented by the action of M\"obius transformations. The latter can for instance be fixed by setting any three of the poles/zeroes to specific values. This leaves a total of $2M+n-2$ free parameters.\\

We now turn to the elliptic case, for which $z$ was seen as a coordinate on the complex plane modulo the equivalence relation $z \sim z + 2\ell_i$. In this formulation, the allowed transformations of coordinates are the dilations and translations
\begin{equation}
    z \longmapsto \zt = h(z) = a z +b\,,
\end{equation}
with parameters $a\in\R_{> 0}$ and $b\in\R$. This new coordinate $\zt$ is then considered up to the equivalence relation $\zt \sim \zt + 2\widetilde{\ell}_i$, where $(\widetilde{\ell}_1,\widetilde{\ell}_2)=a(\ell_1,\ell_2)$. The transformations with $a\neq 1$ thus change the values of the half-periods $\ell_i$. However, the combination
\begin{equation}\label{eq:tau}
    \tau = \frac{\ell_2}{\ell_1}
\end{equation}
is left invariant and thus corresponds to an intrinsic parameter characterising the complex structure of $C$, called the modulus of the torus.

In this elliptic case and for a given choice of pole multiplicities $\m=(m_1,\dots,m_n)$, one can then describe the moduli $(C,\omega)$ by the choice of the half-periods $(\ell_1,\ell_2)$, a global factor and the positions of the poles and zeroes\footnote{Note that these positions are not completely arbitrary: they should be such that the sum of zeroes minus the sum of poles (with multiplicities) lies in the lattice $\Gamma$.}, quotiented by the dilations and translations. This leaves $2M+n$ free parameters. One standard way to fix the dilation freedom, often considered in the literature on elliptic functions, is to set the periods of the torus to be $1$ and $\tau$: one is then free to use the translation freedom to set one of the zeroes/poles to a specific value. Another possibility is to fix two zeroes/poles: in that case, the half-periods $(\ell_1,\ell_2)$ are then left as free parameters.

\paragraph{Moduli space of split abelian differentials.} Let us note that the geometric data considered earlier does not exactly correspond to points $(C,\omega)$ in the moduli space of abelian differentials. First of all, we restricted ourselves to the case where $\omega$ has an even number of zeroes which are all simple. Secondly, we introduced an additional structure, namely the splitting of these zeroes into two equal-size subsets $\Zh^\pm$ -- see equation \eqref{eq:zeroes} -- which will later correspond to a distribution of these points as poles of the two light-cone components $\Lc_\pm(z)$ of the Lax connection. Motivated by this observation, we introduce a slightly different but more adapted space of parameters $\Dc_{g,\m}$, which we will call the \textit{moduli space of split abelian differentials} and which depends on $g\in\lbrace 0,1 \rbrace$ and integers $\m=(m_1,\dots,m_n)\in \Z^n_{\geq 1}$ such that $\sum_r m_r$ is even. Explicitly, a point in $\Dc_{g,\m}$ is defined as the data $(C,\omega,\Zh^\pm)$ of:\vspace{-1pt}
\begin{itemize}\setlength\itemsep{2pt}
    \item a Riemann surface $C$ of genus $g=0$ or $g=1$ ;
    \item a meromorphic 1-form $\omega$ on $C$, which has $n$ poles with multiplicities $\m=(m_1,\dots,m_n)$ and an even number $2M=\sum_{r=1}^n m_r + 2(g-1)$ of simple zeroes ;
    \item the partition of these zeroes into two subsets $\Zh^\pm$ of size $M$.
\end{itemize}
Moreover, we ask that this data respects the reality conditions discussed earlier. The moduli space $\Dc_{g,\m}$ has (real) dimension $\sum_{r=1}^n m_r+n+4(g-1)$. In practice, it can be parameterised as in the previous paragraph for the data $(C,\omega)$ only, but additionally keeping track of the labels $\pm$ attached to the zeroes $\ha{z}^{\,\pm}_r \in \Zh^\pm$. This structure will play a crucial role throughout the paper. 

\subsection[The integrable \texorpdfstring{$\sigma$}{sigma}-model and its Lax connection]{The integrable \texorpdfstring{$\bm\sigma$}{sigma}-model and its Lax connection}
\label{subsec:SigmaModel}

 We will now describe the integrable $\sigma$-model built from the geometric data $(C,\omega,\Zh^\pm)\in \Dc_{g,\m}$ introduced in the previous subsection. In particular, we will explain which additional ingredients are needed for its construction and will discuss its Lax connection and $\Rc$-matrix. We note that the content of this subsection is technically not needed to formulate the main results of the paper in sections \ref{sec:GeoRG} and \ref{sec:DPCM}: the readers eager to hear about those or already familiar with the construction can thus skip this part. However, we still include this review as we feel it can provide some useful context and intuition for the rest of the paper.

\paragraph{Defining data.} Recall from subsection \ref{subsec:Basics} that a $\s$-model is characterised by its target space $\Tc$, with metric $G$ and B-field $B$. The integrable model under consideration corresponds to a very specific choice of $(\Tc,G,B)$, built from the following ingredients \vspace{-2pt}
\begin{itemize}
    \item the geometric data $(C,\omega,\ha{\Z}^\pm)\in \Dc_{g,\m}$ described in the previous subsection ; \vspace{-2pt}
    \item the choice of a Lie algebra $\g$, which can be any real simple Lie algebra in the rational case but is restricted to $\slNR$ in the elliptic one ; \vspace{-2pt}
    \item the choice of a maximally isotropic subalgebra $\mathfrak{k}$ of the so-called \textit{defect Lie algebra}.
\end{itemize}
 A complete description of this last point would require properly introducing the notion of defect Lie algebra. For the purposes of this paper, we will not need the details of this construction nor the general definition of $(\Tc,G,B)$ and will thus stay quite brief for simplicity. Suffice it to say, this defect Lie algebra is canonically built from the data of the simple Lie algebra $\g$ and the pole structure of $\omega$ and the construction of the integrable $\sigma$-model further requires the choice of a subalgebra $\mathfrak{k}$, satisfying some properties of isotropy and maximality\footnote{To give some intuition, let us consider the example where the poles $\hat{\mathbb{P}} = \lbrace \hat{p}_r \rbrace_{r=1}^n$ of $\omega$ are all real and simple. The defect algebra is then defined as the direct sum $\g^{\oplus n}$, equipped with an invariant bilinear form $(X,Y) \mapsto \sum_{r=1}^n \res_{\hat{p}_r} \omega\, \langle X_r,Y_r\rangle$, where $\langle\cdot,\cdot\rangle$ is the invariant pairing on $\g$. The additional data that is needed for the construction is then a subalgebra $\kf\subset \g^{\oplus n}$, which is maximally isotropic with respect to this bilinear form. Further explanations on the nature and role of the defect Lie algebra and its isotropic subalgebra can be found in the appendix \ref{app:4dCS}, where the origin of the integrable $\s$-model from 4D Chern-Simons theory is reviewed.}. Given some fixed $\omega$ and $\g$, there can in general exist several admissible choices for this $\mathfrak{k}$, which lead to different $\sigma$-models: however, these theories turn out be deeply related\footnote{In particular, this non-uniqueness underlies the phenomenon of Poisson-Lie T-dualities~\cite{Klimcik:1995ux,Klimcik:1995dy} between these integrable $\sigma$-models -- see~\cite{delduc_unifying_2020,lacroix_integrable_2021,Liniado:2023uoo}.}. Importantly, the aspects that we will discuss in this paper turn out to be independent of this choice and rely only on the data of $(C,\omega,\ha{\Z}^\pm)$ and $\g$.

Although we will not review the explicit construction of the target space geometry $(\Tc,G,B)$ from the data $(C,\omega,\ha{\Z}^\pm,\g,\mathfrak{k})$, we stress here that it is systematic and naturally follows from the unifying frameworks of 4D-CS theory and affine Gaudin models. In particular, the expression of the metric and B-field $(G,B)$ on $\Tc$ strongly depends on the data $(C,\omega,\ha{\Z}^\pm)\in \Dc_{g,\m}$ associated with the geometry of the spectral parameter (for instance the positions of the poles and zeroes of $\omega$ as well as the periods of the torus $C$ in the elliptic case). We will see an explicit illustration of this in subsection \ref{subsec:IntDPCM}, with the example of an elliptic integrable deformation of the Principal Chiral Model. We finally note that the dimension of the target space $\Tc$ is given by $d=M\,\dim\g$, where we recall that $M=|\ha{\Z}^\pm|$ is half the number of zeroes of $\omega$.

\paragraph{Lax connection.} We now describe the integrable structure of this model, in particular its Lax connection $\Lc_\pm(z)$, which is also built from the defining data $(C,\omega,\ha{\Z}^\pm,\g,\mathfrak{k})$. To start with, the choice of $\g$ simply prescribes the Lie algebra $\g^\C$ in which $\Lc_\pm(z)$ is valued. Furthermore, the analytic structure of $\Lc_\pm(z)$ strongly depends on the geometric data $(C,\omega,\ha{\Z}^\pm)\in\Dc_{g,\m}$. In agreement with the notations of subsection \ref{subsec:Basics}, $\Lc_\pm(z)$ has simple poles at the zeroes $\Zh^\pm = \lbrace \ha{z}^{\,\pm}_1,\dots, \ha{z}^{\,\pm}_{M} \rbrace$ of $\omega$:
\begin{equation}\label{eq:PolesLax}
    \Lc_\pm(z) = \frac{U_{\pm,r}}{z-\ha{z}^{\,\pm}_r} + O\bigl( (z-\ha{z}^{\,\pm}_r)^0 \bigr)\,,
\end{equation}
for some $\g$-valued currents $U_{\pm,r} : \Sigma \to \g$. In particular, this uses the separation of these zeroes into the two equal-size subsets $\Zh^\pm$ (see equation \eqref{eq:zeroes} and the surrounding discussion), which is ultimately related to the relativistic invariance of the theory. In the next two paragraphs, we will describe in more detail the structure of $\Lc_\pm(z)$ as a meromorphic function, distinguishing the rational and elliptic cases. In both of these, the zeroes $\ha{z}^{\,\pm}_r$ and the associated currents $U_{\pm,r}$ will turn out to encode all the information on the Lax connection. 

The poles of $\omega$, the associated defect Lie algebra and the corresponding choice of maximally isotropic subalgebra $\mathfrak{k}$ also enter the construction of the Lax connection $\Lc_\pm(z)$ and in particular its expression in terms of the fields $\phi^{\,i}$ of the $\sigma$-model. More precisely, this data prescribes a relation between these fields and the object $\bigl( \p_z^k \Lc_\pm(\ha{p}_r) \bigr)_{r=1,\dots,n}^{k=0,\dots,m_r-1}$, which is built from evaluations of the Lax connection and its derivatives at the poles of $\omega$ and is naturally interpreted as an element of the defect Lie algebra. Solving this relation, one finds that the currents $U_{\pm,r}$ appearing as residues of $\Lc_\pm(z)$ can be expressed as linear combinations of the derivatives $\p_\pm \phi^{\,i}$, with coefficients depending in a complicated way on the parameters of $\omega$ and the fields $\phi^{\,i}$, but not their derivatives.

To summarise, the fundamental fields $\lbrace \phi^{\,i}\rbrace_{i=1}^d$ of the $\s$-model are naturally attached to the poles of $\omega$, the Lax connection is conveniently expressed in terms of the currents $\lbrace U_{\pm,r} \rbrace_{r=1}^M$ associated with the zeroes and the two are related by some ``interpolation'' mechanism controlled by the choice of $\kf$. We note that the number of components contained in these currents (for a given chirality $\pm$) is equal to the dimension $d=M\,\dim\g$ of the target space: this reflects the fact that the flatness condition \eqref{eq:flatness} encodes all of the equations of motion \eqref{eq:EoMSigma} of the $\sigma$-model fields $\lbrace \phi^{\,i} \rbrace_{i=1}^d$.

\paragraph{The rational Lax connection.} For this paragraph, we consider the rational case $C=\CP$. In that setup, the Lax connection $\Lc_\pm(z)$ can be essentially characterised as the unique rational function of $z\in\CP$ with the singularities \eqref{eq:PolesLax} and no other poles. Namely, we have
\begin{equation}\label{eq:RationalLax}
    \Lc_\pm(z) = \sum_{r=1}^M \frac{U_{\pm,r}}{z-\ha{z}^{\,\pm}_r} + U_{\pm,0}\,,
\end{equation}
where the constant term $U_{\pm,0}$ is related\footnote{The precise form of this relation depends on the choice of $\omega$ and $\mathfrak{k}$ and will not be needed here.} to the other currents $\lbrace U_{\pm,r} \rbrace_{r=1}^{M}$ and thus should not be interpreted as a new independent ingredient in the construction. For simplicity, we supposed here that all the zeroes $\ha{z}^{\,\pm}_r$ of $\omega$ are located in the finite complex plane $\C\subset\CP$. If the point $z=\infty$ was a zero of $\omega$, one would get a polynomial term of degree 1 in the equation \eqref{eq:RationalLax} instead of a simple fraction. One can always go back to the case of finite zeroes by a M\"obius transformation of $z$: here and in what follows, we always suppose that we are in this situation, to simplify the presentation.

\paragraph{The elliptic Lax connection.} Let us now turn to the elliptic case $C=\C/\Gamma$, which is slightly more subtle. Recall from the beginning of this subsection that in this setup, the Lie algebra $\g^\C$ can only be chosen as $\slNC$. To proceed, we will need to pick an appropriate choice of basis of this algebra, which is called the \textit{Belavin basis} $\lbrace T_\alpha \rbrace_{\alpha\in\Ab}$. To avoid getting into technicalities, we gather the explicit definition and the main properties of this basis in the appendix \ref{app:Belavin} and only recall some basic facts here. The elements $T_\alpha$ of this basis are labelled by non-vanishing couples $\alpha=(\alpha_1,\alpha_2)$ of integers modulo $N$, \textit{i.e.} by the set $\Ab \equiv \Z_N \times \Z_N \setminus \lbrace (0,0) \rbrace$ (in particular $|\Ab|=N^2-1=\dim\slNC$, as expected). In addition to the Belavin basis, we will need to use a family of meromorphic functions $\lbrace r^\alpha(z) \rbrace_{\alpha\in\Ab}$ also labelled by $\Ab$, defined in terms of the Weierstrass $\s$-function and closely related to the so-called \textit{Kronecker function} associated with the torus $C=\C/\Gamma$. To unclutter the discussion, we also review the definition and the properties of these various functions in an appendix \ref{app:elliptic} and will call upon some of these results as we progress in the main text. Using these various ingredients, we can express the elliptic Lax connection as~\cite{Lacroix:2023qlz} 
\begin{equation}\label{eq:EllipticLax}
    \Lc_\pm(z) = \sum_{r=1}^M \sum_{\alpha\in\Ab} r^\alpha(z-\ha{z}^{\,\pm}_r)\, U^\alpha_{\pm,r}\,T_\alpha\,,
\end{equation}
where we decomposed the currents $U_{\pm,r}= \sum_\alpha U_{\pm,r}^\alpha\,T_\alpha$ along the Belavin basis. The first property of the functions $\lbrace r^\alpha(z) \rbrace$ that we shall need is their behaviour $r^\alpha(z) \sim \frac{1}{z}$ around $z=0$. In particular, this ensures that the above Lax connection has the required singularity \eqref{eq:PolesLax} at the zeroes $\ha{z}^{\,\pm}_r$ of $\omega$.\\

Another useful result about the functions $\lbrace r^\alpha(z) \rbrace_{\alpha\in\Ab}$ is the fact that they are not periodic under shifts of $z$ by $2\ell_i$, but rather satisfy the quasi-periodicity condition \eqref{eq:QuasiR}. Combined with the ``grading'' property \eqref{eq:GradingT} of the Belavin basis $\lbrace T_\alpha \rbrace_{\alpha\in\Ab}$, this translates to the following quasi-periodicity of the Lax connection:
\begin{equation}\label{eq:equivL}
    \Lc_\pm(z+2\ell_i) = \Ad_{\Xi_i}\,\Lc_\pm(z)\,.
\end{equation}
Here, $\Xi_1$ and $\Xi_2$ are specific $N\times N$ matrices defined in equation \eqref{eq:defofXi}: we refer to the appendix \ref{app:Belavin} for more details about them. The relation \eqref{eq:equivL} means that, technically, the Lax connection $\Lc_\pm(z)$ does not reduce to an elliptic function on the torus $C=\C/\Gamma$. However, an important property of the matrices $\Xi_i$ is their cyclicity of order $N$, \textit{i.e.} the fact that they satisfy  $\Xi_1^N = \Xi_2^N = \Id$. In particular, this means that although $\Lc_\pm(z)$ is only quasi-periodic under shifts of $z$ by $2\ell_i$, it is properly periodic under shifts by $2N\ell_i$:
\begin{equation}
    \Lc_\pm(z+2N\ell_i) = \Lc_\pm(z)\,.
\end{equation}
One can thus think of $\Lc_\pm(z)$ as an elliptic function on another torus $\C/\Lambda$, defined using the sublattice
\begin{equation}\label{eq:Lambda}
    \Lambda = 2N\ell_1\,\Z \oplus 2N\ell_2\,\Z\, \subset \Gamma\,.
\end{equation}

Let us note that the translations $z\mapsto z + 2n_1\ell_1 + 2n_2\ell_2$ define an action of the cyclic group $\Z_N \times \Z_N$ on the torus $\C/\Lambda$ and that the initial curve $C=\C/\Gamma$ can be then thought of as the quotient of $\C/\Lambda$ with respect to this action. The quasi-periodicity condition \eqref{eq:equivL} then states that the Lax connection $\Lc_\pm(z)$, seen as a meromorphic function on $\C/\Lambda$, is \textit{equivariant} with respect to the translational action of $\Z_N \times \Z_N$ on $\C/\Lambda$ and its action by the adjoint automorphisms $\Ad_{\Xi_i}$ on the Lie algebra $\slNC$ in which $\Lc_\pm$ is valued. In fact, one can argue that the right-hand side of \eqref{eq:EllipticLax} is the unique function with this equivariance property and the singularities \eqref{eq:PolesLax} (see~\cite{Lacroix:2023qlz} for further details).

\paragraph{Maillet bracket and $\bm\Rc$-matrix.} As explained in subsection \ref{subsec:Basics}, in addition to the existence of a flat Lax connection, integrability requires the Poisson-commutation of the associated charges, which is ensured if the Lax matrix satisfies a Maillet bracket~\cite{maillet_kac-moody_1985,maillet_new_1986}. In particular, the latter depends on a $\Rc$-matrix $\Rc(z_1,z_2) \in \g^\C\otimes\g^\C$, which then characterises the Hamiltonian structure of the theory. For the integrable $\sigma$-model under consideration, associated with the data $(C,\omega,\ha{\Z}^\pm,\g,\mathfrak{k})$, it was proven in the rational case~\cite{vicedo_integrable_2019,vicedo_4d_2021} and conjectured in the elliptic one~\cite{Lacroix:2023qlz} that the Lax matrix satisfies such a Maillet bracket. The corresponding $\Rc$-matrix takes a factorised form
\begin{equation}\label{eq:RTwist}
    \Rc(z_1,z_2) = \Rc^0(z_2-z_1)\,\vp(z_2)^{-1}\,,
\end{equation}
where $\Rc^0(z_2-z_1)$ is a skew-symmetric \textit{seed $\Rc$-matrix}~\cite{belavin_solutions_1982} depending only on the difference of the spectral parameters and $\vp(z)$ is the meromorphic function appearing in the 1-form $\omega=\vp(z)\,\dd z$.  In this context, $\vp(z)$ is often referred to as the twist function~\cite{maillet_hamiltonian_1986,Reyman:1988sf,Vicedo:2010qd}. One then sees that the 1-form $\omega$ also plays a crucial role in the Hamiltonian integrable structure of the model. 

The seed $\Rc$-matrix $\Rc^0(z)$ is a solution of the classical Yang-Baxter equation, which ensures that $\Rc$ also satisfies this identity. Moreover, it depends only on the choice of $C$ and $\g$. In the rational case $C=\CP$, it is given by the \textit{Yangian $\Rc$-matrix}
\begin{equation}
    \Rc^0_{\text{Yang}}(z) = \frac{1}{z} \sum_{\alpha} T^\alpha \otimes T_\alpha\,.
\end{equation}
Here $\lbrace T_\alpha \rbrace$ is a basis of the Lie algebra $\g^\C$ and $\lbrace T^\alpha \rbrace$ is its dual basis with respect to the invariant bilinear form $\langle\cdot,\cdot\rangle=-\Tr(\cdot)$ on $\g^\C$, where the trace is taken in the fundamental representation.

In the elliptic case $C=\C/\Gamma$, recall that the Lie algebra $\g^\C$ can only be chosen to be $\slNC$. We introduced its Belavin basis $\lbrace T_\alpha \rbrace_{\alpha\in\Ab}$ as well as the family of meromorphic functions $\lbrace r^\alpha(z) \rbrace_{\alpha\in\Ab}$ above equation \eqref{eq:EllipticLax} (see appendices \ref{app:Belavin} and \ref{app:ralpha} for details). These are exactly the ingredients needed to define the (conjectured) seed $\Rc$-matrix of the elliptic case, which is called the \textit{Belavin $\Rc$-matrix}~\cite{belavin_discrete_1981}:
\begin{equation}\label{eq:RBel}
    \Rc^0_{\text{Bel}}(z) = \sum_{\alpha\in\Ab} r^\alpha(z)\, T^\alpha \otimes T_\alpha\,.
\end{equation}
This matrix is doubly periodic with period $2N\ell_i$ and can thus be seen as an elliptic function on the torus $\C/\Lambda$.

\paragraph{Origin from affine Gaudin models and 4D Chern-Simons.} We finally note that the Maillet bracket with twist function discussed in the previous paragraph is the key ingredient in the formulation of integrable $\s$-models as \textit{affine Gaudin models}~\cite{levin_hitchin_2003,feigin_quantization_2009,vicedo_integrable_2019}\footnote{In the elliptic case, the formalism of affine Gaudin models associated with the Belavin $\Rc$-matrix is not yet fully worked out and will be the subject of a future work~\cite{ToAppear:Gaudin}.}. This formalism offers a unifying approach to the study of these theories based on algebraic structures (such as affine Lie algebras) and is deeply rooted in the Hamiltonian formulation. We will not discuss it further in the present paper and refer for instance to the recent lecture notes~\cite{Lacroix:2023gig} for more details.

The affine Gaudin formalism is the Hamiltonian counterpart of another unifying approach, namely the 4D Chern-Simons theory~\cite{costello_gauge_2019}, which is chiefly based on more geometric aspects and is rooted in the Lagrangian formulation. In Appendix \ref{app:4dCS}, we give a brief summary of how the integrable $\sigma$-model associated with the data $(C,\omega,\Zh^\pm,\g,\mathfrak{k})$ can be constructed from this approach. This non-exhaustive review is not a prerequisite for understanding the rest of the paper and is mostly included to provide some intuition behind the integrable model under consideration.

\section{1-loop RG-flow of the geometric data}
\label{sec:GeoRG}

In section \ref{sec:IntSigma}, we reviewed the construction of a large class of classical integrable $\s$-models, whose target spaces $(\Tc,G,B)$ are built from some geometric data $(C,\omega,\Zh^\pm)$ (formally a point in the moduli space of split abelian differentials $\Dc_{g,\m}$) and some additional algebraic ingredients $(\g,\kf)$. We now turn to the discussion of some quantum aspects of these theories, namely their 1-loop renormalisation. We will denote by $\mu$ the energy scale and introduce the Renormalisation-Group (RG) parameter $\tf = \frac{1}{4\pi}\log\mu$. For general $\s$-models, it is well-known that the 1-loop RG-flow takes the form of a generalised Ricci flow of the metric and B-field $(G,B)$~\cite{Ecker:1972bm,Honerkamp:1971sh,Friedan:1980jf,Curtright:1984dz}:
\begin{equation}\label{eq:Ricci}
    \frac{\dd\;}{\dd \tf} (G_{ij}+B_{ij}) = \hbar\, R^+_{ij} + \cdots\,,
\end{equation}
where $R^+$ is the torsionful Ricci tensor associated with the geometry $(G,B)$\footnote{More precisely, $R^+_{ij} = R^{+\,k}\null_{ijk}$, where $R^{+\,k}\null_{ijl}$ is the Riemann tensor measuring the curvature of the connection $\Gamma^{k}\null_{ij} - \frac{1}{2} H^{k}\null_{ij}$, $\Gamma$ are the Christoffel symbols of the metric $G$ and $H=\dd B$ is the torsion arising from the B-field.} and the dots contain higher-order $\hbar$-corrections as well as diffeomorphism terms and shifts of the B-field by exact forms (which can contribute to the 1-loop order but will not play an important role in this article). It is conjectured that the flow \eqref{eq:Ricci} preserves the specific subspace of metrics and B-fields corresponding to the aforementioned integrable $\s$-models and thus that these theories are 1-loop renormalisable. The RG-flow is then simply reabsorbed as a running of the coupling constants of the integrable $\s$-model, \textit{i.e.} the continuous parameters entering its defining data $(C,\omega,\Zh^\pm,\g,\kf)$. Of particular interest for this section are two conjectures, put forward in~\cite{delduc_rg_2021,derryberry_lax_2021}, which predict the form of the RG-flow of the geometric data $(C,\omega,\Zh^\pm)\in\Dc_{g,\m}$. Remarkably, this flow depends in a very minimal way on the choice of the simple Lie algebra $\g$ (more precisely through an overall proportionality factor) and is independent of the last ingredient $\kf$: we will thus focus here on $(C,\omega,\Zh^\pm)$ exclusively.

We will start by describing a conjecture proposed by Costello, which was reported and supported in the work~\cite{derryberry_lax_2021} of Derryberry. This proposal, which we will call the \textit{period-conjecture}, applies directly to both the rational case $C=\CP$ and the elliptic one $C=\C/\Gamma$. In contrast, the second conjecture that we will discuss, which was put forward in~\cite{delduc_rg_2021} was initially phrased for rational models only. We will refer to this as the \textit{$\partial\Psi$-conjecture} and propose a natural extension to the elliptic case; moreover, we will establish its equivalence with the period-conjecture.

\subsection{The period-conjecture}
\label{subsec:Periods}

\paragraph{Periods of $\bm{(C,\omega)}$.} Let us consider the data $(C,\omega)$ of the Riemann surface $C$ and the meromorphic 1-form $\omega$. We will be interested in the associated periods, which are defined as integrals
\begin{equation}\label{eq:periods}
    \int_\gamma \omega
\end{equation}
of $\omega$ over well-chosen paths $\gamma$ in $C$. We will consider two different types of such integrals:\vspace{-1pt}
\begin{itemize}\setlength\itemsep{2pt}
    \item the \textit{absolute periods}, defined as integrals of $\omega$ over closed contours in $C$ ;
    \item the \textit{relative periods}, defined as integrals of $\omega$ between two of its zeroes.
\end{itemize}
These are specific functions of the parameters entering the definitions of $C$ and $\omega$, whose data is essentially equivalent to the moduli $(C,\omega)$. We will flesh this relation out in more detail in a few paragraphs, but let us first give the statement of the period-conjecture.

\paragraph{The conjecture.} To formulate it, let us recall that the geometric data entering the definition of the integrable $\s$-model is not only the choice of $(C,\omega)$ but also the splitting of its zeroes into two equal-size subsets $\Zh^\pm = \lbrace \ha{z}^{\,\pm}_r \rbrace_{r=1}^M$, forming what we called a split abelian differential in the space $\Dc_{g,\m}$. Recall moreover that the model depends on the choice of a simple Lie algebra $\g$: for the purposes of this section, $\g$ will only enter the discussion through its \textit{dual Coxeter number} $c_\g$\footnote{To define this number, let us consider the invariant bilinear pairing $\langle\cdot,\cdot\rangle = -\Tr(\cdot)$ on $\g$, where the trace is taken in the fundamental representation and the minus sign is introduced so that this form is positive definite if $\g$ is compact. As $\g$ is simple, this pairing is proportional to the Killing form, defined by the bilinear trace in the adjoint representation. The dual Coxeter number $c_\g$ is then defined through the corresponding proportionality factor:
\begin{equation}
    \Tr(\ad_{X} \circ \ad_Y) = -2c_\g \langle X, Y \rangle \,\,, \qquad \forall\,X,Y\in\g\,.
\end{equation}
}. In these notations, Costello's period-conjecture~\cite{derryberry_lax_2021} then states that, at 1-loop,\footnote{Technically, this conjecture was proposed in~\cite{derryberry_lax_2021} for models corresponding to a 1-form $\omega$ with only double poles and no equivariance properties akin to~\eqref{eq:equivL}. However, its formulation is manifestly insensitive to having higher-order poles in $\omega$ and we expect that it also applies to equivariant theories, including the elliptic models considered in this paper, without major modifications.}
\begin{equation}
		\frac{\dd}{\dd \mathfrak{t}}\Bigg[\oint\omega\Bigg]=0\,, \qquad
		\frac{\dd}{\dd \mathfrak{t}}\Bigg[\int_{\hat{z}^\pm_r}^{\hat{z}^\pm_s}\omega\Bigg]=0\,, \qquad
		\frac{\dd}{\dd \mathfrak{t}}\Bigg[\int_{\hat{z}^+_r}^{\hat{z}^-_s}\omega\Bigg]=2\hbar\,c_{\hg}\,. \label{eq:RGperiods}
  \end{equation}
In other words, the absolute periods are RG-invariants, whereas the relative periods are invariants if they are between two zeroes in the same set $\Zh^\pm$ and flow linearly otherwise.

\paragraph{The period parametrisation.}  The conjecture \eqref{eq:RGperiods} shows that the periods of $\omega$ are quite natural quantities to describe the 1-loop RG-flow of the theory. In fact, they define a reparametrisation of the moduli $(C,\omega)$ in which this flow is completely trivialised. To explain this, let us first describe these periods in more detail.

We start with the absolute ones, which are obtained from closed paths in $C$. Typical examples of such paths are small circles $\gamma_r$ encircling the poles $\ha{p}_r$ of $\omega$: the corresponding absolute periods then coincide with the residues $\res_{\hat{p}_r} \omega$. We note that these residues sum to zero and thus encode $n-1$ independent quantities. Other examples of closed contours in $C$ are its A- and B-cycles, which are generators of its first homotopy group: for the Riemann sphere $C=\CP$ there are no such cycles, while there are two of them $\alpha$ and $\beta$ for the torus $C=\C/\Gamma$, thus yielding 2 additional absolute periods. All the other closed paths in $C$ are either contractible to a point without crossing any of the poles of $\omega$ or can be deformed into a concatenation of contours $\gamma_r$, $\alpha$ and $\beta$, in which case the corresponding integrals are linear combinations of the absolute periods described above (with integer coefficients). We thus obtain a total of $n-1+2g$ independent absolute periods (where $n$ is the number of poles of $\omega$ and $g$ the genus of $C$), given by
\begin{equation}\label{eq:abs}
    \res_{\hat{p}_r} \omega = \oint_{\gamma_r} \omega\,\quad \text{ for } r\in\lbrace 1,\dots,n-1 \rbrace\,, \qquad \Pi_A = \oint_\alpha \omega \qquad \text{ and} \qquad \Pi_B = \oint_\beta \omega\,,
\end{equation}
where the last two exist only in the elliptic case $g=1$. According to the period-conjeture \eqref{eq:RGperiods}, these quantities are all RG-invariants.\\

We now turn our attention to the relative periods, defined as the integrals of $\omega$ between two of its zeroes. Fixing two such zeroes, there are of course many different paths which connect them: the corresponding integrals are the same if the paths can be deformed smoothly one from another without crossing poles of $\omega$. Up to such deformations, two of these paths can differ only by a combination of the closed contours $\gamma_r$, $\alpha$ and $\beta$ described above. Therefore, although a relative period is not uniquely defined by the choice of the two corresponding zeroes, these non-equivalent definitions yield the same answer up to integral linear combinations of the absolute periods $\res_{\hat{p}_r} \omega$, $\Pi_A$ and $\Pi_B$. In particular, this non-unicity does not pose a problem for the period-conjecture \eqref{eq:RGperiods}, since these absolute periods do not flow. From now on, we fix a choice of path between each pair of zeroes $(\ha{z},\ha{z}\,')$ and use it to define the corresponding integral: by a slight abuse of notation, we do not keep track of this choice and simply use the generic symbol $\int_{\hat{z}}^{\hat{z}'}\omega$ to designate this integral.

We now note that this construction does not define an independent set of periods. Indeed, the sum $\int_{\hat{z}}^{\hat{z}'}\omega + \int_{\hat{z}'}^{\hat{z}''}\omega$ coincides with $\int_{\hat{z}}^{\hat{z}''}\omega$, up to absolute periods. In the end, this leaves only $2M-1$ independent relative periods, where we recall that $2M$ is the number of zeroes of $\omega$. Here, we will consider a particular choice for these ``generators'', which relies on the separation of the zeroes into the two subsets $\Zh^\pm = \lbrace \ha{z}^{\,\pm}_r \rbrace_{r=1}^M$, namely
\begin{equation}
    \Pi^\pm_r = \int_{\hat{z}_r^\pm}^{\hat{z}_M^\pm} \omega \quad \text{ for } r\in\lbrace 1,\dots,M-1 \rbrace \qquad \text{ and } \qquad \Pi_0 = \int_{\hat{z}_M^+}^{\hat{z}_M^-} \omega\,.
\end{equation}

We have thus defined a complete set of independent periods, composed of $n-1+2g$ absolute ones and $2M-1$ relative ones. As it turns out, these define local coordinates on the moduli space $\Dc_{g,\m}$. In other words, they provide enough information to fully characterise $(C,\omega,\Zh^\pm)$, although this relation is in practice not explicit. As a heuristic argument for this claim, we note that the total number of independent periods found above is $2M+n+2(g-1)$: this agrees with the dimension of $\Dc_{g,\m}$, determined in subsection \ref{subsec:GeoTwist} using the parametrisation in terms of zeroes and poles modulo changes of coordinate. The main interest of this \textit{period parameterisation} is that it completely trivialises the 1-loop RG-flow of the theory. Indeed, the quantities
\begin{equation}
    \bigl( \res_{\hat{p}_r} \omega \bigr)_{r=1}^{n-1}, \qquad (\Pi_A,\Pi_B) \qquad \text{ and } \qquad (\Pi_r^\pm)_{r=1}^{M-1}
\end{equation}
are all RG-invariants, while the remaining period $\Pi_0$ has a very simple evolution under the RG:
\begin{equation}
    \Pi_0 = 2\hbar\,c_\g (\tf-\tf_0) = \frac{\hbar\,c_\g}{2\pi} \log\left(\frac{\mu}{\mu_0}\right)\,.
\end{equation}
In particular, this quantity is the only scale-dependent coupling and can thus be used to quantify the dimensional transmutation phenomenon occurring in the quantum model.

\paragraph{Additional remarks.} We end this subsection with a variety of further remarks about the period-conecture.

To start with, note that even before the choice of the 1-form $\omega$, the very first defining ingredient of the model is the data of the Riemann surface $C$. This data is characterised by certain moduli, whose number and nature depend on the genus $g$ of $C$. For this paper, we restricted ourselves to the rational case $g=0$, for which there are no such moduli, and the elliptic case $g=1$, for which there is only one modulus $\tau$, as defined in \eqref{eq:tau}. This modulus can be characterised in terms of the absolute periods of the 1-form $\dd z$ (which, up to a global factor, is the unique abelian differential of the first kind on the torus, \textit{i.e.} a 1-form without any pole). More precisely, we have
\begin{equation}
    \tau =  \raisebox{1pt}{$\displaystyle\oint_\beta \dd z$}\bigg/ \raisebox{-1pt}{$\displaystyle\oint_\alpha \dd z$} \,.
\end{equation}
We note however that the 1-form $\dd z$ is quite different from the twist 1-form $\omega$ associated with integrable $\s$-models. In particular, it is the absolute periods of $\omega$ which are invariants of the RG-flow \eqref{eq:RGperiods}, not those of $\dd z$. This means that the modulus $\tau$ of the torus generally flows non-trivially under renormalisation. In other words, the complex structure underlying the spectral parameter of an elliptic integrable $\s$-model varies with the energy scale at the quantum level.\\

Furthermore, we stress that the period-conjecture \eqref{eq:RGperiods} does not depend only on the data $(C,\omega)$ but also on the separation of the zeroes of $\omega$ into the two subsets $\Zh^\pm$. Together, these data form what we called a split abelian differential $(C,\omega,\Zh^\pm)\in\Dc_{g,\m}$. The equation \eqref{eq:RGperiods} can thus naturally be seen as a flow on the moduli space $\Dc_{g,\m}$, which is intrinsically defined in terms of the data encoded in its points (the only external information required in equation \eqref{eq:RGperiods} is the overall factor $c_\g$, which can be reabsorbed in a rescaling of the flow parameter). This flow is in fact a special case of a more general framework, which appeared in a different context in the mathematics literature, under the name of \textit{Rel-flow} -- see for instance~\cite{Zorich:2006sur,Bainbridge:2016hor,Winsor:2022rel}\footnote{In particular, we refer to the introduction of~\cite{Winsor:2022rel} for the most explicit connection to equation \eqref{eq:Rel-flow}. Note that the Rel-flow considered in the references~\cite{Zorich:2006sur,Bainbridge:2016hor,Winsor:2022rel} is defined on the moduli space of abelian differentials of the first kind, \textit{i.e.} for holomorphic 1-forms without poles. However, this construction can be extended without modifications to the case considered in the present paper, where the 1-form $\omega$ has poles, generally with non-vanishing residues (in the mathematical terminology, this is called an abelian differential of the third kind).}. To describe this, let us consider abelian differentials $(C,\omega)$ with arbitrary structures of poles and zeroes. We denote by $\ha{z}_1,\dots,\ha{z}_{M'}$ the zeroes, which are then not necessarily simple and not split into two subsets $\Zh^\pm$ (hence the absence of $\pm$-labels). In place of this splitting, we associate to each zero $\ha{z}_r$ a number $f_r\in\C$. We then define the Rel-flow on the moduli space of these differentials (with fixed multiplicities of poles/zeroes) by
\begin{equation}
		\frac{\dd}{\dd \mathfrak{t}}\Bigg[\oint\omega\Bigg]=0\qquad \text{ and } \qquad
		\frac{\dd}{\dd \mathfrak{t}}\Bigg[\int_{\hat{z}_r}^{\hat{z}_s}\omega\Bigg]=f_r-f_s\,. \label{eq:Rel-flow}
\end{equation}
The flow of the period-conjecture \eqref{eq:RGperiods} then corresponds to the special case where there are an even number $M'=2M$ of simple zeroes, half of which being associated with the number $f_r = +\hbar\, c_\g$ while the other half has $f_r = -\hbar\, c_\g$. We expect the general Rel-flow \eqref{eq:Rel-flow} to also describe the 1-loop RG-flow of some integrable 2d field theories built from 4d-CS / affine Gaudin models but which however would not be relativistic $\s$-models (this is consistent with the observation made in section \ref{sec:IntSigma} that the splitting of the zeroes into $\Zh^\pm$ is related to the relativistic invariance of the underlying integrable model). We note that the right-hand side of the second equation in the rel-flow \eqref{eq:Rel-flow} is reminiscent of the structure of curious matrices recently discussed in~\cite{Hassler:2023xwn}. This suggests that the relative periods might also play a role in the $\mathcal{E}$-model formulation of these integrable field theories, which is the framework used in~\cite{Hassler:2023xwn}: it would be interesting to further explore these perspectives.\\

We end with a slight reformulation of the period-conjecture \eqref{eq:RGperiods}. We introduce a function $\mathcal{P}(z)$ through the differential equation
\begin{equation}
    \p_z \log \mathcal{P}(z) = \frac{2\pi}{\hbar\,c_\g}\,\vp(z)\,,
\end{equation}
where we recall that  $\vp(z)$ is the twist function, defined through $\omega=\vp(z)\,\dd z$. This function $\mathcal{P}(z)$ is generally multi-valued on $C$: more precisely, it possesses a non-trivial monodromy around the poles $\ha{p}_r$ of $\omega$ and along the A- and B-cycles of $C$ in the elliptic case. The period-conjecture is then equivalent to the statement that these monodromies are RG-invariants, together with
\begin{equation}\label{eq:RGP}
    \frac{\mathcal{P}(\ha{z}^{\,\pm}_s)}{\mathcal{P}(\ha{z}^{\,\pm}_r)} = \text{RG-invariant} \qquad \text{ and } \qquad \frac{\mathcal{P}(\ha{z}^{\,-}_s)}{\mathcal{P}(\ha{z}^{\,+}_r)} = \frac{\mu}{\mu_0^{(rs)}}\,,
\end{equation}
where $\mu_0^{(rs)}$ are constant numbers. It is worth noticing that the use of this function $\mathcal{P}(z)$ is reminiscent of the results of~\cite{Lukyanov:2013wra,Bazhanov:2013cua,Bazhanov:2013oya,Lacroix:2018fhf,Lacroix:2018itd,Gaiotto:2020dhf,Kotousov:2021vih,Kotousov:2022azm,Franzini:2022duf}, which concerned the quantisation of integrable $\sigma$-models and in particular the construction and the diagonalisation of quantum integrals of motion in these theories. We will further comment on this point in the conclusion section \ref{sec:Conc}.

\subsection[The rational \texorpdfstring{$\partial\Psi$}{dpsi}-conjecture]{The rational \texorpdfstring{$\bm{\partial\Psi}$}{dpsi}-conjecture}
\label{subsec:RatDpsi}

We now review another conjectural formulation of the 1-loop RG-flow that we call the $\partial\Psi$-conjecture. It was proposed in~\cite{delduc_rg_2021,Kotousov:2022azm} for rational integrable $\s$-models and was recently proven for a large class of such theories in~\cite{hassler_rg_2021,Hassler:2023xwn}. After reviewing its statement, we will discuss its equivalence with the period-conjecture. We focus here on the rational case $C=\CP$ and will treat the elliptic one in the next subsection.

\paragraph{The functions $\bm{f}$ and $\bm\Psi$.} Let us first introduce the necessary ingredients to state the conjecture. We work with a coordinate $z$ and the corresponding expression of the 1-form as $\omega=\vp(z)\,\dd z$, where $\vp(z)$ is the twist function. It will be instructive to consider the inverse of this function, which takes the form
\begin{equation}\label{eq:TwistInv}
    \vp(z)^{-1} = \sum_{r=1}^M \left( \frac{1}{\vp'(\ha{z}^{\,+}_r)} \frac{1}{z-\ha{z}^{\,+}_r} + \frac{1}{\vp'(\ha{z}^{\,-}_r)} \frac{1}{z-\ha{z}^{\,-}_r} \right) + a_0 + a_1\,z + a_2\,z^2\,,
\end{equation}
where $(a_0,a_1,a_2)$ are real parameters whose expression we will not need. Note that for simplicity, we supposed here that the zeroes $\ha{z}^{\,\pm}_r$ of $\omega$ are all finite (this is always possible, up to a redefinition of $z$). Let us now introduce another function $f(z)$, taking a form very similar to $\vp(z)^{-1}$:
\begin{equation}\label{eq:f}
    f(z) = \hbar\,c_\g \left[ \sum_{r=1}^M \left( \frac{1}{\vp'(\ha{z}^{\,+}_r)} \frac{1}{z-\ha{z}^{\,+}_r} - \frac{1}{\vp'(\ha{z}^{\,-}_r)} \frac{1}{z-\ha{z}^{\,-}_r} \right) + b_0 + b_1\,z + b_2\,z^2 \right] \,,
\end{equation}
where we recall that $c_\g$ is the dual Coxeter number of $\g$ and $(b_0,b_1,b_2)$ are real parameters which for the moment are arbitrary. Up to these parameters and the prefactor $\hbar\,c_\g$, the main difference with $\vp(z)^{-1}$ is the relative sign between the contributions of the zeroes $\ha{z}^{\,+}_r$ and $\ha{z}^{\,-}_r$. The definition of $f(z)$ is thus deeply related to the separation of the zeroes into the two subsets $\Zh^\pm$. It will also be useful to introduce the related functions
\begin{equation}\label{eq:fpm}
    f_\pm(z) = \hbar\,c_\g \left[ \pm 2\sum_{r=1}^M \frac{1}{\vp'(\ha{z}^{\,\pm}_r)} \frac{1}{z-\ha{z}^{\,\pm}_r} + b_0 \pm a_0 + (b_1 \pm a_1)\,z + (b_2 \pm a_2)\,z^2 \right] \,,
\end{equation}
defined such that
\begin{equation}\label{eq:fpm2}
    f(z) = \frac{f_+(z) + f_-(z)}{2} \qquad \text{ and } \qquad \hbar\,c_\g\,\vp(z)^{-1} = \frac{f_+(z) - f_-(z)}{2} \,.
\end{equation}
We note that $f_\pm(z)$ has poles at the points $\Zh^\pm$ but not at the ones $\Zh^\mp$.\\

The main protagonist of the $\partial\Psi$-conjecture is a function $\Psi$, related to $f$ by
\begin{equation}\label{eq:Psi}
    \Psi(z) = -\, \vp(z)\,f(z)\,.
\end{equation}
We note that this function is regular at $\ha{z}^{\,\pm}_r$ despite these points being poles of $f(z)$ since they are also zeroes of $\vp(z)$. The poles of $\Psi(z)$ are thus located at the same points $\ha{p}_r$ as that of the twist function $\vp(z)$ itself. In fact, $\Psi(z)$ can be alternatively defined as the (almost unique) rational function of $z$ which has the same pole structure\footnote{\label{foot:PsiInf}More precisely, $\Psi(z)$ has the same pole structure as $\vp(z)$ on the complex plane $\C$. Around infinity, if $\vp(z)= O(1/z^k)$ for some $k \in \Z$, then $\Psi(z)= O(1/z^{k-2})$. This slight change of behaviour is due to the different nature of $\vp(z)$ and $\Psi(z)$ as geometric objects on $\CP$ and essentially captures the presence of the term $ b_0 + b_1\,z + b_2\,z^2$ in $f(z)$.} (\textit{i.e.} positions and multiplicities) as $\vp(z)$ and such that
\begin{equation}\label{eq:PsiZ}
    \Psi(\ha{z}^{\,\pm}_r) = \mp \hbar\,c_\g\,.
\end{equation}

\paragraph{The conjecture.} Using the ingredients introduced in the previous paragraph, the $\p\Psi$-conjecture~\cite{delduc_rg_2021,Kotousov:2022azm} states that, at 1-loop\footnote{In terms of the twist function, this conjecture becomes $\frac{\dd\;}{\dd\tf} \vp(z) = \p_z\Psi(z) = -\p_z( \vp(z)f(z))$. Up to a change of convention in the prefactor of $f(z)$, this last equality was the original formulation proposed in~\cite{delduc_rg_2021,Kotousov:2022azm}.},
\begin{equation}\label{eq:dPsi}
    \frac{\dd\;}{\dd\tf}\, \omega = \p\Psi = \p_z\Psi(z)\, \dd z\,,
\end{equation}
where $\p$ is the holomorphic Dolbeault operator. This formula thus provides a direct and explicit expression for the 1-loop RG-flow of the twist 1-form $\omega$, which however in practice can be quite complicated when expressed in terms of the defining parameters of $\omega$. This is in contrast with the period-conjecture, which is formulated more implicitly in terms of $\omega$ (depending instead on its integrals) but which essentially trivialises the RG-flow.

The $\p\Psi$-conjecture was first proposed in~\cite{delduc_rg_2021} for models corresponding to twist 1-forms $\omega$ with double poles at infinity, which further had to satisfy a specific technical requirement. If these conditions are met, the conjecture was proven in the recent works~\cite{hassler_rg_2021,Hassler:2023xwn}, using the interpretation~\cite{lacroix_integrable_2021} of these theories as so-called $\mathcal{E}$-models~\cite{Klimcik:1995ux,Klimcik:1995dy}, for which convenient formulations of the 1-loop RG-flow are known~\cite{Valent:2009nv,sfetsos_renormalization_2010}. For rational models with general $\omega$, \textit{i.e.} relaxing the condition mentioned above, the $\p\Psi$-conjecture was formulated in~\cite{Kotousov:2022azm} and has not been proven yet, except for very specific cases\footnote{In the work~\cite{Liniado:2023uoo}, these general theories were reinterpreted as degenerate/gauged $\mathcal{E}$-models~\cite{Klimcik:1996np,Sfetsos:1999zm}, whose 1-loop RG-flow is also known~\cite{Severa:2018pag}. This could offer a potential approach for a general proof of the rational $\p\Psi$-conjecture, in the spirit of~\cite{hassler_rg_2021,Hassler:2023xwn}.}.

\paragraph{Geometric interpretation.} To understand the structure underlying this conjecture, let us compare the twist 1-form $\omega_{\tf}$ at a given value $\tf$ of the RG-parameter and the one $\omega_{\tf+\delta\tf}$ after an infinitesimal shift $\delta\tf$ of this parameter. From equation \eqref{eq:dPsi}, we get
\begin{equation}
    \omega_{\tf+\delta\tf} = \omega_{\tf} - \delta\tf \bigl( \vp_{\tf}'(z) f_{\tf}(z) + \vp_{\tf}(z) f_{\tf}'(z) \bigr) \dd z + O(\delta\tf^2) = \vp_{\tf}\bigl( h^{-1}_{\tf;\delta\tf}(z) \bigr) \p_z h^{-1}_{\tf;\delta \tf}(z)\,\dd z\,
\end{equation}
where
\begin{equation}
    h_{\tf;\delta \tf}(z) = z + \delta\tf\,f_{\tf}(z) + O(\delta\tf^2)\,.\,\label{eq:varinz}
\end{equation}
Comparing with equation \eqref{eq:ChangeOmega}, we thus see that the infinitesimal variation $\tf \mapsto \tf + \delta\tf$ along the RG-trajectory formally acts on $\omega$ as an infinitesimal change of spectral parameter $z \mapsto z + \delta\tf\,f(z)$. This might seem surprising at first sight, since the underlying $\sigma$-model is independent of the choice of coordinate $z$ (see section \ref{sec:IntSigma}): indeed, we might then be tempted to conclude from the above interpretation of the RG-flow that this theory is invariant under renormalisation. This is of course not the case, as one should expect: the caveat in this reasoning is that the $\sigma$-model is invariant only under global biholomorphisms, \textit{i.e.} transformations of $z$ which are holomorphic, invertible and whose inverse is also holomorphic. For the rational case, the only biholomorphisms are the M\"obius transformations (see equation \eqref{eq:Mobius}), which infinitesimally correspond to variations
\begin{equation}\label{eq:InfMob}
    \delta z = \epsilon_0 + \epsilon_1\,z + \epsilon_2\,z^2\,.
\end{equation}
In other words, any infinitesimal variation of $z$ which is not a polynomial of degree at most 2 does not lift to a global biholomorphism. This is the case of the variation $\delta z = \delta\tf\,f(z)$ corresponding to the RG-flow, since the function $f(z)$ contains simple fractions at the points $\ha{z}^{\,\pm}_r$. The induced variation of $\omega$ thus amounts to a non-trivial transformation of the $\sigma$-model and its parameters, as expected.

We note, however, that the polynomial part $b_0 + b_1\,z + b_2\,z^2$ in the definition \eqref{eq:f} of $f(z)$ creates a term in the variation of $z$ \eqref{eq:varinz} which takes the form of an infinitesimal M\"obius transformation \eqref{eq:InfMob}. This part thus encodes the freedom of performing a M\"obius transformation along the RG-flow without affecting the running of the physical coupling constants. This is why the parameters $(b_0,b_1,b_2)$ are left undetermined in the definition of $f(z)$: in principle, they can take any value without changing the physics. More precisely, this is the case if we worked with a redundant parameterisation of $\omega$, where we did not fix the M\"obius freedom. If one has fixed it, for instance by setting three of the poles/zeroes to a given choice of positions, the parameters $(b_0,b_1,b_2)$ are not free anymore and should take very specific values which ensure that this choice is preserved along the RG-flow. 

\paragraph{RG-flow of the poles and zeroes.} As explained in~\cite{delduc_rg_2021}, the $\p\Psi$-conjecture is well-adapted to describe the RG-flow of some of the natural parameters of the model, namely the positions of the poles and zeroes of $\omega$. For instance, a simple analysis of the most singular terms around a pole $z=\ha{p}_r$ on both sides of the formula \eqref{eq:dPsi} yields
\begin{equation}\label{eq:RGpole}
    \frac{\dd\;}{\dd \tf} \, \ha{p}_r = f(\ha{p}_r)\,.
\end{equation}
This is consistent with the interpretation of the RG-flow as the infinitesimal change of spectral parameter $\delta z = \delta\tf \,f(z)$ discussed above. There exists a similar formula for the RG-flow of the zeroes $\ha{z}^{\,\pm}_r$; however, this is slightly more subtle since the function $f(z)$ has a pole at this point. Instead, one has to use the function $f_\mp(z)$ introduced in equation \eqref{eq:fpm}, which is closely related to $f(z)$ but is regular at $\ha{z}^{\,\pm}_r$. Indeed, a careful analysis of the equation \eqref{eq:dPsi} around $z=\ha{z}^{\,\pm}_r$ shows that
\begin{equation}\label{eq:RGzero}
    \frac{\dd\;}{\dd \tf}\, \ha{z}^{\,\pm}_r = f_\mp(\ha{z}^{\,\pm}_r)\,.
\end{equation}
For completeness, we note that the right-hand side of \eqref{eq:RGpole}, \textit{i.e.} the flow of the pole $\ha{p}_r$, can be rewritten as $f_\pm(\ha{p}_r)$ independently of the sign, using equation \eqref{eq:fpm2} and the fact that $\ha{p}_r$ is a zero of $\vp(z)^{-1}$ by construction.

\paragraph{Equivalence with the period-conjecture.} We now prove that the $\p\Psi$-conjecture \eqref{eq:dPsi} is equivalent to the period-conjecture \eqref{eq:RGperiods} in the rational case. This relation was suggested in~\cite[Remark 9.2]{derryberry_lax_2021}: here, we spell out its detailed derivation.

The right-hand side of equation \eqref{eq:dPsi} is a $\p$-exact 1-form and thus has no residues (in other words, the $z$-derivative of a rational function cannot create simple fractions of order one). This means that, assuming the $\p\Psi$-conjecture, the residues of $\omega$ are RG-invariants:
\begin{equation}
    \frac{\dd\;}{\dd \tf} \, \res_{\hat{p}_r} \omega = 0\,.
\end{equation}
This gives back the ``absolute'' part of the period-conjecture, \textit{i.e.} the first equation in \eqref{eq:RGperiods}, since in the rational case, the only absolute periods of $\omega$ are its residues.

Reciprocally, assuming the RG-invariance of the absolute periods means that $\frac{\dd\;}{\dd \tf}\,\omega$ is a meromorphic 1-form on $\CP$ with no residues: it a standard result that it is then equal to $\p\widetilde{\Psi}$ for some rational function $\widetilde{\Psi}(z)$. Moreover, one easily shows that $\widetilde{\Psi}(z)$ then has the exact same pole structure as $\vp(z)$.\footnote{More precisely, $\tilde{\Psi}(z)$ and $\vp(z)$ have the same pole structure on the finite complex plane. Around infinity, one shows that $\tilde{\Psi}(z)= O(1/z^{k-2})$ if $\vp(z)= O(1/z^k)$. This is the same behaviour as $\Psi(z)$, as discussed in footnote \ref{foot:PsiInf}.} To summarise, the ``absolute'' part of the period-conjecture is equivalent to the existence of a rational function $\widetilde{\Psi}(z)$, with the same pole structure as $\vp(z)$, such that
\begin{equation}\label{eq:dPsit}
    \frac{\dd\;}{\dd\tf}\, \omega = \p\widetilde\Psi = \p_z\widetilde\Psi(z)\, \dd z\,.
\end{equation}
Our goal now will be to prove that the ``relative'' part of the period conjecture, \textit{i.e.} the second and third equations in \eqref{eq:RGperiods}, is then equivalent to the function $\widetilde\Psi(z)$ being equal to the one $\Psi(z)$ appearing in the $\p\Psi$-conjecture.\\
 
To do so, let us consider the integral of $\omega$ between two given points $z_1$ and $z_2$. 
Its $\tf$-derivative has two types of contributions, coming respectively from the variation of the integrand $\omega$ and the variation of the endpoints. More precisely, we have
\begin{equation}
    \frac{\dd\;}{\dd \tf} \Bigg[\int_{z_1}^{z_2}\omega\Bigg] = \int_{z_1}^{z_2} \frac{\dd \;}{\dd \tf} \,\omega + \frac{\dd z_2}{\dd \tf}\, \frac{\partial\;}{\partial z_2}\Bigg[\int_{z_1}^{z_2}\omega\Bigg] + \frac{\dd z_1}{\dd \tf} \,\frac{\partial\;}{\partial z_1}\Bigg[\int_{z_1}^{z_2}\omega\Bigg]\,.
\end{equation}
In our setup, the $\tf$-derivative of $\omega$ is given by equation \eqref{eq:dPsit}. The fundamental theorem of complex integration then allows us to compute the three terms in the above equation, simply yielding
\begin{equation}\label{eq:FlowInt}
    \frac{\dd\;}{\dd \tf} \Bigg[\int_{z_1}^{z_2}\omega\Bigg] = \widetilde\Psi(z_2) - \widetilde\Psi(z_1) + \frac{\dd z_2}{\dd \tf}\, \vp(z_2) - \frac{\dd z_1}{\dd \tf}\, \vp(z_1)\,.
\end{equation}

Let us now specialise to $z_1$ and $z_2$ beings zeroes $\ha{z}^{\,\s}_r$ and $\ha{z}^{\,\s'}_s$ of $\omega$, with $r,s\in\lbrace 1,\dots,M\rbrace$ and $\s,\s' \in \lbrace +,- \rbrace$, so that the integral becomes a relative period. The last two terms in the above equation then vanish by construction, leaving
\begin{equation}\label{eq:dRelPsi}
    \frac{\dd\;}{\dd \tf} \Bigg[\int_{\hat{z}^{\s}_r}^{\hat{z}^{\s'}_s}\omega\Bigg] = \widetilde\Psi\bigl(\ha{z}^{\,\s'}_s \bigr) - \widetilde\Psi\bigl(\ha{z}^{\,\s}_r \bigr)\,.
\end{equation}
The second and third equations in the period-conjecture \eqref{eq:RGperiods} are then equivalent to
\begin{equation}\label{eq:ConsPsit}
    \widetilde\Psi\bigl(\ha{z}^{\,\pm}_s \bigr) - \widetilde\Psi\bigl(\ha{z}^{\,\pm}_r \bigr) = 0 \qquad \text{ and } \qquad \widetilde\Psi\bigl(\ha{z}^{\,-}_s \bigr) - \widetilde\Psi\bigl(\ha{z}^{\,+}_r \bigr) = 2\hbar\,c_\g\,.
\end{equation}
A solution to these constraints is given by $\widetilde{\Psi}(\ha{z}^{\,\pm}_r)=\mp \hbar\,c_\g$. This is the same behaviour as $\Psi(z)$, as one can see from equation \eqref{eq:PsiZ}. It was mentioned then that this property, together with having the same pole structure as $\vp(z)$, uniquely characterises the function $\Psi(z)$. In the end, we get $\widetilde{\Psi}(z)=\Psi(z)$, as we wanted to show: either conjecture thus follows from the other.

We note for completeness that $\widetilde{\Psi}(\ha{z}^{\,\pm}_r)=\mp \hbar\,c_\g$ is not the unique solution to the constraints \eqref{eq:ConsPsit}: in fact, one easily shows that the most general solution takes the form $\widetilde{\Psi}(\ha{z}^{\,\pm}_r)=\mp \hbar\,c_\g + \alpha$, where $\alpha$ is a constant (independent of $r$ and the $\pm$-label). This simply corresponds to taking $\widetilde\Psi(z) = \Psi(z) + \alpha$: it is clear that the $\p\Psi$-conjecture stays the same if one shifts $\Psi$ by a constant so that these more general solutions simply correspond to a redundancy in the formulation of this conjecture and not to a different outcome (equivalently, this redundancy amounts to a shift of $f(z)$ by $-\alpha\,\vp(z)^{-1}$). This ends the proof of the equivalence between the period-conjecture and the $\p\Psi$-one for rational models. As mentioned earlier, the latter was shown to hold for a large class of such theories in~\cite{hassler_rg_2021,Hassler:2023xwn}, thus providing an indirect proof of the former in those cases.

\paragraph{General Rel-flow.} We end this subsection with a small digression about the general Rel-flow \eqref{eq:Rel-flow}. The latter was mentioned in the previous subsection as a generalisation of the period-conjecture flow \eqref{eq:RGperiods}, defined for 1-forms $\omega$ having zeroes $\lbrace \ha{z}_r \rbrace_{r=1}^{M'}$ which are not necessarily simple and not split into the two subsets $\Zh^\pm$. This Rel-flow depends on a collection of numbers $( f_r )_{r=1}^{M'}$ attached to the zeroes and reduces to the period-conjecture flow \eqref{eq:RGperiods} when half of these numbers are equal to $+\hbar\,c_\g$ and the other half to $-\hbar\,c_\g$. Going through the reasoning of the previous paragraph, one easily sees that (in the rational case) this generalised flow also admits a ``$\p\Psi$-form'', \textit{i.e.} can be written as $\frac{\dd\;}{\dd\tf}\omega = \p\Psi$ for some $\Psi$. More precisely, the function $\Psi(z)$ is characterised by having the same pole structure as $\vp(z)$ and the evaluations $\Psi(\ha{z}_r) = -f_r$ at its zeroes. When the latter are simple, we can write $\Psi(z)=-\vp(z) f(z)$ where $f(z)$ has a form akin to \eqref{eq:f}: more precisely, $f(z)$ has poles at $z=\ha{z}_r$, with residues $f_r / \vp'(\ha{z}_r)$. This is similar to the setup considered in~\cite{Hassler:2023xwn}.

\subsection[The elliptic \texorpdfstring{$\partial\Psi$}{dpsi}-conjecture]{The elliptic \texorpdfstring{$\bm{\partial\Psi}$}{dpsi}-conjecture}
\label{subsec:EllDpsi}

The $\p\Psi$-conjecture put forward in~\cite{delduc_rg_2021} and reviewed in the previous subsection initially concerned the RG-flow of rational integrable $\s$-models. We now formulate an elliptic version of this proposal and show that it is also equivalent to the period-conjecture.

\paragraph{Elliptic generalisation of the function $\bm{f}$.} A key role in the rational $\p\Psi$-conjecture was played by the function $f(z)$. As a first step towards an elliptic generalisation, we describe its equivalent in the case $C=\C/\Gamma$. Recall the form \eqref{eq:f} of the rational $f(z)$, containing a simple fraction $1/(z-\ha{z}^{\,\pm}_r)$ for each zero $\ha{z}^{\,\pm}_r$ of $\omega$. A natural guess for the elliptic generalisation would be to replace this fraction with an elliptic function having a simple pole at $\ha{z}^{\,\pm}_r$ and no other singularities. However, such a function does not exist (as the sum of its residues would not vanish). The next natural option is the Weierstrass $\zeta$-function $\zeta(z-\ha{z}^{\,\pm}_r)$: we refer to the appendix \ref{app:elliptic} for a review of its definition and properties. In particular, it has the required behaviour $\sim 1/(z-\ha{z}^{\,\pm}_r)$ around the point $\ha{z}^{\,\pm}_r$ but is only quasi-periodic, as seen from equation \eqref{eq:QuasiZeta}. We then define the generalisation of $f(z)$ to the elliptic case as
\begin{equation}\label{eq:fEll}
    f(z) = \hbar\,c_\g \left[ \sum_{r=1}^M \left( \frac{\zeta(z-\ha{z}^{\,+}_r)}{\vp'(\ha{z}^{\,+}_r)}  - \frac{\zeta(z-\ha{z}^{\,-}_r)}{\vp'(\ha{z}^{\,-}_r)} \right) + b_0 + b_1\,z \right] \,,
\end{equation}
where $(b_0,b_1)$ are coefficients which for the moment are arbitrary. These are the equivalent of the parameters $(b_0,b_1,b_2)$ in the rational formula \eqref{eq:f}: their interpretation and the absence of $b_2$ in the elliptic case will be explained later in this subsection. As another motivation for the elliptic version \eqref{eq:fEll} of $f(z)$, recall that the rational one was defined by drawing a parallel with the inverse of the twist function -- see equation \eqref{eq:TwistInv} and the discussion below. In the elliptic case, we have
\begin{equation}
    \vp(z)^{-1} = \sum_{r=1}^M \left( \frac{\zeta(z-\ha{z}^{\,+}_r)}{\vp'(\ha{z}^{\,+}_r)}  + \frac{\zeta(z-\ha{z}^{\,-}_r)}{\vp'(\ha{z}^{\,-}_r)} \right) + a_0\,,
\end{equation}
for some coefficient $a_0$, so that a similar parallel still holds.\\

We finally note that $f$ is not elliptic: instead, the quasi-periodicity \eqref{eq:QuasiZeta} of the Weierstrass $\zeta$-function (written in terms of the constants $L_i$) translates to the following property of $f(z)$:
\begin{equation}\label{eq:QuasiF}
    f(z+2\ell_i) = f(z) - 2\beta_{\ell_i}\,, \qquad \text{ with } \qquad \beta_{\ell_i} = \hbar\,c_\g \left[ L_i \sum_{r=1}^M \left( \frac{1}{\vp'(\ha{z}^{\,-}_r)} - \frac{1}{\vp'(\ha{z}^{\,+}_r)}  \right) + b_1 \ell_i \right]\,.
\end{equation}
$\beta_{\ell_i}$ is a constant and the notation will be justified later in this subsection, when we will identify this quantity as the $\beta$-function of the half-period $\ell_i$. We note that the 2-vectors $\bm\ell=(\ell_1,\ell_2)$ and $\bm{\beta_\ell}=(\beta_{\ell_1},\beta_{\ell_2})$ satisfy
\begin{equation}\label{eq:BetaxEll}
    \bm{\beta_{\ell}} \times \bm\ell  = \beta_{\ell_2} \ell_1 - \beta_{\ell_1} \ell_2 = \frac{\ri\pi\hbar\,c_\g}{2} \sum_{r=1}^M \left( \frac{1}{\vp'(\ha{z}^{\,-}_r)} - \frac{1}{\vp'(\ha{z}^{\,+}_r)} \right)\,.
\end{equation}
Here, we used the ``scalar-valued cross-product'' $\times$ defined in equation \eqref{eq:crossproduct} and the Legendre relation \eqref{eq:Legendre} -- see Appendix \ref{app:elliptic} for further details.

\paragraph{Elliptic generalisation of the function $\bm{\Psi}$.} We now turn our attention to the function $\Psi(z)$. We keep the same definition $\Psi(z)=-\vp(z)f(z)$ as in the rational case \eqref{eq:Psi}. Similarly to that case, one easily checks that this function has the same pole structure (on $\C$) as $\vp(z)$ and the evaluations $\Psi(\ha{z}^{\,\pm}_r) = \mp \hbar\,c_\g$ at its zeroes. In addition, it now satisfies the quasi-periodicity property
\begin{equation}\label{eq:QuasiPsi}
    \Psi(z+2\ell_i) = \Psi(z) - 2\beta_{\ell_i}\,\vp(z)\,,
\end{equation}
which follows from the one \eqref{eq:QuasiF} of $f(z)$ and the elliptic nature \eqref{eq:PhiEll} of $\vp(z)$.

The three properties above (pole structure, evaluations at $\ha{z}^{\,\pm}_r$ and quasi-periodicity) uniquely characterise $\Psi(z)$. To prove this statement, let us consider any function $\Psi(z)$ satisfying these conditions and define $\widetilde{f}(z)=- \Psi(z)/\vp(z)$: we then want to prove that $\widetilde{f}(z)=f(z)$, with $f$ as defined in the previous paragraph. Since $\Psi(z)$ and $\vp(z)$ have the same pole structure, the only poles of $\widetilde{f}(z)$ in $\C$ are the zeroes of $\vp(z)$, \textit{i.e.} the points $\ha{z}^{\,\pm}_r$ and their translates by lattice vectors in $\Gamma$. Moreover, one easily checks from the condition $\Psi(\ha{z}^{\,\pm}_r) = \mp \hbar\,c_\g$ that the residue of $\widetilde{f}(z)$ at these points is $\pm \hbar\,c_\g / \vp'(\ha{z}^{\,\pm}_r)$, exactly as $f(z)$\footnote{Note that this also holds at the translated points $\hat{z}_r^\pm + \Gamma$ due to the quasi-periodicity property \eqref{eq:QuasiF} of $\Psi(z)$, which implies $\Psi(\hat{z}_r^\pm + 2n_1\ell_1+2 n_2\ell_2) = \Psi(\hat{z}_r^\pm) = \mp \hbar\,c_\g$, using the fact that $\hat{z}_r^\pm$ is a zero of $\vp(z)$.}. Therefore, the function $f(z)-\widetilde{f}(z)$ has no poles in the complex plane. Let us now observe that the quasi-periodicity property \eqref{eq:QuasiPsi} of $\Psi(z)$ means that $\widetilde{f}(z)$ has the same behaviour \eqref{eq:QuasiF} as $f(z)$. The difference $f(z)-\widetilde{f}(z)$ is therefore properly periodic. In conclusion, $f(z)-\widetilde{f}(z)$ is an elliptic function with no poles: it is a well-known result that such a function is simply a constant. We then obtain $\widetilde{f}(z)=f(z)$, as required, by reabsorbing this constant in the arbitrary parameter $b_0$ in \eqref{eq:fEll}.

To be more precise, the result stated in this paragraph is then that the aforementioned properties of $\Psi(z)$ uniquely characterise its expression as $-\vp(z)f(z)$ for some choice of constant coefficient $b_0$ in $f(z)$. Note that the linear coefficient $b_1$ is already fixed here by the quasi-periodicity condition \eqref{eq:QuasiPsi}, as it enters the expression \eqref{eq:QuasiF} of $\beta_{\ell_i}$. Alternatively, one can think of a situation where we require the quasi-periodicity \eqref{eq:QuasiPsi} to hold for some unspecified numbers $\beta_{\ell_i}$: the statement is then that $\Psi(z)=-\vp(z)f(z)$ for some constant coefficient $b_0$ in $f(z)$ and a linear coefficient $b_1$ defined from $\beta_{\ell_i}$ through the second equation in \eqref{eq:QuasiF}.\footnote{Note that for this to work, the relation \eqref{eq:QuasiF} should hold for both $\beta_{\ell_1}$ and $\beta_{\ell_2}$, with the same $b_1$: this is equivalent to requiring that the numbers $\beta_{\ell_1}$ and $\beta_{\ell_2}$ satisfy the identity \eqref{eq:BetaxEll}. This is however not an additional independent constraint that needs to be included in our hypotheses on $\Psi(z)$: indeed, it follows from its meromorphicity and its quasi-periodicity \eqref{eq:QuasiPsi}. To see that, let us consider the integral of $\Psi(z)$ along the contour encircling the fundamental cell of $\C/\Gamma$, which is a parallelogram with sides $2\ell_1$ and $2\ell_2$. This integral can be easily evaluated using the quasi-periodicity \eqref{eq:QuasiPsi}: indeed, taking into account the orientation, most of the terms coming from parallel sides cancel, leaving only $4(\beta_{\ell_1}\ell_2 - \beta_{\ell_2}\ell_1)$ in the end. Up to the factor 4, this is the left-hand side of the desired identity \eqref{eq:BetaxEll}. We then obtain the equality with the right-hand side by writing the integral as the sum of the residues of $2\ri\pi\,\Psi(z)$ inside the fundamental cell.}

\paragraph{The conjecture and its geometric interpretation.} The elliptic $\p\Psi$-conjecture takes the exact same form \eqref{eq:dPsi} as in the rational case, which we recall here for the reader's convenience:
\begin{equation}\label{eq:dPsiEll}
    \frac{\dd\;}{\dd\tf}\, \omega = \p\Psi = \p_z\Psi(z)\, \dd z\,.
\end{equation}
The only difference in the elliptic setup is the definition of the function $\Psi(z)$, which is explained in detail in the previous paragraph. Partly to justify this generalisation of the rational $\p\Psi$ conjecture, we will check it for a specific example of elliptic integrable $\sigma$-model in section \ref{sec:DPCM}.\\

As in the rational case, the transformation of $\omega$ under an infinitesimal variation of the RG-scale $\tf \mapsto \tf+\delta\tf$ takes the same symbolic form as a change of spectral parameter $z \mapsto z+\delta\tf\,f(z)$. However, this does not correspond to a biholomorphic transformation of $z$ and thus has a non-trivial effect on the physical parameters contained in $\omega$, as one should expect: we refer to the corresponding discussion in the rational subsection \ref{subsec:RatDpsi} for a detailed explanation. The only transformations of $z$ that leave the model completely invariant are the global biholomorphic ones, which in the elliptic case are the dilations and translations, whose infinitesimal form is $\delta z = \epsilon_0 + \epsilon_1\,z$. Thus the coefficients $(b_0,b_1)$ in the definition \eqref{eq:fEll} of $f(z)$ encode the freedom of performing such dilations and translations along the RG-flow without affecting the physical parameters. We note that contrarily to the rational case, there are no special conformal transformations of $z$ among the biholomorphisms here, corresponding to the absence of a term $b_2\, z^2$ in $f(z)$. If the dilation and translation freedom have been fixed, the coefficients $(b_0,b_1)$ have to take specific values so that this fixing is preserved along the RG-flow. Otherwise, they can be arbitrary, reflecting the presence of a redundancy in the parameterisation.

\paragraph{RG-flow of the poles, zeroes and half-periods.} Let us now analyse the consequences of the $\p\Psi$-conjecture \eqref{eq:dPsiEll}. It is well-adapted to describe the RG-flow of certain natural parameters of the model. For instance, the flow of the poles $\lbrace \ha{p}_r \rbrace_{r=1}^n$ and zeroes $\lbrace \ha{z}^{\,\pm}_r \rbrace_{r=1}^M$ of $\omega$ can be derived from this conjecture, as done for the rational models in subsection \ref{subsec:RatDpsi}. In the end, we find that the same formulae \eqref{eq:RGpole} and \eqref{eq:RGzero} still apply in the present case, but with $f(z)$ now taking the elliptic form \eqref{eq:fEll} (and where the related functions $f_\pm(z)$ are defined through \eqref{eq:fpm2}).\\

In addition to these poles and zeroes, natural quantities used to parameterise the elliptic moduli $(C,\omega)$ are the half-periods $\bm\ell=(\ell_1,\ell_2)$ of the torus. Their RG-flow can also be extracted from the $\p\Psi$-conjecture \eqref{eq:dPsiEll}. To do so, we consider the identity $\vp(z+2\ell_i)=\vp(z)$ and impose that it is preserved along the flow $\frac{\dd\;}{\dd\tf}\vp(z)=\p_z\Psi(z)$. A few straightforward manipulations show that this is equivalent to
\begin{equation}
    \p_z\Psi(z+2\ell_i) + 2\frac{\dd\ell_i}{\dd\tf} \p_z\vp(z) = \p_z\Psi(z) \,.
\end{equation}
Comparing to the quasi-periodicity property \eqref{eq:QuasiPsi} of $\Psi(z)$, we thus get
\begin{equation}\label{eq:RGell}
    \frac{\dd\ell_i}{\dd\tf} = \beta_{\ell_i}\,,
\end{equation}
where $\beta_{\ell_i}$ is the quantity defined in equation \eqref{eq:QuasiF}. As anticipated then, $\beta_{\ell_i}$ is therefore identified with the $\beta$-function of the half-period $\ell_i$.

Note that the expression \eqref{eq:QuasiF} of $\beta_{\ell_i}$ contains the parameter $b_1$. This is to be expected: indeed, as explained in the previous paragraph, this coefficient encodes the freedom of performing dilations of the spectral parameter along the RG-flow. As mentioned in subsection \ref{subsec:GeoTwist}, the half-periods $\ell_i$ homogeneously scale under such transformations: it is thus normal that their flow depends on $b_1$. Recall however that the ratio \eqref{eq:tau} of these half-periods forms the elliptic modulus $\tau$, which is invariant under dilations and is an intrinsic geometric characteristic of the torus. The RG-flow of this quantity is easily derived from that of $\ell_i$, yielding
\begin{equation}
    \frac{\dd\tau}{\dd\tf} =  \frac{\beta_{\ell_2} \ell_1 - \beta_{\ell_1} \ell_2}{\ell_1^2}\,.
\end{equation}
We recognise in the numerator the specific combination considered in the identity \eqref{eq:BetaxEll}. We then obtain the $\beta$-function of the elliptic modulus as
\begin{equation}
   \frac{\dd\tau}{\dd\tf} = \beta_\tau =\frac{\ri\pi\hbar\,c_\g}{2\ell_1^2} \sum_{r=1}^M \left( \frac{1}{\vp'(\ha{z}^{\,-}_r)} - \frac{1}{\vp'(\ha{z}^{\,+}_r)} \right)\,.
\end{equation}
As one should expect, the $b_1$-dependence completely dropped out from this expression, in agreement with the fact that $\tau$ is invariant under dilations.

\paragraph{Equivalence with the period-conjecture.} We now show the equivalence of the elliptic $\p\Psi$-conjecture \eqref{eq:dPsiEll} with the period-conjecture \eqref{eq:RGperiods}. The latter is divided into two parts, describing respectively the flow of the absolute periods of $\omega$ and the flow of the relative ones. In the elliptic case, there are two types of absolute periods: the residues $\res_{\hat{p}_r}\omega$ and the A- and B-cycles integrals (see subsection \ref{subsec:Periods}). The latter are the new protagonists appearing in the elliptic setup, compared to the rational one. The analysis of the flow of the residues and the relative periods can be performed exactly as in the rational case: we will not detail this again here and will simply use the results obtained in the rational Subsection \ref{subsec:RatDpsi}. In particular, the main statement is as follows. The part of the period-conjecture which concerns residues and relative periods is equivalent to the flow of $\omega$ taking the form $\frac{\dd\;}{\dd\tf}\omega=\p\widetilde\Psi$, for some meromorphic function $\widetilde\Psi(z)$ on $\C$, having the same pole structure as $\vp(z)$ and satisfying $\widetilde\Psi(\ha{z}^{\,\pm}_r) = \mp \hbar\,c_\g$ at the zeroes of $\omega$.\\

We now work in this setup and want to prove that the rest of the period-conjecture is then equivalent to having $\widetilde\Psi(z)=\Psi(z)$, where $\Psi(z)$ is the specific function appearing in the elliptic $\p\Psi$-conjecture. By ``the rest of the period-conjecture'', we mean the part that does not concern the flow of residues or relative periods: this is the novelty of the elliptic case, namely the RG-invariance of the A- and B-periods $\Pi_A$ and $\Pi_B$. These quantities are defined in equation \eqref{eq:abs} as integrals of $\omega$ over the fundamental cycles $\alpha$ and $\beta$ of the torus $C=\C/\Gamma$. In terms of the coordinate $z$ on $\C$, these cycles can be realised as segments $[z_0,z_0+2\ell_1]$ and $[z_0,z_0+2\ell_2]$, where $z_0$ is any reference point. We then represent the A- and B-periods as
\begin{equation}
    \Pi_A = \int_{z_0}^{z_0+2\ell_1} \omega \qquad \text{ and } \qquad \Pi_B = \int_{z_0}^{z_0+2\ell_2} \omega\,.
\end{equation}
We can compute the flow of these quantities induced by $\frac{\dd\;}{\dd\tf}\omega=\p\widetilde\Psi$, using the general result \eqref{eq:FlowInt} derived in the previous subsection. We then get:
\begin{equation}
    \frac{\dd\;}{\dd \tf} \Pi_{A/B} = \widetilde\Psi(z_0+2\ell_i) - \widetilde\Psi(z_0) + \frac{\dd (z_0+2\ell_i)}{\dd \tf}\, \vp(z_0+2\ell_i) - \frac{\dd z_0}{\dd \tf}\, \vp(z_0)\,,
\end{equation}
where the indices $A$ and $B$ correspond respectively to $i=1$ and $i=2$. The twist function is elliptic so that $\vp(z_0+2\ell_i)=\vp(z_0)$. We thus get various simplifications in the above equation. In the end, we find that the integrals $\Pi_{A/B}$ are RG-invariant if and only if $\widetilde\Psi(z)$ satisfies the quasi-periodicity condition
\begin{equation}
    \widetilde\Psi(z+2\ell_i) = \widetilde\Psi(z) - 2\frac{\dd \ell_i}{\dd \tf} \vp(z)\,.
\end{equation}
This is the same property \eqref{eq:QuasiPsi} as the function $\Psi(z)$. As explained in detail earlier in this subsection, this condition, together with having the same pole structure as $\vp(z)$ and evaluations $\mp \hbar\,c_\g$ at $\ha{z}^{\,\pm}_r$, uniquely characterises $\Psi(z)$. We thus find that $\Pi_{A/B}$ are RG-invariant if and only if $\widetilde\Psi(z) = \Psi(z)$. Altogether, this proves that the period-conjecture is equivalent to the $\p\Psi$-conjecture in the elliptic case as well. We note that for the last step of this proof to work, we need to identify the parameter $\beta_{\ell_i}$ appearing in the quasi-periodicity property \eqref{eq:QuasiPsi} of $\Psi(z)$ with the flow $\frac{\dd \ell_i}{\dd \tf}$ of the half-period $\ell_i$: this gives an alternative confirmation of the formula \eqref{eq:RGell} derived earlier.

\section{1-loop renormalisation of the elliptic integrable deformed PCM}
\label{sec:DPCM}

In the previous section, we have discussed various results and conjectures on the 1-loop renormalisation of a large class of integrable $\s$-models. As mentioned there, the conjectures have been checked for various rational integrable theories. In contrast, the elliptic case is much less explored. The simplest example in the elliptic family is a specific deformation of the Principal Chiral Model (PCM) on $\SLNR$, introduced in~\cite{Lacroix:2023qlz}. In this section, we compute explicitly its 1-loop RG-flow and use this result to check the validity of these conjectures for this new class of models. We start by reviewing the definition of this model.

\subsection{The elliptic integrable deformed PCM}
\label{subsec:IntDPCM}

\paragraph{Twist 1-form.} As explained in Section \ref{sec:IntSigma}, the integrable $\s$-models considered in this article depend on the choice of a twist 1-form $\omega$, which is a meromorphic 1-form on a Riemann surface $C$. Since this section concerns an elliptic model, we will take this surface to be the torus $C=\C/\Gamma$, with half-periods $\bm\ell = (\ell_1,\ell_2) \in \R\times \ri\,\R$. Moreover, we consider a specific twist 1-form, defined as~\cite{Lacroix:2023qlz}
\begin{equation}\label{eq:TwistDPCM}
\omega = \vp(z) \, \dd z = \rho\,\bigl\lbrace \wp(z) - \wp(\ha{z}) \bigr\rbrace \,\dd z \,, 
\end{equation}
where $\rho$ and $\ha{z}$ are real non-zero parameters and $\wp(z)=\wp(z\,;\bm\ell)$ is the Weierstrass $\wp$-function associated with the torus $C=\C/\Gamma$. We refer to the appendix \ref{app:Weierstrass} for a detailed review of this function and its properties. In particular, it is elliptic, even and has a double pole at the origin:
\begin{equation}
    \wp(z+2\ell_i)=\wp(z)\,, \qquad \wp(-z) = \wp(z)\,, \qquad \wp(z) = \frac{1}{z^2} + O(z^2)\,.
\end{equation}
On the torus, the twist 1-form \eqref{eq:TwistDPCM} then has one pole at $0$ (of multiplicity 2) and two simple zeroes at $\pm \ha{z}$. In the notations of section \ref{sec:IntSigma}, this means that
\begin{equation*}
    \text{Poles:} \qquad n=1\,, \qquad \ha{p}_1=0\,, \qquad m_1 = 2\,,
\end{equation*}
\begin{equation}
    \text{Zeroes:} \qquad M=1\,, \qquad  \ha{z}^{\,\pm}_1 = \pm \ha{z} \,.
\end{equation}
In particular, we made a choice of partition of the zeroes into two equal-size subsets $\Zh^\pm = \lbrace \ha{z}^{\,\pm}_1 \rbrace = \lbrace \pm \ha{z} \rbrace$, as required in the general formalism of section \ref{sec:IntSigma}.\\

The moduli $(C,\omega)$ is described here using 4 parameters $(\ell_1,\ell_2,\rho,\ha{z})$. This parametrisation admits a redundancy, corresponding to the freedom of rescaling the spectral parameter $z$: namely, the associated model will be invariant under the transformation $(\ell_1,\ell_2,\rho,\ha{z}) \mapsto a(\ell_1,\ell_2,\rho,\ha{z})$ with $a \neq 0$. This thus leaves a total of 3 physical parameters. Note that there is no translation freedom in this parametrisation: it has been removed by fixing the pole of $\omega$ to $z=0$.

\paragraph{Target space and action.} The target space of the elliptic integrable deformed PCM is the Lie group $\Tc=\SLNR$: the theory is therefore described by a $\SLNR$-valued field\footnote{This is the equivalent of the field $\phi : \Sigma \to \Tc$ considered in subsection \ref{subsec:Basics} for a general $\s$-model with target space $\Tc$. In the present section, we use the notation $g : \Sigma \to \SLNR$, which is standard for $\s$-models on group manifolds.} $g(x^+,x^-)$. It will be useful to introduce the Maurer-Cartan currents
\begin{equation}
    j_\pm = g^{-1}\p_\pm g\,,
\end{equation}
valued in the Lie algebra $\slNR$, and to equip the latter with the non-degenerate invariant bilinear pairing $\langle\cdot,\cdot\rangle=-\Tr(\cdot)$. We then consider an action of the form
\begin{equation}\label{eq:DPCMActionHighRank}
		S_{\text{DPCM}}[g]=\iint_\Sigma \dd x^+ \, \dd x^-\,\left\langle j_+,D[j_-] \right\rangle\,.
\end{equation}
where $D: \slNR \to \slNR$ is a constant linear operator on the Lie algebra, which we call the \textit{deformation operator}. More precisely, the integrable deformed PCM that we are interested in corresponds to a very specific choice for $D$, defined in terms of the parameters $(\ell_1,\ell_2,\rho,\ha{z})$ of the geometric moduli $(C,\omega)$, in a way that ensures its integrability.\\

To introduce this choice of deformation operator $D$, we will need a few additional ingredients. Recall from subsection \ref{subsec:SigmaModel} that elliptic integrable $\s$-models are naturally described in terms of a specific basis of the Lie algebra $\slNC$ called the Belavin basis $\lbrace T_\alpha \rbrace_{\alpha\in\Ab}$, labelled by elements of $\Ab \equiv \Z_N \times \Z_N \setminus \lbrace (0,0) \rbrace$. We refer to appendix \ref{app:Belavin} for more details. The specific integrable operator $D$ acts diagonally on the Belavin basis, \textit{i.e.}
\begin{equation}\label{eq:DBel}
    D[T_\alpha] = D_\alpha\,T_\alpha\,.
\end{equation}
To complete the definition of the model, we now need to give the expression of the coefficients $\lbrace D_\alpha \rbrace_{\alpha\in\Ab}$ in terms of the parameters $(\ell_1,\ell_2,\rho,\ha{z})$. For that, we will use the family of meromorphic elliptic functions $\lbrace r^\alpha(z)\rbrace_{\alpha\in\Ab}$ appearing in the Belavin $\Rc$-matrix \eqref{eq:RBel}, whose definition and properties are reviewed in Appendix \ref{app:ralpha}. We then have
\begin{equation}\label{eq:EllipticDeformationsHighRank}
		D_\alpha=-\rho\frac{r^{\alpha\prime}(\ha{z})}{r^\alpha(\ha{z})}=\rho\,\Big\{Q_\alpha + \zeta(\ha{z})-\zeta(\ha{z}+q_\alpha)\Big\}\,.
\end{equation}
The last equality in this formula is an equivalent rewriting of $D_\alpha$ in terms of the Weierstrass $\zeta$-function \eqref{eq:Zeta} and the numbers $q_\alpha$ and $Q_\alpha$ defined in equation \eqref{eq:qalphadef} --- see appendix \ref{app:elliptic} for further details.

As expected from the previous paragraph, one checks that the coefficient $D_\alpha$ in \eqref{eq:EllipticDeformationsHighRank} is invariant under a rescaling $(\ell_1,\ell_2,\rho,\ha{z}) \mapsto a(\ell_1,\ell_2,\rho,\ha{z})$ of the parameters and thus effectively depends only on three physical coupling constants. This redundancy can for instance be fixed by setting the zero $\ha{z}$ of $\omega$ to a specific value, say $\ha{z}=1$: in this ``fixed-zero parameterisation'', the remaining parameters $(\ell_1,\ell_2,\rho)$ are then all physical. Another possibility is the ``modular parameterisation'', in which we fix the elliptic periods to $2\ell_1=1$ and $2\ell_2=\tau$, with $\tau$ the torus modulus: the physical parameters of the model are then $(\tau,\rho,\ha{z})$. We refer to~\cite[Subsection 4.2]{Lacroix:2023qlz} for a more detailed discussion of these aspects. Here, we will however keep this freedom unfixed, as this will allow us to illustrate some of the general ideas put forward in the previous section.

We note that in the limit $\ell_1\to +\infty$ and $\ell_2 \to +\ri\,\infty$, which corresponds to a decompactification of the torus $C=\C/\Gamma$ to the Riemann sphere $\CP$, the coefficients $D_\alpha$ all tend to the same number so that the operator $D$ becomes proportional to the identity: this is the undeformed limit, in which one recovers the standard PCM. Finally, let us mention that for $N=2$, and up to reality conditions, the model \eqref{eq:DPCMActionHighRank} coincides with the celebrated Cherednik anisotropic integrable $\s$-model on SU(2)~\cite{cherednik_relativistically_1981}.

\paragraph{Equations of motion and Lax connection.} The equations of motion obtained by varying the action \eqref{eq:DPCMActionHighRank} read
\begin{equation}\label{eq:EoM}
    D[\p_+ j_-] + D^t[\p_- j_+] + \bigl[ j_+, D[j_-] \bigr] - \bigl[ D^t[j_+], j_- \bigr] = 0\,,
\end{equation}
where $D^t$ is the transpose of the operator $D$, defined by $\langle X,D[Y] \rangle = \langle D^t[X],Y \rangle$. We note that in addition to this equation of motion, the currents $j_\pm$ satisfy the Maurer-Cartan identity
\begin{equation}\label{eq:MC}
    \p_+ j_- - \p_- j_+ + \bigl[ j_+, j_- \bigr] = 0\,,
\end{equation}
which holds off-shell, as a direct consequence of the definition $j_\pm=g^{-1}\p_+ g$.\\

For the particular choice \eqref{eq:DBel}--\eqref{eq:EllipticDeformationsHighRank} of deformation operator $D$, the equation of motion \eqref{eq:EoM} and the Maurer-Cartan identity \eqref{eq:MC} are equivalent to the flatness of the following Lax connection:
\begin{equation}\label{eq:HighRankLax}
    \Lc_\pm(z)=\sum_{\alpha\in\mathbb{A}}\frac{r^\alpha(z\mp\ha{z})}{r^\alpha(\mp \ha{z})}\,j^\alpha_\pm \,T_\alpha\,.
\end{equation}
Here, we used the decomposition $j_\pm = \sum_{\alpha\in\Ab} j_\pm^\alpha\,T_\alpha$ of the currents in the Belavin basis. The light-cone component $\Lc_\pm(z)$ of this Lax connection is elliptic with periods $2N\ell_i$ and has a simple pole at $\ha{z}^{\,\pm}_1=\pm \ha{z}$ and its translates by $\Gamma$, which are the zeroes of the twist 1-form \eqref{eq:TwistDPCM}. This is in agreement with the general formalism of subsection \ref{subsec:SigmaModel} and in particular the equation \eqref{eq:EllipticLax}.

The specific deformation operator \eqref{eq:DBel}--\eqref{eq:EllipticDeformationsHighRank} and the corresponding Lax connection \eqref{eq:HighRankLax} were derived in~\cite{Lacroix:2023qlz} using equivariant elliptic 4d-CS theory, which automatically ensures the flatness of $\Lc_\pm(z)$. The latter can also be checked by direct computation, starting from the equation of motion \eqref{eq:EoM} and using various properties of the functions $\lbrace r^\alpha(z)\rbrace_{\alpha\in\Ab}$, in particular the Fay identity \eqref{eq:Fay}. Finally, we note that the spatial component of this Lax connection satisfies a Maillet bracket, with the Belavin $\Rc$-matrix \eqref{eq:RBel} as seed and the twist function \eqref{eq:TwistDPCM}.

\subsection{1-loop RG-flow of the deformation operator}

We now turn to the 1-loop renormalisation of the elliptic integrable deformed PCM introduced in the previous subsection. We will proceed in two main steps. In this subsection, we will start by determining the renormalisation of the deformation operator $D$ characterising the model, using Ricci-flow techniques. Subsequently, we will prove in the next subsection that this renormalisation can be reabsorbed in a flow of the defining parameters $(\ell_1,\ell_2,\rho,\ha{z})$ which enter the expression of $D$, thereby proving the renormalisability of the theory. The computation of the RG-flow of the deformation operator $D$ will itself be decomposed in three steps: we will first describe it for a general operator $D$, not yet specialising to the specific choice making the theory integrable\footnote{We note that this RG-flow with arbitrary deformation operator was already obtained in~\cite{Sfetsos:2014jfa}. We include its derivation in the present paper for completeness.}; secondly, we will restrict to the case of an operator \eqref{eq:DBel} which is diagonal in the Belavin basis; and finally we will consider the explicit integrable choice \eqref{eq:EllipticDeformationsHighRank} of $D$. Various proofs and technical computations are relegated to appendix \ref{sec:RGAppendix}.

\paragraph{General deformation operator.} We start by considering the $\s$-model \eqref{eq:DPCMActionHighRank} with a generic deformation operator $D$. The latter can be characterised by its matrix entries
\begin{equation}
    D_{\alpha\beta} = \bigl\langle T_\alpha, D[T_\beta] \bigr\rangle
\end{equation}
in a basis\footnote{Note that for this paragraph, $\lbrace T_\alpha \rbrace$ can in principle be any basis of $\slNC$, not necessarily the Belavin one. To lighten the conventions and as we will eventually be using the Belavin basis specifically, we do not introduce a different notation for the basis elements $T_\alpha$. In the present paragraph, $\alpha$ can then be thought of as an abstract index labelling this basis and not necessarily an element of the set $\Ab \equiv \Z_N \times \Z_N \setminus \lbrace (0,0) \rbrace$ as for the Belavin one. For such a generic case, we assume implicit summations over repeated indices: once we will restrict to explicit labels in $\Ab$, we will always indicate the sums explicitly.} $\lbrace T_\alpha \rbrace$ of $\slNC$. To describe the geometry underlying this $\s$-model, it is useful to introduce a set of (local) coordinates $(\phi^{\,i})_{i=1}^d$ on the target space, which in the present case is the group manifold $\SLNR$: in practice, this means that we think of the group-valued field $g$ as a matrix in $\SLNR$ whose entries are expressed in terms of $d=N^2-1$ real scalar fields $\phi^{\,i}$. The currents $j_\pm$ can then be re-expressed in terms of these fields as
\begin{equation}
    j_\pm = \e\alpha i(\phi)\,\p_\pm \phi^{\,i}\, T_\alpha\,,
\end{equation}
where $\e\alpha i(\phi)$ is the component of $g^{-1}\frac{\p g}{\p\phi^{\,i}}$ along $T_\alpha$, which is then a function of the coordinate fields $(\phi^{\,i})_{i=1}^d$ but not of their derivatives. With these notations, the action \eqref{eq:DPCMActionHighRank} can then be rewritten as
\begin{equation}
    S_{\text{DPCM}}[g]=\iint_\Sigma \dd x^+ \, \dd x^-\,D_{\alpha\beta}\,\e\alpha i(\phi)\,\e\beta j(\phi)\,\p_+\phi^{\,i}\, \p_-\phi^{j}\,.
\end{equation}
Compared to the general form \eqref{eq:action} of a $\s$-model action, we then see that the metric and B-field of this theory are identified through
\begin{equation}\label{eq:GBVielbeins}
    G_{ij}(\phi) + B_{ij}(\phi) = D_{\alpha\beta}\,\e\alpha i(\phi)\,\e\beta j(\phi)\,.
\end{equation}
In this context, the object $\e\alpha i(\phi)$ is called a \textit{vielbein} -- see appendices \ref{subsec:RTsigma} and \ref{subsec:RTArbitrary} for more details. It completely captures the dependence of the metric and B-field on the coordinates $(\phi^{\,i})_{i=1}^d$, while the dependence on the coupling constants/parameters is encoded in the coefficients $D_{\alpha\beta}$. We will denote the inverse of $\e\alpha i(\phi)$ by $\e i\alpha(\phi)$.\\

We now come to the question of the renormalisation of this theory. For a general $\s$-model, it is well-known~\cite{Ecker:1972bm,Honerkamp:1971sh,Friedan:1980jf,Curtright:1984dz} that the 1-loop renormalisation of the metric and B-field takes the form of the Ricci-flow \eqref{eq:Ricci}, controlled by the torsionful Ricci tensor $R^+_{ij}$. For the present form \eqref{eq:GBVielbeins} of the metric and B-field, this equation simply translates to a flow of the coefficients $D_{\alpha\beta}$:
\begin{equation}
    \frac{\dd\;}{\dd \tf} D_{\alpha\beta} = \hbar\, R^+_{\alpha\beta}\,,
\end{equation}
where $R^+_{\alpha\beta}=R^+_{ij}\,\e i\alpha \e j\beta$ is the torsionful Ricci tensor expressed in ``flat-space indices'' $\alpha,\beta$. The explicit value of this tensor was computed in~\cite[Subsection 5.2]{Sfetsos:2014jfa}. For completeness and self-containment, we rederive it here in appendix \ref{subsec:RTArbitrary}. In the end, one has
\begin{equation}
	R^+_{\alpha\beta}= - \Omega_{\rho\alpha}^{-\phantom{\eta}\sigma}\Omega_{\sigma\beta}^{+\phantom{\lambda}\rho}.\label{eq:RiccifromSpin}
\end{equation}
 where $\Omega_{\rho\sigma}^{\pm \phantom{\alpha}\alpha}$ are the so-called spin connections, which read
\begin{subequations}\label{eq:Omega}
	\begin{equation}
		\Omega_{\rho\sigma}^{+\phantom{\alpha}\alpha}=I^{\alpha\lambda} \, \Big\{f_{\rho\sigma}^{\phantom{\rho\sigma}\eta}D_{\eta\lambda}-f_{\rho\lambda}^{\phantom{\rho\lambda}\eta}D_{\eta\sigma}-f_{\sigma\lambda}^{\phantom{\sigma\lambda}\eta}D_{\rho\eta}\Big\},
	\end{equation}
	\begin{equation}
		\Omega_{\rho\sigma}^{-\phantom{\alpha}\alpha}=I^{\alpha\lambda}\,\Big\{f_{\rho\sigma}^{\phantom{\rho\sigma}\eta}D_{\lambda\eta}-f_{\rho\lambda}^{\phantom{\rho\lambda}\eta}D_{\sigma\eta}-f_{\sigma\lambda}^{\phantom{\sigma\lambda}\eta}D_{\eta\rho}\Big\}\,.
	\end{equation}
\end{subequations}
In this equation, we used the structure constants of the basis $\lbrace T_\alpha \rbrace$, defined by $[T_\alpha,T_\beta] = f_{\alpha\beta}^{\phantom{\alpha\beta}\gamma}\,T_\gamma$, as well as the symmetric tensor $I^{\alpha\beta}$, defined as the inverse of $D_{\alpha\beta}+D_{\beta\alpha}$.

\paragraph{Diagonal deformations.} We now focus more specifically on deformations of the form \eqref{eq:DBel}, \textit{i.e.} that are diagonal in the Belavin basis $\lbrace T_\alpha \rbrace_{\alpha\in\Ab}$. We refer to the appendix \ref{app:Belavin} for more details about this basis. In particular, recall that its elements $T_\alpha$ are labelled by couples $\alpha=(\alpha_1,\alpha_2)$ in $\Ab \equiv \Z_N \times \Z_N \setminus \lbrace (0,0) \rbrace$. For our purposes, we will need their commutation relations \eqref{eq:Tcommutator}, as well as the expression \eqref{eq:slcnkappa} of the form $\langle\cdot,\cdot\rangle$ in this basis. In the present language, these imply that the deformation coefficients and structure constants take the form
\begin{equation}\label{eq:DDiagonal}
    D_{\alpha\beta} = -\delta_{\alpha+\beta,0}\,\xi^{-\beta_1\beta_2}\,D_\beta \qquad \text{ and } \qquad f_{\alpha\beta}^{\phantom{\alpha\beta}\gamma} = \frac{\delta_{\alpha+\beta,\gamma} }{\sqrt{N}}\left(\xi^{\alpha_1\beta_2}-\xi^{\beta_1\alpha_2}\right)\,,
\end{equation}
where we recall that $\xi=\exp\left(\frac{2\ri\pi}{N}\right)$.

Substituting the above expressions of $D_{\alpha\beta}$ and $f_{\alpha\beta}^{\phantom{\alpha\beta}\gamma}$ in equations \eqref{eq:RiccifromSpin}--\eqref{eq:Omega}, one can determine the form of the Ricci tensor $R^+_{\alpha\beta}$ for such diagonal deformations. The details of this computation are given in appendix \ref{subsec:RTDiagonal}. In the end, we show that
\begin{equation}
	R^+_{\alpha\beta}= -\delta_{\alpha+\beta,0}\,\xi^{-\beta_1\beta_2}\,R^+_\beta\,,\label{eq:RDiagonal}
\end{equation}
where
\begin{equation}\label{eq:RalphaDiag}
	R^+_{\alpha}=\frac{4}{N} \sum_{\substack{\theta,\sigma\in\mathbb{A}\\ \sigma-\theta=\alpha}}\sin^2\left(\pi\frac{\sigma\times\alpha}{N}\right)\frac{\left\{D_\alpha+D_{\theta}-D_\sigma\right\}\left\{D_\alpha+D_{-\sigma}-D_{-\theta}\right\}}{\{D_\theta+D_{-\theta}\}\{D_\sigma+D_{-\sigma}\}},
\end{equation}
written with the "cross-product" $\times$ defined in \eqref{eq:crossproduct}.
Notice in particular that the form of $R^+_{\alpha\beta}$ in equation \eqref{eq:RDiagonal} is the same as that of the deformation coefficients $D_{\alpha\beta}$ in \eqref{eq:DDiagonal}. Thus, diagonal deformations lead to diagonal Ricci tensors and are therefore stable under 1-loop RG-flow, with the $\beta$-function of $D_\alpha$ being given by $\hbar\,R^+_\alpha$.

\paragraph{Elliptic integrable deformation.} We finally specialise to the elliptic integrable deformation, for which $D_\alpha$ is given by \eqref{eq:EllipticDeformationsHighRank}, in terms of the parameters $(\ell_1,\ell_2,\rho,\ha{z})$. Reinserting this expression in equation \eqref{eq:RalphaDiag}, we derive in appendix \ref{subsec:RTIntegrable} the expression of $R_\alpha^+$ in the integrable case. This computation requires various manipulations and in particular (sufficiently regularised versions of) the Fay identity \eqref{eq:Fay} obeyed by the functions $\lbrace r^\alpha(z) \rbrace_{\alpha\in\Ab}$. The final result reads
\begin{equation}\label{eq:Rr}
	R_\alpha^+=-2N\frac{r^{\alpha\prime}(2\ha{z})}{r^\alpha(\ha{z})^2}\,.
\end{equation}
Despite the compactness of this form, it will be useful for extracting the $\beta$-functions of the theory to re-express this coefficient solely in terms of the Weierstrass $\wp$- and $\zeta$-functions. This is also done in detail in appendix \ref{subsec:RTIntegrable}, using various addition identities obeyed by these functions and yields
\begin{equation}
	R_\alpha^+=\frac{N}{\wp'(\ha{z})}\left[2\Big\{\wp(\ha{z}+q_\alpha)-\wp(\ha{z})\Big\}\Big\{Q_\alpha+\zeta(\ha{z})-\zeta(\ha{z}+q_\alpha)+\frac{\wp''(\ha{z})}{2\wp'(\ha{z})}\Big\}-\Big\{\wp'(\ha{z}+q_\alpha)-\wp'(\ha{z})\Big\}\right]\,.\label{eq:RicciFlowofD}
\end{equation}

\subsection{Extracting the RG-flow}\label{subsec:FindingtheRG-Flow}

From the results of the previous subsection, we find that the 1-loop RG-flow of the elliptic integrable deformed PCM is governed by
\begin{equation}
    \frac{\dd D_{\alpha}}{\dd \mathfrak{t}} = \hbar\,R^+_{\alpha}\,,\label{eq:RGD}
\end{equation}
where $D_\alpha$ and $R^+_\alpha$ are respectively given by \eqref{eq:EllipticDeformationsHighRank} and \eqref{eq:RicciFlowofD}. We will now show that we can reabsorb this flow of $D_\alpha$ into a running of the parameters $(\ell_1,\ell_2,\rho,\ha{z})$ specifying the $\sigma$-model.\footnote{In the case $N=2$, this $\s$-model coincides, up to reality conditions, with the Cherednik model on SU(2)~\cite{cherednik_relativistically_1981}. The 1-loop RG-flow of this theory was derived in~\cite{Cvetic:2001zx} (using a different parameterisation than the one considered here).} Thus, we need to prove that we can write the RG-flow \eqref{eq:RGD} in the form
\begin{equation}
	\frac{\dd D_\alpha}{\dd\mathfrak{t}}=\beta_{\ell_1}\frac{\partial D_\alpha}{\partial \ell_1}+\beta_{\ell_2}\frac{\p D_\alpha}{\p \ell_2}+\beta_{\rho}\frac{\partial D_\alpha}{\partial \rho}+\beta_{\hat{z}}\frac{\partial D_\alpha}{\partial \hat{z}}\,,\label{eq:Dflowlotsofder}
\end{equation}
where the coefficients $\beta_{\bullet}$ are the same for all $\alpha\in\mathbb{A}$ and are then interpreted as the $\beta$-functions of the parameters.  Using various identities, we can calculate the derivatives of $D_\alpha$ appearing in the right-hand side of the above equation: this is done in appendix \ref{subsec:MatchingRG}. One then finds that it is possible to match \eqref{eq:Dflowlotsofder} with \eqref{eq:RicciFlowofD}, leading to the following set of $\beta$-functions:
\begin{equation}
    \beta_{\ell_i}=\hbar N\left( \frac{2 L_i}{\rho\,\wp'(\ha{z})} + b\,\ell_i \right)\,,\quad
    \beta_\rho = \hbar N\left( -\frac{2\wp(\ha{z})}{\wp'(\ha{z})}+ b\, \rho \right)\,,\quad \beta_{\hat{z}}=\hbar N\left(\frac{2\zeta(\ha{z})\,\wp'(\ha{z})+\wp''(\ha{z})}{\rho\,\wp'(\ha{z})^2}+b\,\ha{z} \right)\,.\label{eq:betafunctions}
\end{equation}
Here, $b$ is an unconstrained variable: in practice, it appears in the computation because the comparison of \eqref{eq:Dflowlotsofder} with \eqref{eq:RicciFlowofD} leads to 3 linear equations for 4 unknown $\beta$-functions. This is to be expected: indeed, it is clear from the above formula that the effect of $b$ on the RG-flow is to shift $(\ell_1,\ell_2,\rho,\ha{z})$ by $b\hbar N(\ell_1,\ell_2,\rho,\ha{z})$ and thus corresponds to performing a homogeneous dilation of these parameters along the flow. As discussed in subsection \ref{subsec:IntDPCM}, such a transformation simply corresponds to a rescaling of the spectral parameter and leaves the theory unchanged.\\

To describe the RG-flow purely in terms of physical parameters, we have to fix a choice of parametrisation. Recall the two different possibilities discussed in subsection \ref{subsec:IntDPCM}, namely the fixed-zero parametrisation and the modular parametrisation. In the first one, we fix $\ha{z}=1$. To conserve this property along the flow, we thus choose $b$ such that $\beta_{\hat{z}}=0$. We then find the following set of $\beta$-functions for the remaining variables:
\begin{equation}\label{eq:fixedzerobeta}
    \beta_{\ell_i}=\frac{2\hbar N}{\rho\,\wp'(1)}\left[L_i-\ell_i\Big\{\zeta(1)+\frac{\wp''(1)}{2\wp'(1)}\Big\}\right]\,,\qquad \beta_\rho=-\frac{2\hbar N}{\wp'(1)}\left[\wp(1)+\zeta(1)+\frac{\wp''(1)}{2\wp'(1)}\right]\,.
\end{equation}
Alternatively, if we consider the modular parameterization, \textit{i.e.} fix $2\ell_1=1$, denote $2\ell_2=\tau$ and choose $b$ such that $\beta_{\ell_1}=0$, we find instead
\begin{equation}
    \beta_\tau=\frac{4\ri\pi \,\hbar N}{\rho\,\wp'(\ha{z})}\,,\qquad \beta_{\rho}=-\frac{2\hbar N}{\wp'(\ha{z})}\Big[\wp(\ha{z})-2\eta_W\Big]\,, \qquad \beta_{\hat{z}}=\frac{2\hbar N}{\rho\,\wp'(\ha{z})}\left[\zeta(\ha{z})+\frac{\wp''(\ha{z})}{2\wp'(\ha{z})}-2\ha{z}\,\eta_W\right]\,, \label{eq:modularbeta}
\end{equation}
where $\eta_W(\tau)=L_1(1/2,\tau/2)$ denotes the \textit{Weierstrass $\eta$-function}\footnote{Not to be confused with the more familiar Dedekind $\eta$-function. Note that the appearance of $\pi$ in the first equation of \eqref{eq:modularbeta} is due to the use of the Legendre identity \eqref{eq:Legendre}.}. We note that $\tau$ has a non-trivial $\beta$-function, meaning that the torus $\C/(\mathbb{Z}+\tau\mathbb{Z})$ is changing along the RG-flow.

\subsection[RG-Flow of \texorpdfstring{$\omega$}{omega} and Check of the Conjectures]{RG-Flow of \texorpdfstring{$\bm\omega$}{omega} and check of the conjectures}
\label{subsec:ConjDPCM}

We now want to use the results obtained above to check the general conjectures discussed in section \ref{sec:GeoRG}, which concerned the 1-loop RG-flow of the geometric data $(C,\omega)$ from which integrable $\s$-models are built. For the elliptic deformed PCM considered here, recall that $C=\C/\Gamma$ and that the twist 1-form $\omega$ was defined as \eqref{eq:TwistDPCM}, in terms of the parameters $(\ell_1,\ell_2,\rho,\ha{z})$.

\paragraph{The $\bm{\p\Psi}$-conjecture.} The RG-flow of the twist 1-form is given by
\begin{equation}
	\frac{\dd\;}{\dd\tf}\, \omega = \Bigl\lbrace \beta_{\ell_1}\frac{\partial \varphi(z)}{\partial \ell_1}+\beta_{\ell_2}\frac{\p \varphi(z)}{\p \ell_2}+\beta_{\rho}\frac{\partial \varphi(z)}{\partial \rho}+\beta_{\hat{z}}\frac{\partial \varphi(z)}{\partial \hat{z}} \Bigr\rbrace\, \dd z\,,
\end{equation}
with $\vp(z)=\rho\lbrace\wp(z)-\wp(\ha{z})\rbrace$. The partial derivatives in the above equation can be computed using techniques similar to the ones of appendix \ref{subsec:MatchingRG} while the $\beta$-functions were found explicitly in \eqref{eq:betafunctions}. After various manipulations and the application of many functional elliptic identities, that we shall not detail here, we find
\begin{equation}
	\frac{\dd\;}{\dd\tf}\, \omega =-\hbar N\frac{\dd}{\dd z}\left[\bigg\{\frac{\zeta(z-\ha{z})}{\wp'(\ha{z})}-\frac{\zeta(z+\ha{z})}{\wp'(-\ha{z})}+bz\,\rho\bigg\}\bigg\{\wp(z)-\wp(\ha{z})\bigg\}\right]\dd z\,.\label{eq:flowofelliptictwist}
\end{equation}
This flow takes the exact same form as the $\p\Psi$-conjecture \eqref{eq:dPsiEll}, with the identification
\begin{equation}
    \Psi(z) = -f(z)\,\vp(z)\,, \qquad f(z) = \hbar N \bigg\{\frac{\zeta(z-\ha{z})}{\rho\,\wp'(\ha{z})}-\frac{\zeta(z+\ha{z})}{\rho\,\wp'(-\ha{z})}+bz\bigg\}\,.
\end{equation}
Using that the dual Coxeter number of $\g=\slNR$ is $c_{\g}=N$, one easily checks that this expression for $f(z)$ agrees with the one \eqref{eq:fEll} appearing in the general elliptic $\p\Psi$-conjecture, if one sets $b_0=0$ and $b_1=b$. The fact that $b_0$ is fixed to a specific value here is natural. Indeed, as explained in subsection \ref{subsec:EllDpsi}, $b_0$ encodes the freedom of performing translations of the spectral parameter along the flow. For the deformed PCM, we have fixed this freedom by setting the pole of $\omega$ at $z=0$: preserving this choice under the RG-flow \eqref{eq:dPsiEll} is equivalent to setting $b_0=0$. In contrast, as discussed around \eqref{eq:betafunctions} and in agreement with its interpretation in subsection \ref{subsec:EllDpsi}, the coefficient $b_1=b$ corresponds to rescalings of $z$ and is left unconstrained here as the dilation freedom was not fixed in the parameterisation $(\ell_1,\ell_2,\rho,\ha{z})$.

\paragraph{The period-conjecture.} In section \ref{sec:GeoRG}, we have proven that the $\p\Psi$-conjecture is equivalent to the period-conjecture \eqref{eq:RGperiods} of Costello. The latter should thus also hold for the elliptic integrable deformed PCM. Let us check this explicitly to illustrate some of the notions discussed in subsection \ref{subsec:Periods}. In particular, recall that this conjecture is formulated in terms of the periods of $\omega$, which are integrals of this 1-form along well-chosen paths in $C=\C/\Gamma$. The simplest examples of absolute periods are the residues of $\omega$ at its poles: in the present case, the twist 1-form \eqref{eq:TwistDPCM} of the deformed PCM has only one pole at $z=0$, with vanishing residue. This residue is then trivially conserved under the RG-flow, in agreement with the period-conjecture \eqref{eq:RGperiods}.

Since we are considering an elliptic model, we have two additional absolute periods, namely the integrals of $\omega$ along the A- and B-cycles of the torus. In terms of the coordinate $z$, these can be computed as the integrals of $\vp(z) = \rho\lbrace \wp(z) - \wp(\ha{z}) \rbrace$ over $z\in[z_0,z_0+2\ell_i]$, where $z_0$ is any reference point. Using the relation $\wp(z)=-\zeta'(z)$ and the quasi-periodicity property \eqref{eq:QuasiZeta} of the Weierstrass $\zeta$-function, one easily finds
\begin{equation}
    \Pi_{A/B} = - 2\rho\,\bigl\lbrace L_i  + \ell_i\,\wp(\ha{z}) \bigr\rbrace\,,
\end{equation}
where $A$ and $B$ correspond respectively to $i=1$ and $i=2$. As expected these are independent of the reference point $z_0$. The RG-flow of these quantities can be computed from the $\beta$-functions \eqref{eq:betafunctions} of $(\ell_1,\ell_2,\rho,\ha{z})$ determined earlier, using various elliptic identities and techniques similar to the ones developed in appendix \ref{subsec:MatchingRG}. In the end, we find that the absolute periods $\Pi_{A/B}$ are RG-invariants, in agreement with the period-conjecture \eqref{eq:RGperiods}.

Finally, let us turn our attention to the relative periods, which are integrals of $\omega$ between two of its zeroes. In the present case, there are two such zeroes $\pm \ha{z}$, so there exists only one relative period $\Pi_0$, defined as the integral of $\vp(z) = \rho\lbrace \wp(z) - \wp(\ha{z}) \rbrace$ from $+\ha{z}$ to $-\ha{z}$. Explicitly, we find
\begin{equation}
    \Pi_0 = 2\rho\bigl\lbrace \zeta(\ha{z}) + \ha{z}\,\wp(\ha{z}) \bigr\rbrace\,.
\end{equation}
The RG-flow of this quantity can also be computed using the $\beta$-functions \eqref{eq:betafunctions}, various identities obeyed by the Weierstrass functions and the methods of appendix \ref{subsec:MatchingRG}. This yields
\begin{equation}
    \frac{\dd\;}{\dd\tf} \Pi_0 = 2\hbar\,N\,, \qquad \text{ hence} \qquad  \frac{\Pi_0}{2\hbar\,c_\g} = \tf - \tf_0 = \frac{1}{4\pi} \log\left( \frac{\mu}{\mu_0} \right)\,,
\end{equation}
where we recall that $\mu$ is the RG-scale and $\mu_0$ is a reference energy scale. This means that the last part of the period-conjecture \eqref{eq:RGperiods} also holds for the deformed PCM, as expected.

These results allow us to illustrate the strength of this conjecture. Indeed, starting from the expression \eqref{eq:betafunctions} of the $\beta$-functions, it would be a rather arduous task to guess which combinations of $(\ell_1,\ell_2,\rho,\ha{z})$ are RG-invariants or which one can be identified with the RG-parameter $\tf$. Remarkably, the period-conjecture provides a very simple way of constructing such quantities, thus trivialising the RG-flow of the theory. We stress here that such a process will also apply (conjecturally) to integrable $\s$-models with much more complicated twist 1-forms $\omega$, whose target space geometry and RG-flow can in general be quite involved.

\section{Conclusion and perspectives}
\label{sec:Conc}

\paragraph{Summary.} The main topic of this article was the 1-loop renormalisation of a large class of rational and elliptic integrable $\s$-models obtained from the formalisms of 4-dimensional Chern-Simons theory and affine Gaudin models. A crucial role was played by the geometric data $(C,\omega,\Zh^\pm)$, which is part of the defining ingredients of these theories and is deeply related to the analytic structure of their Lax connection as a function of the spectral parameter (see section \ref{sec:IntSigma}). In particular, this data canonically encodes the continuous parameters of these models which flow under the renormalisation group. In this context, we discussed in detail two conjectural formulae for the 1-loop RG-flow of this geometric data, which we called the $\p\Psi$- and period-conjectures and which were first proposed in~\cite{delduc_rg_2021,derryberry_lax_2021} and partially proven in~\cite{hassler_rg_2021,Hassler:2023xwn}. In this article, we established the equivalence of these conjectures, after also proposing a generalisation of the former from the rational to the elliptic case.

These two formulations each have their advantages. The $\p\Psi$-one gives a direct and explicit expression for the 1-loop RG-flow of the twist 1-form $\omega$ and can be used to extract the $\beta$-functions of some of the natural parameters of the models, such as the zeroes and poles of this 1-form. However, the resulting differential equations obeyed by these parameters are in general highly coupled and non-linear and it is thus not apparent how to solve them explicitly. Remarkably, an answer to this problem is offered by the period-conjecture. The latter is formulated more implicitly, in terms of certain well-chosen integrals of $\omega$ called periods, and completely trivialises the 1-loop RG-flow. Indeed, these periods are either RG-invariants or grow linearly with the logarithm of the RG scale $\mu$. 

The final part of the paper concerned the renormalisation of an explicit example of elliptic integrable $\s$-model recently introduced in~\cite{Lacroix:2023qlz} and which takes the form of a deformation of the Principal Chiral Model on $\SLNR$. More precisely, we proved the 1-loop renormalisability of this model, computed the corresponding $\beta$-functions and used this result to check the validity of the $\p\Psi$- and period-conjectures for this example.

\paragraph{Towards a general proof.} The most natural perspective of the present article is to prove the period- and $\p\Psi$-conjectures in full generality. In the recent works~\cite{hassler_rg_2021,Hassler:2023xwn}, they were established (in the $\p\Psi$-formulation) for a class of rational models whose twist 1-form $\omega$ has a double pole at infinity. The rational case with an arbitrary pole structure at infinity and the elliptic case are for the moment open questions, although they have been checked in various examples. One possible approach towards a general proof would be to follow the same strategy as in~\cite{hassler_rg_2021,Hassler:2023xwn}, using the formalism of so-called $\mathcal{E}$-models~\cite{Klimcik:1995ux,Klimcik:1995dy,Valent:2009nv,sfetsos_renormalization_2010,lacroix_integrable_2021,Liniado:2023uoo}. As an alternative, we expect that these conjectures can be proven using the ``universal divergences'' approach recently put forward in~\cite{levine_universal_2023,Levine:2023wvt}. This will be the subject of a future work~\cite{ToAppear:Universal}.

\paragraph{Extension to other models.} It would also be interesting to study the period- and $\p\Psi$-conjectures for more general classes of integrable $\s$-models. For instance, the family considered in this paper does not include $\s$-models on symmetric spaces or more generally on $\Z_T$-cosets~\cite{Young:2005jv}. These theories, as well as deformations~\cite{delduc_classical_2013,hoare_integrable_2022} and generalisations~\cite{Arutyunov:2020sdo} thereof, can also be described using the formalisms of affine Gaudin models and 4D Chern-Simons, by introducing appropriate $\Z_T$-equivariance conditions on the twist 1-form $\omega$ and on the fields (see for instance~\cite{vicedo_integrable_2019,costello_gauge_2019,Arutyunov:2020sdo,schmidtt_symmetric_2021}). We expect the period- and $\p\Psi$-conjectures to also hold for the 1-loop renormalisation of these models. This can be checked on various examples and we hope that it will be possible to study the general case using similar techniques as the ones discussed in the previous paragraph and our upcoming work~\cite{ToAppear:Universal}.

Another possible extension is to consider the integrable $\sigma$-models introduced in~\cite[Section 15]{costello_gauge_2019} and~\cite{derryberry_lax_2021}, whose spectral parameter is valued in a higher-genus Riemann surface $C$. The metric and B-field of these models are defined implicitly in terms of a geometric object called the Szegö kernel, but it is in general difficult to find an explicit description of this quantity, making the study of these theories more complicated. It was conjectured in~\cite{derryberry_lax_2021} that their 1-loop renormalisation also follows the period-conjecture, which is easily formulated for a twist 1-form $\omega$ defined on an arbitrary compact Riemann surface $C$. An equivalent of the $\p\Psi$-conjecture for these higher-genus models has not been studied yet and forms a natural perspective for future developments.

\paragraph{Identification of the RG-parameter.} One of the remarkable advantages of the period-conjecture is that it completely trivialises the 1-loop RG-flow. More precisely, it allows us to find well-chosen combinations of the parameters of the model which are all RG-invariants, except for one which is simply identified with the RG-parameter $\tf-\tf_0 = \frac{1}{4\pi} \log(\mu/\mu_0)$ (see subsection \ref{subsec:Periods}).

We note that such a parameterisation was found earlier for certain specific integrable $\sigma$-models using a more direct approach, for instance the sausage model in~\cite[Equation (3.7)]{Fateev:1992tk} and the Fateev model in~\cite[Equation (73)]{Fateev:1996ea}.  These examples fall into the framework of affine Gaudin models / 4D Chern-Simons and correspond to specific twist 1-forms $\omega$~\cite{delduc_classical_2013,Delduc:2015xdm,vicedo_integrable_2019}. A direct computation shows that the quantity found in~\cite{Fateev:1992tk,Fateev:1996ea} as the RG-parameter coincides (up to the prefactor $2\hbar\,c_\g)$ with a period of $\omega$ appearing in the period-conjecture. However, we stress that the latter applies quite more generally and thus allows the identification of the RG-parameter for all models with twist 1-forms.  In the references~\cite{Fateev:1992tk,Fateev:1996ea}, this identification of $\tf-\tf_0$ in terms of the coupling constants of the metric/B-field was used to study various properties of the models through Lagrangian perturbation theory, including, for instance, the computation of ground state energies and minisuperspace approximations\footnote{In particular, these results can be compared with the ones obtained from the conjectured Factorised Scattering and Thermodynamic Bethe Ansatz, when available, providing important support for these conjectures. See also the non-exhaustive list of references~\cite{Hasenfratz:1990zz,Evans:1995dn,Appadu:2018ioy,Fateev:2018yos,Fateev:2019xuq} for similar developments in various other models with twist 1-forms.}. It would be interesting to explore whether similar techniques can be applied to integrable $\s$-models corresponding to more general twist 1-forms.

\paragraph{Extension to higher-loops.} Another natural perspective is the generalisation of the $\p\Psi$- and period-conjectures at higher-loop. The study of the RG-flow of integrable $\s$-models beyond one-loop has been an active subject of research in recent years, see for instance~\cite{hassler_rg_2021,Hoare:2019ark,Georgiou:2019nbz,Levine:2021fof,Alfimov:2021sir}. In particular, it was argued in these references that renormalisability at higher-loop generally requires adding quantum corrections to the metric and B-field. It would be interesting to see if these corrections and the higher-loop RG-flow can be understood in a systematic way using the geometric language of twist 1-forms, as done at 1-loop with the $\p\Psi$- and period-conjectures. An interesting example to explore these aspects is the sausage model, for which conjectural expressions for the all-loop metric and RG-flow were proposed in~\cite{Hoare:2019ark} and~\cite{Alfimov:2021sir} (in two different schemes). In particular, the work~\cite{Alfimov:2021sir} discussed the explicit solution of this RG-flow and found a combination of the coupling constants which coincides with the RG-parameter $\tf-\tf_0$ at all loop (see~\cite[Footnote 3]{Alfimov:2021sir}). This is an extension of the 1-loop result of~\cite{Fateev:1992tk}, which as discussed in the previous paragraph can be reinterpreted as the specialisation of the period-conjecture for the sausage model. It would be interesting to reinterpret the all-loop result of~\cite{Alfimov:2021sir} in terms of the twist 1-form of this model and explore potential generalisations to wider classes of theories.

\paragraph{Conformal limits.} Having found a general description of the (1-loop) RG-flow of integrable $\s$-models, a logical next step is to enquire about its fixed points, which will define conformal (or at least scale-invariant) theories. This can be studied by considering the UV or IR limit $\tf \to \pm \infty$ of this flow. In particular, a natural question in this context is to understand the effect of this limit on the twist 1-form of the theory. This was studied in some rational examples in the work~\cite{Kotousov:2022azm}, showing that it generally requires a careful treatment: indeed, the conformal limit corresponds to a rather degenerate process, in which the poles and zeroes of $\omega$ can collide together and/or grow infinitely far apart. Extracting a meaningful conformal limit $\omega_{\text{CFT}}$ of the twist 1-form typically requires a well-chosen rescaling of the spectral parameter $z$: more precisely, one has to consider two different rescaled spectral parameters, leading to two different limits of $\omega$ in which various poles/zeroes survive, while others decouple. These two limits correspond to the two chiral halves of the conformal point, one describing left-moving fields, the other right-moving ones. This was checked in~\cite{Kotousov:2022azm} for various examples. It is an alluring perspective to study whether such a behaviour can be established in more generality using the period- or $\p\Psi$-conjecture.

Let us also make a brief remark concerning the conformal points in the elliptic case. In the example of the elliptic deformation of the Principal Chiral Model studied in section \ref{sec:DPCM}, one checks that the UV limit sends the torus moduli $\tau$ to $+\ri\infty$, thus corresponding to a trigonometric limit, in which the torus decompactifies to a cylinder. It would be interesting to study if such a phenomenon occurs for all elliptic models or if some of them retain their elliptic nature in the conformal limit.

\paragraph{Comments on the function $\bm{\mathcal{P}(z)}$.} We observed at the end of subsection \ref{subsec:Periods} that the period-conjecture can be re-expressed in terms of a certain function $\mathcal{P}(z)$, defined by $\p_z \log \mathcal{P}(z) = \frac{2\pi}{\hbar c_\g}\,\vp(z)$  (up to a constant). In particular, at 1-loop and as seen in equation \eqref{eq:RGP}, the RG-scale $\mu$ of the model is directly related with the quantity $\mathcal{P}(\ha{z}^{\,+})/\mathcal{P}(\ha{z}^{\,-})$, where $\ha{z}^{\,\pm} \in \Zh^\pm$ are zeroes of $\omega$.

It is worth noticing that this function is expected to also play an important role in the study of quantum integrable structures at the conformal points of these models -- see~\cite{Lukyanov:2013wra,Bazhanov:2013cua,Bazhanov:2013oya,Lacroix:2018fhf,Lacroix:2018itd,Gaiotto:2020dhf,Kotousov:2021vih,Kotousov:2022azm,Franzini:2022duf}. More precisely, the conformal limit\footnote{See the previous paragraph for a discussion of the subtleties arising when taking this limit. In particular, in what follows, $\omega_{\text{CFT}}$ and $\mathcal{P}_{\text{CFT}}$ will denote the conformal limit of $\omega$ and $\mathcal{P}$ in one choice of rescaling of the spectral parameter, corresponding to either of the chiral sectors of the conformal model.} $\mathcal{P}_{\text{CFT}}(z)$ of this function appears in various conjectures concerning the construction of commuting conserved charges in these models and in the so-called ODE/IQFT (or ODE/IM) correspondence describing their spectrum. Remarkably, in the case of the Fateev model\footnote{Here, we use the terminology Fateev model to describe the integrable $\s$-model introduced in~\cite{Fateev:1996ea} and whose target space is a two-parameter deformation of the 3-sphere. We note that this name is also often used for a quantum field theory with exponential interactions, which is dual to the $\s$-model. The relation between the function $\mathcal{P}_{\text{CFT}}(z)$ appearing in the works~\cite{Lukyanov:2013wra,Bazhanov:2013cua,Bazhanov:2013oya} and the twist 1-form in the language of the present article was established in~\cite{Kotousov:2022azm}.}, a deformation of this correspondence away from the conformal point has been proposed in the works~\cite{Lukyanov:2013wra,Bazhanov:2013cua,Bazhanov:2013oya}, in which the function $\mathcal{P}_{\text{CFT}}(z)$ still plays an important role. From the point of view of the renormalisation group flow, this deformation corresponds to a relevant perturbation, which introduces a dependence of the model on the RG-scale $\mu$. As mentioned above, at 1-loop, this scale can be extracted from the specific combination $\mathcal{P}(\ha{z}^{\,+})/\mathcal{P}(\ha{z}^{\,-})$ of the parameters, built from the function $\mathcal{P}(z)$. It would be interesting to explore potential relations between these results and understand in more generality the role of the function $\mathcal{P}(z)$ in the quantum integrable structure of $\sigma$-models with twist 1-forms (at and away from their conformal point).

Let us end with a remark on the geometric nature of the objects $\omega$ and $\mathcal{P}$. In the construction of classical integrable $\s$-models, as reviewed in section \ref{sec:IntSigma}, $\omega = \vp(z)\,\dd z$ is seen as a meromorphic 1-form on $C$. In particular, under a change of coordinate $z \mapsto \zt = h(z)$ on $C$, we have $\omega = \vpt(\zt)\,\dd \zt$ with $\vp(z) = \vpt(h(z))\,h'(z)$. On the other hand, the conformal limit $\mathcal{P}_{\text{CFT}}$ of $\mathcal{P}$ is expected~\cite{Lacroix:2018fhf,Kotousov:2022azm} to behave, at the fully quantum level, as a multi-valued section of the canonical line bundle $L_C$ over $C$. This suggests that the conformal limit $\omega_{\text{CFT}} = \frac{\hbar\,c_\g}{2\pi} \p\log \mathcal{P}_{\text{CFT}}$ of $\omega$ receives quantum corrections which change its geometric nature, from a meromorphic 1-form on $C$ to a connection on the line bundle $L_C^{\hbar\,c_\g/2\pi}$. More precisely, under a change of coordinate $z \mapsto \zt = h(z)$, we expect $\omega_{\text{CFT}} = \vp_{\text{CFT}}(z)\,\dd z$ to transform according to
\begin{equation}
    \vp_{\text{CFT}}(z) = \vpt_{\text{CFT}}\bigl(h(z)\bigr) \, h'(z) + \frac{\hbar\,c_\g}{2\pi} \frac{h''(z)}{h'(z)}\,.
\end{equation}

\paragraph{Integrable $\bm\s$-models from twistor space.} Let us consider a particularly simple class of rational integrable $\s$-models, for which the twist 1-form $\omega$ has only two simple zeroes in $\CP$ (this includes for instance the Principal Chiral Model with Wess-Zumino term as well as the $\eta-$ and $\lambda-$models). In the past few years, there have been various progresses~\cite{Costello:2020twi,Bittleston:2020hfv,Penna:2020uky,He:2021xoo,Cole:2023umd} towards a reformulation of these models in terms of a 6-dimensional holomorphic Chern-Simons theory on twistor space\footnote{In some cases like the $\lambda$-model, this construction comes with various subtleties (see~\cite{He:2021xoo,Cole:2023umd} for details).}. In that formulation, the data of the twist 1-form $\omega$ on $\CP$ is replaced by that of a $(3,0)$-form $\Omega$ on twistor space and a symmetry reduction. This offers an interesting alternative perspective on the geometric structures underlying these models, their integrability and their spectral parameters. A natural question in this context is whether their 1-loop renormalisation also has a simple interpretation in terms of the $(3,0)$-form $\Omega$ on twistor space, in analogy with the $\p\Psi$- and period-conjectures on the 1-form $\omega$. More generally, it would be interesting to explore the potential relations between twistor space, the geometry of the spectral parameter and renormalisation for a wider class of integrable $\s$-models, corresponding to more general twist 1-forms $\omega$.

\section*{Acknowledgements} 
	
We are grateful to Niklas Beisert, Johannes Broedel, Kevin Costello, Ryan Cullinan, Alex Eskin, Falk Hassler, Ben Hoare, Shota Komatsu, Gleb Kotousov, Nat Levine, Daniel Thompson and Beno\^it Vicedo for useful and interesting discussions. We also thank Nat Levine for valuable comments on the draft and for a related ongoing collaboration~\cite{ToAppear:Universal}. This work is partly based on the Master thesis of one of the authors (A.W.) prepared at ETH Z\"urich. We thank Niklas Beisert for acting as co-advisor for this thesis and for helpful interactions and comments during its preparation. The work of S.L. is supported by Dr. Max R\"ossler, the Walter Haefner Foundation and the ETH Z\"urich Foundation.

\newpage

\appendix

\section[The Belavin basis]{The Belavin basis}
\label{app:Belavin}

In this appendix, we review the definition and properties of the Belavin basis of $\slNC$, which plays a crucial role in the elliptic integrable $\s$-models considered in this paper. 

\paragraph{Matrices $\bm{\Xi_i}$.} The main ingredient that we will need is a pair of $N\times N$ matrices, defined as
\begin{equation}\label{eq:defofXi}
	\Xi_1\equiv\begin{pmatrix}
		0&1&&&\\
		0&0&\ddots&&\\
		&\ddots&\ddots&\ddots&\\
		&&\ddots&0&1\\
		1&&&0&0
	\end{pmatrix},\qquad\;\;
	\Xi_2\equiv\begin{pmatrix}
		1&0&&&\\
		0&\xi&\ddots&&\\
		&\ddots&\ddots&\ddots&\\
		\phantom{\xi^{N-2}}&\phantom{\xi^{N-2}}&\ddots&\xi^{N-2}&0\\
		&&&0&\xi^{N-1}
	\end{pmatrix}\,,
\end{equation}
where $\xi=\exp\left(\frac{2\ri\pi}{N}\right)$ is a $N^\text{th}$-root of unity. One easily checks that these matrices satisfy the following algebraic relations:
\begin{equation}
    \Xi_1^N = \Xi_2^N = \Id \qquad \text{ and } \qquad \Xi_1 \cdot \Xi_2 = \xi\;\Xi_2 \cdot \Xi_1\,.
\end{equation}
In particular, the adjoint automorphisms $\Ad_{\Xi_1}$ and $\Ad_{\Xi_2}$ are cyclic of order $N$ and commute with each other: they thus generate an action of the group $\Z_N \times \Z_N$ on the Lie algebra $\slNC$.

\paragraph{The Belavin basis.} Let $\alpha=(\alpha_1,\alpha_2) \in \Z_N \times \Z_N$ be a couple of integers modulo $N$. Since the matrices $\Xi_i$ are cyclic of order $N$, the quantity
\begin{equation}
    T_\alpha = \frac{1}{\sqrt{N}} \, \Xi_1^{-\alpha_2}\,\Xi_2^{\alpha_1} \label{eq:defofT}
\end{equation}
is well-defined, \textit{i.e.} is independent of the choice of representatives for $\alpha_1$ and $\alpha_2$ (the minus sign and the exchange of labels in the exponents are introduced for future convenience). The matrix $T_{(0,0)}$ is simply proportional to the identity, while the matrices $T_{\alpha}$ for $\alpha \neq (0,0)$ are traceless and linearly independent. Introducing $\Ab = \Z_N \times \Z_N \setminus \lbrace (0,0) \rbrace$, the family $\lbrace T_\alpha \rbrace_{\alpha\in\Ab}$ then forms a basis of $\slNC$ (since $|\Ab|=N^2-1=\dim\slNC$): we call it the Belavin basis.

\paragraph{Algebraic properties.} The commutation relations of the Belavin basis $\lbrace T_\alpha \rbrace_{\alpha\in\Ab}$ read
\begin{equation}\label{eq:Tcommutator}
	\left[T_\alpha,T_\beta\right]=f_{\alpha\beta}\,T_{\alpha+\beta} \qquad \text{ with } \qquad f_{\alpha\beta}=\frac{1}{\sqrt{N}}\left(\xi^{\alpha_1\beta_2}-\xi^{\beta_1\alpha_2}\right)\,.
\end{equation}
Moreover, it is composed of eigenvectors of the $(\Z_N \times \Z_N)$--action generated by the adjoint automorphisms $\Ad_{\Xi_1}$ and $\Ad_{\Xi_2}$. More precisely, we have
\begin{equation}\label{eq:GradingT}
    \Ad_{\Xi_1}^{\beta_1} \Ad_{\Xi_2}^{\beta_2}\, T_\alpha = \xi^{{\alpha}\cdot{\beta}}\, T_\alpha\,, \qquad \text{ with } \qquad \alpha\cdot\beta = \alpha_1 \beta_1 + \alpha_2 \beta_2\,.
\end{equation}

Consider the non-degenerate invariant bilinear form $\langle\cdot,\cdot\rangle=-\Tr(\cdot)$ on $\slNC$. We will denote the dual of the Belavin basis with respect to this form by $\{T^\alpha\}_{\alpha\in\Ab}$, using upper indices. We then have
\begin{equation}
	\left\langle T_\alpha,T_\beta\right\rangle =-\delta_{\alpha+\beta,0}\,\xi^{-\beta_1\beta_2} \qquad \text{ and thus } \qquad T^\alpha=-\xi^{\alpha_1\alpha_2}T_{-\alpha}\,. \label{eq:slcnkappa}
\end{equation}

\section{Elliptic and quasi-elliptic functions}
\label{app:elliptic}

This appendix is a review of elliptic and quasi-elliptic functions, in particular the Weierstrass family and their relatives $r^\alpha(z)$. We say that a meromorphic function $f(z)$ is elliptic with half-periods $\bm\ell=(\ell_1,\ell_2)$ if it satisfies the double-periodicity property
\begin{equation}
    f(z+2\ell_1) = f(z+2\ell_2) = f(z)\,.
\end{equation}
If that is the case, $f$ reduces to a function on the torus $\C/\Gamma$, where
$\Gamma=2\ell_1\,\Z\oplus 2\ell_2\,\Z$ is the period 2D-lattice in $\C$.

\subsection{The Weierstrass functions}
\label{app:Weierstrass}

\paragraph{The Weierstrass $\bm\wp$-function.} The simplest example of a non-trivial elliptic function is arguably the Weierstrass $\wp$-function, defined by
\begin{equation}
	\wp(z\,;\bm\ell)\equiv\frac{1}{z^2}+\sum_{\substack{\bm{n}\in \mathbb{Z}^2\\\bm{n}\neq(0,0)} }\left\{\frac{1}{\left(z-2\bm{n}\cdot\bm{\ell}\right)^2}-\frac{1}{(2\bm{n}\cdot\bm{\ell})^2} \right\}\,.\label{eq:defofwp}
\end{equation}
In this expression, the second term in the infinite sum is there to ensure its convergence. The sum makes it explicit that this satisfies the desired double-periodicity property $\wp(z+2\ell_i\,;\bm\ell) = \wp(z\,;\bm\ell)$. To lighten the notations, we will often omit the dependence on the half-periods $\bm\ell$ and simply write the Weierstrass function as $\wp(z)$.  Schematically, it can be seen as the elliptic version of $1/z^2$. More precisely, it is the unique elliptic function with the behaviour
\begin{equation}
    \wp(z) = \frac{1}{z^2} + O(z^2)\,,
\end{equation}
and no poles outside of $z=0$ and its translates $z\in\Gamma$. Among other useful properties, we note that the Weierstrass function is even, \textit{i.e.} $\wp(-z)=\wp(z)$, and satisfies the differential equation
\begin{equation}
	\wp'(z)^2=4\wp(z)^3-g_2\,\wp(z)-g_3\,.
\end{equation}
Here, $(g_2,g_3)$ are the \textit{Weierstrass invariants}, which are related to the half-periods $\bm\ell=(\ell_1,\ell_2) $ by
\begin{equation}\label{eq:WeiInv}
    g_2 = 60 \sum_{\substack{\bm{n}\in \mathbb{Z}^2\\\bm{n}\neq(0,0)} } \frac{1}{(2\bm{n}\cdot\bm{\ell})^4} \qquad \text{ and } \qquad g_2 = 140 \sum_{\substack{\bm{n}\in \mathbb{Z}^2\\\bm{n}\neq(0,0)} } \frac{1}{(2\bm{n}\cdot\bm{\ell})^6}\,.
\end{equation}
Finally, these parameters can be used to define the \textit{modular discriminant}
\begin{equation}\label{eq:ModDisc}
    \Delta = g_2^3 - 27 g_3^2\,.
\end{equation}

\paragraph{The Weierstrass $\bm\zeta$-function.} The Weierstrass $\zeta$-function is defined as
\begin{equation}\label{eq:Zeta}
    \zeta(z\,;\bm\ell)\equiv\frac{1}{z}+\sum_{\substack{\bm{n}\in \mathbb{Z}^2\\\bm{n}\neq(0,0)} }\left\{\frac{1}{z-2\bm{n}\cdot\bm{\ell}}+\frac{1}{2\bm{n}\cdot\bm{\ell}}+\frac{z}{(2\bm{n}\cdot\bm{\ell})^2} \right\}\,.
\end{equation}
It can equivalently be characterised by the properties
\begin{equation}\label{eq:zetafromwp}
    \zeta'(z) = -\wp(z) \qquad \text{ and } \qquad \zeta(z) = \frac{1}{z} + O(z)\,.
\end{equation}
This function is odd, \textit{i.e.} $\zeta(-z)=-\zeta(z)$, and has simple poles exactly at the lattice points $z\in\Gamma$. We stress here that it is not elliptic: indeed, one can show that there are no elliptic functions on $\C/\Gamma$ with only one simple pole. However, it still satisfies a simple quasi-periodicity property with respect to shifts by lattice vectors in $\Gamma$, namely
\begin{equation}\label{eq:QuasiZeta}
    \zeta(z+2\ell_i) = \zeta(z) + 2 L_i\,, \qquad \text{ with } \qquad \bm{L} = (L_1,L_2) \equiv \bigl(\zeta(\ell_1), \zeta(\ell_2) \bigr)\,.
\end{equation}
Indeed, we see that this is compatible with \eqref{eq:zetafromwp} and the double periodicity of $\wp(z)$.
\paragraph{The Weierstrass $\bm\sigma$-function.} 
The Weierstrass $\sigma$-function is defined as
\begin{equation}
    \sigma(z\,;\bm\ell) \equiv z \prod_{\substack{\bm{n}\in \mathbb{Z}^2\\\bm{n}\neq(0,0)} }\left(1-\frac{z}{2\bm{n}\cdot\bm{\ell}}\right) \exp\left( \frac{z}{2\bm{n}\cdot\bm{\ell}} + \frac{1}{2} \left(\frac{z}{2\bm{n}\cdot\bm{\ell}}\right)^2 \right)\,.
\end{equation}
It can equivalently be characterised by the properties
\begin{equation}
    \sigma'(z) = \zeta(z)\,\sigma(z) \qquad \text{ and } \qquad \sigma(z) = z + O(z^2)\,.
\end{equation}
This function is odd, \textit{i.e.} $\sigma(-z)=-\sigma(z)$, and has simple zeroes exactly at the lattice points $z\in\Gamma$. Moreover, it satisfies the quasi-periodicity property
\begin{equation}\label{eq:QuasiSigma}
    \sigma(z+2\ell_i)=-\exp\Big(2L_i\left[z+\ell_i\right]\Big)\,\sigma(z)\,.
\end{equation}

\paragraph{Cross-product and Legendre relation.} We introduce a $\C$-valued ``cross-product'' $\times$ on $\C^2$ by letting
\begin{equation}
	(a_1,a_2)\times (b_1,b_2) \equiv a_1b_2-b_1a_2\,. \label{eq:crossproduct}
\end{equation}
The half-periods $\bm \ell = (\ell_1,\ell_2)$ and the related numbers $\bm L =\bigl(\zeta(\ell_1),\zeta(\ell_2)\bigr)$ then satisfy the so-called \textit{Legendre relation}
\begin{equation}\label{eq:Legendre}
   \bm L \times \bm\ell = \frac{\ri\,\pi}{2}\,.
\end{equation}

\subsection[The functions \texorpdfstring{$r^\alpha(z)$}{ralpha(z)}]{The functions \texorpdfstring{$\bm{r^\alpha(z)}$}{ralpha(z)}}
\label{app:ralpha}

Let us consider a pair $\alpha=(\alpha_1,\alpha_2)$ of complex numbers. We introduce
\begin{equation}
	q_\alpha\equiv \frac{2}{N}\alpha\times \bm{\ell}
	\,,\qquad Q_\alpha\equiv \frac{2}{N}\alpha\times \bm{L}\,,\label{eq:qalphadef}
\end{equation}
using the cross product \eqref{eq:crossproduct}. When $q_\alpha$ is not in the lattice $\Gamma$, the following function is well-defined:
\begin{equation}
	r^\alpha(z)\equiv\exp(-Q_\alpha z)\,\frac{\sigma(z+q_\alpha)}{\sigma(z)\,\sigma(q_\alpha)}\,.\label{eq:defofralpha}
\end{equation}
Up to an exponential prefactor, this coincides with the so-called \textit{Kronecker function} associated with the torus $\C/\Gamma$. It satisfies $r^\alpha(-z)=-r^{-\alpha}(z)$. Moreover, $r^\alpha(z)$ has simple zeroes on the lattice $\Gamma - q_\alpha $ and simple poles on the lattice $\Gamma$. In particular, around $z=0$, it admits the simple behaviour $r^\alpha(z) \sim \frac{1}{z}$.\\

Combining the property \eqref{eq:QuasiSigma} of the $\s$-function with the Legendre relation \eqref{eq:Legendre}, one finds a simple quasi-periodicity result for $r^\alpha(z)$:
\begin{equation}\label{eq:QuasiR}
	r^\alpha(z+2\ell_i)=\exp(\frac{2 \ri\pi}{N}\,\alpha_i)\,r^\alpha(z)\,.
\end{equation}
The most important property of the functions $r^\alpha(z)$ is the functional identity
\begin{equation}
	r^\alpha(z_1)\,r^\beta(z_2)=r^{\alpha+\beta}(z_1)\,r^\beta(z_2-z_1)+r^{\alpha}(z_1-z_2)\,r^{\alpha+\beta}(z_2)\,,\label{eq:Fay}
\end{equation}
which holds whenever $q_\alpha, q_\beta, q_{\alpha+\beta}\notin \Gamma$ and which follows from the so-called \textit{Fay identity} of elliptic $\sigma$-functions (or equivalently, Jacobi $\theta$-functions)~\cite{fay_theta_1973}.\\

Let us now restrict to the case where the parameters $\alpha=(\alpha_1,\alpha_2)$ are integers in $\Z\times \Z$. Due to the property \eqref{eq:QuasiSigma}, one checks that the function $r^\alpha(z)$ is invariant under a shift of these integers by $N$. We are thus led to consider the couple $\alpha=(\alpha_1,\alpha_2)$ modulo $N$, \textit{i.e.} as an element of $\Z_N \times \Z_N$ (by a slight abuse of notation, we use the same notation for $\alpha\in \Z\times \Z$ and its equivalence class in $\Z_N \times \Z_N$). Discarding the element $\alpha=(0,0)$, for which $r^\alpha(z)$ is not well-defined, this yields a family of $N^2-1$ functions $\lbrace r^\alpha(z) \rbrace_{\alpha\in\Ab}$, labelled by $\Ab = \Z_N \times \Z_N \setminus \lbrace (0,0) \rbrace$. These functions play a crucial role in the description of the elliptic integrable $\sigma$-models considered in this paper.

\section{Origin From 4D Chern-Simons theory} 
\label{app:4dCS}

The subject matter of this article is the integrable $\sigma$-models described in subsection \ref{subsec:SigmaModel}, associated with the data $(C,\omega,\Zh^\pm,\g,\mathfrak{k})$. The goal of this appendix is to briefly review how this data leads to a consistent construction of the corresponding theory by way of 4-dimensional Chern-Simons (4D-CS) theory. 

Before starting, let us discuss the literature on which this review is based, as a guide to further references for the interested readers. The construction of integrable $\sigma$-models from 4D-CS theory was first proposed in the seminal work~\cite{costello_gauge_2019}. In the language of subsection \ref{subsec:SigmaModel}, that work focused mostly on models with double poles in $\omega$ and simple choices of the isotropic subalgebra $\mathfrak{k}$. A more systematic study of the admissible choices for $\mathfrak{k}$ was initiated in~\cite{delduc_unifying_2020} for double and simple poles in $\omega$, while the case with arbitrary pole structure was treated in~\cite{benini_homotopical_2022,lacroix_integrable_2021,Liniado:2023uoo}, leading to a larger class of models. We also refer to the lecture notes~\cite{lacroix_4-dimensional_2022} for a more thorough review. The works~\cite{delduc_unifying_2020,benini_homotopical_2022,lacroix_integrable_2021,Liniado:2023uoo} focused on rational theories (\textit{i.e.} with $C=\CP)$. The references~\cite{costello_gauge_2019,derryberry_lax_2021} also contain some higher-genus models but of a slightly different nature than the elliptic ones considered in the present paper: the latter were introduced in the recent work~\cite{Lacroix:2023qlz} and are based on an equivariant version of the elliptic 4D-CS theory. Although~\cite{Lacroix:2023qlz} focused on the case of doubles poles in $\omega$, we include in the present paper the equivariant elliptic models corresponding to arbitrary pole structures, which can be constructed by combining the approaches of~\cite{benini_homotopical_2022} and~\cite{Lacroix:2023qlz}.

\subsection{The 4D-CS theory}

As its name suggests, this theory is defined on a 4D space-time manifold: more precisely, this manifold is taken to be $\Sigma \times C$, combining the worldsheet of the $\s$-model with the Riemann surface of its spectral parameter. The fundamental field of this theory is a gauge field 
\begin{equation}
    \Ac = \Ac_+ \, \dd x^+ + \Ac_- \, \dd x^- + \Ac_z \, \dd z + \Ac_{\bar{z}} \, \dd \ol{z}\,,
\end{equation}
seen as a $\g^\C$-valued 1-form on $\Sigma\times C$. It satisfies certain specific properties related to the data $(C,\omega,\Zh^\pm,\g,\mathfrak{k})$, which we describe below.
\begin{enumerate}
    \item\label{C1} The component $\Ac_\pm$ is allowed to have singularities of order one at the points $\Zh^\pm = \lbrace \ha{z}^{\,\pm}_r \rbrace_{r=1}^M$ ;
    \item\label{C2} The component $\Ac_\pm$ satisfies certain boundary conditions at the poles $\Ph =  \lbrace \ha{p}_r \rbrace_{r=1}^{n}$ of $\omega$. More precisely, the object $\bigl( \p_z^k \Ac_\pm(\ha{p}_r) \bigr)_{r=1,\dots,n}^{k=0,\dots,m_r-1}$ is naturally interpreted as an element of the defect Lie algebra and is required to be valued in its maximally isotropic subalgebra $\mathfrak{k}$ ;
    \item\label{C3} In the elliptic case, $\Ac$ satisfies the quasi-periodicity property\footnote{Technically, this means that $\Ac$ does not define a smooth field on $C=\C/\Gamma$, as it is only quasi-periodic under translations by lattice vectors in $\Gamma$. Another point of view, based on the observation that $\Ac$ is doubly-periodic with periods $2N\ell_i$, is to see it as defined on another torus $\C/\Lambda$ with $\Lambda=N\Gamma$, similarly to what was discussed for the Lax connection earlier (see equation \eqref{eq:Lambda} and the surrounding discussion). In that case, the quasi-periodicity $\Ac(z+2\ell_i) = \Ad_{\Xi_i}\,\Ac(z)$ is interpreted as a condition of equivariance of the gauge field with respect to the group $\Z_N\times\Z_N$. This is the point of view adopted in~\cite{Lacroix:2023qlz}, where the theory was called \textit{equivariant elliptic 4D-CS}.} $\Ac(z+2\ell_i) = \Ad_{\Xi_i}\,\Ac(z)$.
\end{enumerate}
The action of the theory is defined as~\cite{nekrassov_four_1996,costello_supersymmetric_2013}\footnote{In the case where $\omega$ has higher-order poles, one generally needs to consider a regularised version of this action to ensure its finiteness, see~\cite{benini_homotopical_2022}. We also note that in the formulation of the equivariant elliptic 4D-CS theory in~\cite{Lacroix:2023qlz}, the integral was taken over $\C/\Lambda$ rather than $C=\C/\Gamma$. The latter is the quotient of the former by the equivariance group $\Z_N \times \Z_N$ and due to the equivariance condition $\Ac(z+2\ell_i) = \Ad_{\Xi_i}\,\Ac(z)$, the two integrals coincide up to the overall combinatorial factor $|\Z_N \times \Z_N|=N^2$ added in~\cite[Equation (5.4)] {Lacroix:2023qlz}.}
\begin{equation}\label{eq:S4d}
    S_{4D-CS} [\Ac] = \frac{\ri}{4\pi} \iiiint_{\Sigma\times C} \omega \wedge \mathcal{CS}[\Ac]\,,
\end{equation}
where $\mathcal{CS}[\Ac]$ is the so-called Chern-Simons 3-form of $\Ac$. In particular, note that the 1-form $\omega$ appears very explicitly in the definition of this action. Since $\omega$ is along the $\dd z$-direction, the component $\Ac_z$ of the gauge field decouples completely from the theory and does not define a physical degree of freedom. Moreover, the action \eqref{eq:S4d} is invariant under gauge symmetries $\Ac \mapsto u^{-1}\Ac u + u^{-1}\dd u$, where $u$ is a $G^\C$-valued function of $(x^+,x^-,z,\ol{z})$, satisfying certain appropriate boundary conditions at the poles of $\omega$ and, in the elliptic case, the quasi-periodicity property $u(z+2\ell_i) = \Ad_{\Xi_i}^{-1}\,u(z)$. Finally, varying the action \eqref{eq:S4d} with respect to $\Ac$, we obtain the equation of motion of the theory:
\begin{equation}\label{eq:4dEoM}
    \omega \wedge \mathcal{F}[\Ac] = 0\,, \qquad \text{where} \qquad \mathcal{F}[\Ac] = \dd \Ac + \Ac \wedge \Ac
\end{equation}
is the curvature of $\Ac$.

\subsection[Extracting the 2D integrable \texorpdfstring{$\sigma$}{sigma}-model]{Extracting the 2D integrable \texorpdfstring{$\bm\sigma$}{sigma}-model}

To understand the link with a 2D integrable model, we reparametrise the gauge field components as $\Ac_{\bar z} = - (\p_{\bar z} \gh)\gh^{\,-1}$ and $\Ac_\pm = \gh\,\Lc_\pm\,\gh^{\,-1}- (\p_{\pm} \gh)\gh^{\,-1}$ where $\gh$ is a $G^\C$-valued field\footnote{In the elliptic case, the existence of the field $\hat{g}$, such that $\Ac_{\bar z} = - (\p_{\bar z} \hat{g})\hat{g}^{-1}$, strongly relies on the equivariance property $\Ac_{\bar z}(z+2\ell_i) = \Ad_{\Xi_i}\Ac_{\bar z}(z)$. This is related to the notion of rigid elliptic bundles discussed in ~\cite[Section 10.2]{costello_gauge_2018} and~\cite[Section 9]{costello_gauge_2018-1}. We refer to~\cite[Section 5.3]{Lacroix:2023qlz} for details.} and $\Lc_\pm$ are $\g^\C$-valued fields. In terms of these new degrees of freedom, the equation of motion \eqref{eq:4dEoM} implies
\begin{equation}
    \p_+ \Lc_- - \p_- \Lc_+ +[\Lc_+,\Lc_-] = 0 \qquad \text{ and } \qquad \p_{\bar z} \Lc_\pm = 0 \,, \qquad \text{ for } z\in C\setminus \bigl(\Zh^+ \!\sqcup \Zh^-\bigr)\,,
\end{equation}
\textit{i.e.} away from the zeroes of $\omega$. The first of these equations ensures that $\Lc_\pm$ defines a flat 2D-connection on $\Sigma$, while the second means that it depends holomorphically on $z\in C\setminus (\Zh^+ \!\sqcup \Zh^-)$. These are exactly the defining characteristics of the Lax connection of an integrable 2D field theory on $\Sigma$, with spectral parameter $z$ -- see equation \eqref{eq:flatness} and the associated discussion. This is the key mechanism underlying the 4D-CS theory as a unifying framework for 2D integrable $\sigma$-models.

We note that the holomorphicity of $\Lc_\pm(z)$ only holds away from the zeroes $\Zh^+\! \sqcup \Zh^-$ of $\omega$. In fact, one can argue that the equations of motion of 4D-CS allow for simple poles at these zeroes so that $\Lc_\pm(z)$ can be seen as a meromorphic function of $z$. Combined with \hyperref[C1]{condition 1.} imposed earlier on the gauge field, we find that $\Lc_\pm(z)$ has a simple pole at the points $\Zh^\pm = \lbrace \ha{z}^{\,\pm}_r \rbrace_{r=1}^M$ (forming half of these zeroes), thus recovering the expected behaviour \eqref{eq:PolesLax}. Moreover, in the elliptic case, the quasi-periodicity \hyperref[C3]{condition 3.} imposed on the gauge field implies the similar property \eqref{eq:equivL} of the Lax connection $\Lc_\pm(z)$. As mentioned in subsection \ref{subsec:SigmaModel}, the pole structure of $\Lc_\pm(z)$, together with this quasi-periodicity condition, is enough to completely fix the form of $\Lc_\pm(z)$, which then reads \eqref{eq:RationalLax} in the rational case and \eqref{eq:EllipticLax} in the elliptic one.\\

The next step in the construction is to extract the 2D integrable $\sigma$-model corresponding to this Lax connection from the above 4D setup. The fundamental fields of this model are obtained as the gauge-invariant degrees of freedom contained in $\gh$. Indeed, one shows that the field $\gh$ can be gauged away in any region away from the poles of $\omega$, \textit{i.e.} on $C \setminus \Ph$. The situation at the poles $\Ph$ is more subtle as the gauge transformations are restricted by specific boundary conditions at these points. A careful analysis shows that, in the end, $\gh$ contains only a finite number of 2D gauge-invariant degrees of freedom, extracted from the evaluations\footnote{Note that these evaluations are still subject to residual 2D gauge symmetries and are thus not all physical.} $\bigl( \p_z^k \hspace{1pt}\gh\hspace{1pt}(\ha{p}_r) \bigr)_{r=1,\dots,n}^{k=0,\dots,m_r-1}$ at the poles. These degrees of freedom can be seen as functions on $\Sigma$ and can then be combined into a 2D field $\phi : \Sigma \to \Tc$ valued in a manifold $\Tc$, which is interpreted as the target space of the $\sigma$-model.

The boundary conditions of \hyperref[C2]{point 2.} above constrain the values of $\Ac_\pm = \gh\,\Lc_\pm\,\gh^{\,-1}- (\p_{\pm} \gh)\gh^{\,-1}$ and some of its $z$-derivatives at the poles $\Ph$ of $\omega$. In practice, this translates to a relation between $\bigl( \p_z^k \hspace{1pt}\Lc_\pm\hspace{1pt}(\ha{p}_r) \bigr)_{r=1,\dots,n}^{k=0,\dots,m_r-1}$ and the 2D field $\phi$ extracted earlier from $\bigl( \p_z^k \hspace{1pt}\gh\hspace{1pt}(\ha{p}_r) \bigr)_{r=1,\dots,n}^{k=0,\dots,m_r-1}$. Combined with the aforementioned results \eqref{eq:RationalLax}--\eqref{eq:EllipticLax} on the analytic structure of the Lax connection $\Lc_\pm(z)$, this allows to completely express this connection in terms of the 2D field $\phi$ and the parameters contained in $\omega$, so that $\phi$ is the only physical degree of freedom remaining. Finally, reinserting this expression of $\Lc_\pm(z)$ in terms of $\phi$ into the action \eqref{eq:S4d} and performing the integral over $z\in C$, one eventually rewrites this action as a 2D functional $S[\phi]$ on $\Sigma$.\footnote{For the case with arbitrary pole structure in $\omega$ and arbitrary choice of boundary conditions, this 2D action is most efficiently obtained using the formalism of edge modes, as done in~\cite{benini_homotopical_2022}.} More precisely, one can bring this functional into the $\sigma$-model form \eqref{eq:action}, from which we can read off the metric and B-field $(G,B)$ of the target space $\Tc$, which depend in a complicated way on the starting data $(C,\omega,\Zh^\pm,\g,\mathfrak{k})$. By construction, the equations of motion of $\phi$ derived from this action are equivalent to the flatness of the Lax connection $\Lc_\pm(z)$, signalling the integrability of the model.

\section{1-loop renormalisation of the elliptic DPCM using Ricci flow}\label{sec:RGAppendix}

This appendix provides further details on the calculations performed in section \ref{sec:DPCM}, concerning the 1-loop renormalisation of the elliptic integrable deformed PCM. We will first focus on the torsionful Ricci tensor, reviewing its connection to the RG-flow of $\sigma$-models.
Then we will perform the explicit computation of this tensor in three steps, first considering an arbitrary choice of deformation (as one might expect, this leads to a rather complicated expression); we will then focus on deformations that are diagonal in the Belavin basis; and lastly, we will further specialise to the specific elliptic deformation discussed in subsection \ref{subsec:IntDPCM}, ending up with the result \eqref{eq:RicciFlowofD}. Afterwards, we will show how this result allows us to prove the 1-loop renormalisability of the integrable deformed PCM and the extraction of the $\beta$-functions of its parameters $(\ell_1,\ell_2,\rho,\ha{z})$. 

\subsection[RG-flow of \texorpdfstring{$\sigma$-Models}{sigma-models} and the Ricci tensor]{RG-Flow of \texorpdfstring{$\bm\sigma$-Models}{Sigma-Models} and the Ricci tensor}\label{subsec:RTsigma}

\paragraph{Generalised Ricci flow.} Consider a general $\sigma$-model with worldsheet $\Sigma$ and target space $\Tc$. We will denote a choice of coordinates on this space by $\bigl( \phi^{\,i} \bigr)_{i=1}^d$, the metric by $G_{ij}$ and the $B$-field by $B_{ij}$. The action of the model is then given by
\begin{equation}
    S[\phi] = \iint_\Sigma \dd x^+\,\dd x^- \,\bigl( G_{ij}(\phi) + B_{ij}(\phi) \bigr)\,\p_+\phi^{\,i}\,\p_-\phi^{\,j}\,.
\end{equation}
It is a well-known fact \cite{Ecker:1972bm,Honerkamp:1971sh,Friedan:1980jf,Curtright:1984dz} that the 1-loop RG-flow of this $\sigma$-model is given by the generalised Ricci flow\footnote{One could in general also add a diffeomorphism term and an exact 2-form term on the right-hand side. These will not be needed for the cases at hand.}
\begin{equation}
    \frac{\dd}{\dd \ft}\bigl( G_{ij}(\phi)+B_{ij}(\phi) \bigr)= \hbar\, R^+_{ij}(\phi)\,,\label{eq:RicciandRG}
\end{equation}
where $R^+_{ij}$ is the torsionful Ricci tensor associated with the target space geometry $(\Tc,G,B)$. This tensor is defined in terms of the curvature of the Levi-Civita connection associated with the metric $G$, to which we add the torsion $H=\dd B$ induced by the B-field. We will not detail this general definition here, but will later give a useful way of computing this tensor for the class of models that we are interested in.

\paragraph{Vielbein.} Let us now suppose that we are given a vielbein $\e i \alpha (\phi)$ (with inverse $\e\alpha i(\phi)$) that makes the metric and $B$-field
\begin{equation}
    G_{\alpha\beta}= G_{ij}(\phi)\,\e i \alpha(\phi)\,\e j\beta(\phi)\, \qquad \text{ and } \qquad B_{\alpha\beta}= B_{ij}(\phi)\,\e i \alpha(\phi)\,\e j\beta(\phi)
\end{equation}
independent of the coordinates $\phi$. In that case, we then sometimes refer to the labels $\alpha,\beta,\dots$ as ``flat-space indices'', in contrast with the initial coordinate indices $i,j,\dots$ used to describe the curved geometry. In these notations, the $\s$-model action simply becomes
\begin{equation}\label{eq:vielbeinaction}
    S[\phi] = \iint_\Sigma \dd x^+\,\dd x^- \,\bigl( G_{\alpha\beta} + B_{\alpha\beta} \bigr)\,\e\alpha i (\phi) \, \e\beta j(\phi)\,\p_+\phi^{\,i}\,\p_-\phi^{\,j}\,.
\end{equation}

Here, we will further suppose that the vielbein is independent of the parameters of the theory. Under this assumption, the RG-flow \eqref{eq:RicciandRG} can be rewritten as
\begin{equation}
    \frac{\dd}{\dd\ft}\bigl( G_{\alpha\beta}+B_{\alpha\beta} \bigr) = \hbar\,R^+_{\alpha\beta}\,,\label{eq:vielbeinRG}
\end{equation}
where $R^+_{\alpha\beta} = R^+_{ij}\,\e i \alpha\,\e j \beta$ is the torsionful Ricci tensor in flat indices. The main interest of this reformulation is that the left-hand side of this equation does not depend on the coordinates of the target space. A first necessary condition for the renormalisability of the theory is therefore that the tensor $R^+_{\alpha\beta}$ also does not depend on these coordinates.

\paragraph{Expression of $\bm{R^+_{\alpha\beta}}$ in terms of spin connections.} It is a standard result of Riemannian geometry that the torsionful Ricci tensor can be expressed as
\begin{equation}
	R^+_{\alpha\beta} = \partial_\gamma\Omega^{+\phantom{\alpha}\gamma}_{\alpha\beta} -\partial_\alpha\Omega^{+\phantom{\beta}\gamma}_{\gamma \beta}+\Omega_{\gamma \delta}^{+\phantom{\gamma}\gamma}\,\Omega_{\alpha\beta}^{+\phantom{\beta}\delta}-\Omega_{\gamma \alpha}^{-\phantom{\gamma}\delta}\,\Omega_{\delta\beta}^{+\phantom{a}\gamma}.\label{eq:RiccifromOmega}
\end{equation}
where $\p_{\gamma} = \e i \gamma\frac{\p\;}{\p\phi^{\,i}}$ and $\Omega^{\pm\phantom{\alpha}\gamma}_{\alpha\beta}$ is the so-called \textit{torsionful spin connection}. In practice, we will not need the intrinsic geometric definition of this object. Rather, we will simply use the fact that it can be extracted from the equations of motion (EoMs) of the $\s$-model. Indeed, one shows that these equations, which read \eqref{eq:EoMSigma} in coordinate indices, can always be rewritten as
\begin{subequations}\label{eq:vielbeinEOM}
\begin{align}
    \p_+\left[\e\alpha i \,\p_-\phi^i\right]+\Omega_{\beta\gamma}^{+\phantom{\beta}\alpha}\,\e \beta j\, \e \gamma k\,\p_+ \phi^{\hspace{1pt}j}\,\p_-\phi^k = 0\,, \\
    \p_-\left[\e\alpha i \,\p_+\phi^i\right]+\Omega_{\beta\gamma}^{-\phantom{\beta}\alpha}\,\e \beta j\, \e \gamma k\,\p_- \phi^{\hspace{1pt}j}\,\p_+\phi^k = 0\,,
\end{align}
\end{subequations}
in terms of the vielbein.

\subsection{Ricci tensor for arbitrary deformations}\label{subsec:RTArbitrary}

\paragraph{Setup.} We will now focus on $\sigma$-models that arise from linear deformations of PCMs. Thus, let the target space $\Tc$ be a Lie group $G$, whose Lie algebra $\g$ is semi-simple. We consider a deformed PCM, with a field $g\in G$ and an arbitrary choice of constant, linear deformation operator $D:\g \to \g$. The corresponding action then takes the form \eqref{eq:DPCMActionHighRank}, which we recall here for the reader's convenience:
\begin{equation}
	S_{\text{DPCM}}[g]=\int_{\Sigma}\dd x^+\dd x^-\bigl\langle j_+,D[j_-]\bigr\rangle\,,\label{eq:RGDPCMACtion}
\end{equation}
written in terms of the Maurer-Cartan currents $j_\pm=g^{-1}\p_\pm g$ and the invariant pairing $\langle\cdot,\cdot\rangle$ on $\g$. We now choose a basis $\lbrace T_\alpha \rbrace$ for the Lie algebra $\g$. We can then consider the structure constants and the entries of the bilinear form and deformation operator in this basis, defined through
\begin{equation}\label{eq:Basis}
    \left[T_\alpha,T_\alpha\right] = f_{\alpha\beta}^{\phantom{\alpha\beta}\gamma}\,T_\gamma\,, \qquad \kappa_{\alpha\beta}\equiv \langle T_\alpha,T_\beta \rangle\,, \qquad D_{\alpha\beta}\equiv \langle T_\alpha,D[T_\beta] \rangle\,,
\end{equation}
where summation over repeated indices is implied.

Moreover, as in the main text, we write the $G$-valued field $g$ in terms of coordinate fields $(\phi^{\,i})_{i=1}^d$ and introduce $\e\alpha i(\phi)$ as the component of $T_\alpha$ in $g^{-1}\frac{\p g}{\p\phi^{\,i}}$. This allows us to decompose the currents as
\begin{equation}\label{eq:je}
    j_\pm = j_\pm^\alpha\,T_\alpha = \e\alpha i(\phi)\,\p_\pm \phi^{\,i}\, T_\alpha
\end{equation}
and rewrite the action \eqref{eq:RGDPCMACtion} as
\begin{equation}
    S_{\text{DPCM}}[g]=\iint_\Sigma \dd x^+ \, \dd x^-\,D_{\alpha\beta}\,\e\alpha i(\phi)\,\e\beta j(\phi)\,\p_+\phi^{\,i}\, \p_-\phi^{j}\,.
\end{equation}
Comparing this equation to \eqref{eq:vielbeinaction}, we see that $\e\alpha i(\phi)$ defines a vielbein for the deformed PCM and that the metric and $B$-field in flat indices are given by
\begin{equation}
    G_{\alpha\beta}=\frac{1}{2}\left[D_{\alpha\beta}+D_{\beta\alpha}\right]\qquad \text{ and } \qquad B_{\alpha\beta}=\frac{1}{2}\left[D_{\alpha\beta}-D_{\beta\alpha}\right].
\end{equation}
As expected, these quantities are independent of the coordinates $\phi^{\,i}$.\\

We note that the vielbein $\e\alpha i$ is independent of the parameters of the $\sigma$-model. According to the previous section, the RG-flow then takes the form \eqref{eq:vielbeinRG}, which in the present case reads
\begin{equation}
    \frac{\dd D_{\alpha\beta}}{\dd \ft} = \hbar\, R^+_{\alpha\beta}.\label{eq:RGvielbeinandD}
\end{equation}

\paragraph{Extracting the spin connections from the EoMs.} As explained in equation \eqref{eq:RiccifromOmega}, the tensor $R^+_{\alpha\beta}$ can be expressed in terms of the torsionful spin connections $\Omega^{\pm\phantom{\alpha}\alpha}_{\beta\gamma}$. Moreover, recall that the latter can be easily read off the EoMs of the $\s$-model when written in the form \eqref{eq:vielbeinEOM}. In the present case, we see from the identity \eqref{eq:je} that $\e \alpha i\,\p_\pm\phi ^{\,i} = j_\pm^\alpha$. The equations \eqref{eq:vielbeinEOM} then take the following simple form in terms of the torsionful spin connections and the currents:
\begin{equation}
    \p_\pm j^\alpha_\mp+\,\Omega^{\pm\phantom{\alpha}\alpha}_{\beta\gamma} j^\beta_\pm\, j^\gamma_\mp = 0\,.\label{eq:Extractspin}
\end{equation}
To extract $\Omega^{\pm\phantom{\alpha}\alpha}_{\beta\gamma}$, this is to be compared with the explicit EoMs obtained by varying the action \eqref{eq:RGDPCMACtion} with respect to $g$. As explained in the main text, these take the form \eqref{eq:EoM}. Using the basis decomposition $j_\pm = j_\pm^\alpha\,T_\alpha$ and equation \eqref{eq:Basis}, they become
\begin{equation}
    D_{\alpha\beta}\,\partial_+j_-^\beta + D_{\beta\alpha}\,\partial_-j_+^\beta -\left\lbrace f_{\beta\alpha}^{\phantom{\beta\alpha}\delta}\,D_{\delta\gamma}+f_{\gamma\alpha}^{\phantom{\gamma\alpha}\,\delta}D_{\beta\delta}\right\rbrace\,j_+^\beta\,j_-^\gamma = 0 \,.\label{eq:eompre}
\end{equation}
These EoMs are not quite in the desired form \eqref{eq:Extractspin}, since they contain derivatives of both $j_+$ and $j_-$. To solve this problem, we use the Maurer-Cartan identity \eqref{eq:MC}, which gives
\begin{equation}
	\partial_-j^\alpha_+ = \partial_+j^\alpha_-+f^{\phantom{\beta\gamma}\alpha}_{\beta\gamma}\,j^\beta_+\,j^\gamma_-
\end{equation}
when written in components. This makes it possible to write the EoMs as
\begin{equation}
	\{D_{\alpha\beta}+D_{\beta\alpha}\}\,\partial_+j^\beta_-+\left\{f^{\phantom{\beta\delta}\delta}_{\beta\gamma}D_{\delta\alpha}-f^{\phantom{\beta\alpha}\delta}_{\beta\alpha}D_{\delta\gamma}-f^{\phantom{\alpha\beta}\delta}_{\gamma\alpha}D_{\beta\delta}\right\}j^\beta_+\,j^\gamma_-=0 \,.
\end{equation}
To finally put this in the form \eqref{eq:Extractspin}, we take the contraction with the tensor $I^{\alpha\beta}$ defined as the inverse of $D_{\alpha\beta}+D_{\beta\alpha}$, thus allowing us to read off $\Omega^+$.
We can similarly substitute $\partial_+j^\alpha_- = \partial_-j_+^\alpha +f^{\phantom{\beta\gamma}\alpha}_{\beta\gamma}\,j_-^\beta\, j_+^\gamma$ in \eqref{eq:eompre} to extract $\Omega^-$. In the end, we are left with
\begin{subequations}\label{eq:OmegaApp}
\begin{equation}
	\Omega_{\alpha \beta}^{+\phantom{\alpha}\gamma}=I^{\gamma\delta}\left\{f^{\phantom{\alpha\beta}\varepsilon}_{\alpha\beta}D_{\varepsilon\delta}-f^{\phantom{\alpha\delta}\varepsilon}_{\alpha \delta}D_{\varepsilon\beta}-f^{\phantom{\beta\delta}\varepsilon}_{\beta \delta}D_{\alpha \varepsilon}\right\},
\end{equation}
\begin{equation}
	\Omega_{\alpha\beta}^{-\phantom{\alpha}\gamma}=I^{\gamma\delta}\left\{f^{\phantom{\alpha\beta}\varepsilon}_{\alpha\beta}D_{\delta \varepsilon}-f^{\phantom{\alpha\delta}\varepsilon}_{\alpha\delta}D_{\beta \varepsilon}-f^{\phantom{\beta\delta}\varepsilon}_{\beta\delta}D_{\varepsilon\alpha}\right\}.
\end{equation}
\end{subequations}
This is the equation \eqref{eq:Omega} of the main text. Moreover, it coincides with~\cite[Equation (5.16)]{Sfetsos:2014jfa}, up to differences in conventions.

\paragraph{Calculating the Ricci tensor.} Having obtained an expression for the spin connection, we can calculate the Ricci tensor by using \eqref{eq:RiccifromOmega}. Since we assumed $D$ to be constant (in the sense of independent of the coordinates $\phi^{\,i}$), the same is true for $\Omega^+$, such that the first two terms in \eqref{eq:RiccifromOmega} trivially vanish. For the third contribution, we need to calculate
\begin{equation}
	\Omega_{\beta\alpha}^{+\phantom{\beta}\beta} = -I^{\beta\gamma}f_{\beta\gamma}^{\phantom{\beta\gamma}\delta}D_{\delta\alpha}+I^{\beta\gamma}f_{\beta\alpha}^{\phantom{\beta\alpha}\delta}\left\{D_{\gamma\delta}+D_{\delta\gamma}\right\}.
\end{equation}
The first term is a contraction between the symmetric tensor $I$ and the antisymmetric $f$ and thus vanishes. Using the definition of $I$, the second term becomes $f_{\beta\alpha}^{\phantom{\beta\alpha}\beta}=-\Tr_{\g}(\ad_{T_\alpha})$, which is zero due to the \textit{unimodularity} of the semi-simple Lie algebra $\g$. Thus, only the fourth term in \eqref{eq:RiccifromOmega} contributes and we are left with 
\begin{equation}\label{eq:RicciApp}
R^+_{\alpha\beta}=-\Omega_{\gamma\alpha}^{-\phantom{\gamma}\delta}\,\Omega_{\delta\beta}^{+\phantom{\alpha}\gamma}\,,
\end{equation}
as claimed in equation \eqref{eq:RiccifromSpin} of the main text. Explicitly, this tensor then reads
\begin{equation}
	R^+_{\alpha\beta}=-I^{\gamma\eta}I^{\delta\varepsilon}\left\{f_{\gamma\alpha}^{\phantom{\gamma\alpha}\rho}D_{\varepsilon\rho}-f_{\gamma\varepsilon}^{\phantom{\gamma\varepsilon}\rho}D_{\alpha\rho}-f_{\alpha\varepsilon}^{\phantom{\alpha\varepsilon}\rho}D_{\rho
		\gamma}\right\}\left\{f_{\delta\beta}^{\phantom{\delta\beta}\sigma}D_{\sigma\eta}-f_{\delta\eta}^{\phantom{\delta\eta}\sigma}D_{\sigma\beta}-f_{\beta \eta}^{\phantom{\beta \eta}\sigma}D_{\delta \sigma}\right\}\,,
\end{equation}
in agreement with~\cite[Equation (5.17)]{Sfetsos:2014jfa}. This is a closed-form expression for $R^+$. However, note that it includes a total of 6 internal indices and is thus hard to compute in general.

\subsection{Diagonal deformations}\label{subsec:RTDiagonal}

This is as far as we can get without specifying the form of $D$. We now take $\hg=\mathfrak{sl}_\mathbb{C}(N)$ and we assume that $D$ is diagonal in the Belavin basis \eqref{eq:defofT}. In particular, from now on, we take the $T_\alpha$'s to be elements of this specific basis, labelled by couples $\alpha=(\alpha_1,\alpha_2)$ in $\Ab \equiv \Z_N \times \Z_N \setminus \lbrace (0,0) \rbrace$. In this case, the bilinear form \eqref{eq:slcnkappa} and deformations coefficients read
\begin{equation}\label{eq:DfDiagonalApp}
    \kappa_{\alpha\beta} = -\delta_{\alpha+\beta,0}\,\xi^{-\beta_1\beta_2} \qquad \text{ and } \qquad D_{\alpha\beta} = -\delta_{\alpha+\beta,0}\,\xi^{-\beta_1\beta_2}\,D_\beta\,,
\end{equation}
with $\xi=\exp\left(\frac{2\ri\pi}{N}\right)$. Suspending the Einstein summation convention, the torsionful spin connection \eqref{eq:OmegaApp} then greatly simplifies and can be expressed as
\begin{equation}
    \Omega^{\pm\phantom{\sigma}\alpha}_{\theta\sigma} =f_{\theta\sigma}^{\phantom{\theta\sigma}\alpha}\left\{\frac{D_{\mp\alpha}+D_{\pm\sigma}-D_{\mp\theta}}{D_{\alpha}+D_{-\alpha}}\right\}\,,
\end{equation}
which in turn implies that the Ricci tensor \eqref{eq:RicciApp} can be written in the form
\begin{equation}
	R^+_{\alpha\beta}=-\sum_{\theta,\sigma\in\mathbb{A}}f_{\sigma\alpha}^{\phantom{\sigma\alpha}\theta}f_{\theta\beta}^{\phantom{\theta\beta}\sigma}\frac{\left\{D_\theta+D_{-\alpha}-D_\sigma\right\}\left\{D_{-\sigma}+D_\beta-D_{-\theta}\right\}}{\{D_\theta+D_{-\theta}\}\{D_\sigma+D_{-\sigma}\}}\label{eq:Rbeforeff}.
\end{equation}
Furthermore, the structure constants in the Belavin basis can be easily extracted from the commutation relations \eqref{eq:Tcommutator} and read
\begin{equation}
     f_{\alpha\beta}^{\phantom{\alpha\beta}\gamma} = \frac{\delta_{\alpha+\beta,\gamma} }{\sqrt{N}}\left(\xi^{\alpha_1\beta_2}-\xi^{\beta_1\alpha_2}\right)\,.
\end{equation}
This expression can be used to prove the following identity for the product of structure constants, where no summation over $\theta$ or $\sigma$ is implied (the cross-product was defined in \eqref{eq:crossproduct}):
\begin{equation}
    f_{\sigma\alpha}^{\phantom{\sigma\alpha}\theta} f_{\theta\beta}^{\phantom{\theta\beta}\sigma} = -\frac{4}{N}\sin^2\left(\pi\frac{\sigma\times\beta}{N}\right)\delta_{\theta+\beta-\sigma,0}\,\kappa_{\alpha\beta}.
\end{equation}
Notice that $\kappa_{\alpha\beta}$ vanishes unless $\alpha+\beta=0$ and we can thus trade all $\beta$ in \eqref{eq:Rbeforeff} for $-\alpha$. Then we obtain the answer
\begin{equation}
	R^+_{\alpha\beta}=\kappa_{\alpha\beta}\,R_\beta^+
\end{equation}
with
\begin{equation}\label{eq:RicciDPCM}
    R_\alpha^+\equiv\frac{4}{N}\sum_{\substack{\theta,\sigma\in\mathbb{A}\\ \sigma-\theta=\alpha}}\sin^2\left(\pi\frac{\sigma\times\alpha}{N}\right)\frac{\left\{D_\alpha+D_{\theta}-D_\sigma\right\}\left\{D_\alpha+D_{-\sigma}-D_{-\theta}\right\}}{\{D_\theta+D_{-\theta}\}\{D_\sigma+D_{-\sigma}\}}.
\end{equation}
These are the equations \eqref{eq:RDiagonal} and \eqref{eq:RalphaDiag} of the main text.\\

Importantly, we see that $R^+_{\alpha\beta}=\kappa_{\alpha\beta}\,R_\beta^+$ has the same ``diagonal'' form as $D_{\alpha\beta}=\kappa_{\alpha\beta}\,D_\beta$. Thus, the renormalisation \eqref{eq:RGvielbeinandD} of a diagonal deformation does not introduce non-diagonal terms at 1-loop and we get the following flow for the coefficients $D_\alpha$:
\begin{equation}\label{eq:RGDalpha}
    \frac{\dd D_\alpha}{\dd \ft}=\hbar\,R_\alpha^+.
\end{equation}

\subsection{Elliptic integrable deformations}\label{subsec:RTIntegrable}

In \eqref{eq:RicciDPCM}, we have reduced the contraction over 6 internal indices to a single sum. To compute it, we now finally specialise to the elliptic integrable deformed PCM, for which the coefficients $D_\alpha$ are given by \eqref{eq:EllipticDeformationsHighRank}. We will first simplify the fraction appearing in \eqref{eq:RicciDPCM} to make the calculation tractable. For that, we will rely heavily on certain properties of the elliptic functions $r^\alpha(z)$, which enter the expression \eqref{eq:EllipticDeformationsHighRank} of $D_\alpha$ and are defined in \eqref{eq:defofralpha}. We refer to the appendix \ref{app:ralpha} for more details about these functions. Here, we will need the following four identities:
\begin{enumerate}
	\item The functions $r^\alpha$ have a simple behaviour under reflection, namely
	\begin{equation}
		r^{\alpha}(z)=-r^{-\alpha}(-z)\label{eq:risodd}.
	\end{equation}
	\item The functions $r^\alpha$ satisfy the Fay identity
	\begin{equation}
		r^\alpha(z_1)\,r^{\beta}(z_2)=r^{\alpha+\beta}(z_1)\,r^\beta(z_2-z_1)+r^{\alpha+\beta}(z_2)\,r^\alpha(z_1-z_2)\label{eq:risacomp}.
	\end{equation}
	\item Taking the derivative with respect to $z_1$ of the above equation, we obtain a related identity:
	\begin{equation}
		r^{\alpha\prime}(z_1)\,r^\beta(z_2)=r^{\alpha+\beta\prime}(z_1)\,r^\beta(z_2-z_1)-r^{\alpha+\beta}(z_1)\,r^{\beta\prime}(z_2-z_1)+r^{\alpha\prime}(z_1-z_2)\,r^{\alpha+\beta}(z_2)\label{eq:rprimeisacomp}.
	\end{equation}
	\item Lastly, by carefully considering limits, one can show:
	\begin{equation}
		\frac{\dd}{\dd z}\bigl[r^\alpha(z)r^{\alpha}(-z)\bigr]=-\wp'(z)\label{eq:rcangivewp}\,,
	\end{equation}
    where crucially the right-hand side is independent of $\alpha$.\vspace{3pt}
\end{enumerate}
We will now compute the numerator and denominator of \eqref{eq:RicciDPCM} separately.

\paragraph{Computing the numerator.} First, consider the term $D_\theta-D_\sigma$ appearing in the numerator:
\begin{equation}
	D_\theta-D_\sigma=\rho\frac{r^{\sigma\prime}(\ha{z})}{r^\sigma(\ha{z})}-\rho\frac{r^{\theta\prime}(\ha{z})}{r^\theta(\ha{z})}.
\end{equation}
We use \eqref{eq:risodd} to change $\theta$ to $-\theta$ and then \eqref{eq:rprimeisacomp} to simplify the expression, remembering that the sum in \eqref{eq:RicciDPCM} enforces $\sigma-\theta=\alpha$:
\begin{equation}
	D_\theta-D_\sigma=\rho\,\frac{r^{\sigma\prime}(\ha{z})\,r^{-\theta}(-\ha{z})+r^{-\theta\prime}(-\ha{z})\,r^{\sigma}(\ha{z})}{r^{\sigma}(\ha{z})\,r^{-\theta}(-\ha{z})}=\rho\,\frac{r^{\alpha\prime}(-\ha{z})\,r^{\sigma}(2\ha{z})+r^{\alpha\prime}(\ha{z})\,r^{-\theta}(-2\ha{z})}{r^{\sigma}(\ha{z})\,r^{-\theta}(-\ha{z})}.
\end{equation}
Adding also the contribution from $D_\alpha$, expanding $r^{\sigma}(\ha{z})\,r^{-\theta}(-\ha{z})$ using \eqref{eq:risacomp} in the numerator to cancel terms and using \eqref{eq:rcangivewp} to simplify the rest, we find that the left bracket in the numerator in \eqref{eq:RicciDPCM} takes the form
\begin{equation}
	D_\alpha+D_\theta-D_\sigma=\rho\,\frac{\wp'(\ha{z})\,r^\sigma(2\ha{z})}{r^\alpha(\ha{z})\,r^{\sigma}(\ha{z})\,r^{-\theta}(-\ha{z})}.
\end{equation}
The right bracket gives a similar contribution, with $\sigma\leftrightarrow -\theta$. Using \eqref{eq:risodd} one last time, the numerator in \eqref{eq:RicciDPCM} finally evaluates to
\begin{equation}
	\left\{D_\alpha+D_{\theta}-D_\sigma\right\}\left\{D_\alpha+D_{-\sigma}-D_{-\theta}\right\}=\frac{\rho^2\,\wp'(\ha{z})^2}{r^{\sigma}(\ha{z})\,r^{\sigma}(-\ha{z})\,r^{\theta}(\ha{z})\,r^{\theta}(-\ha{z})}\frac{r^\sigma(2\ha{z})\,r^{-\theta}(2\ha{z})}{r^\alpha(\ha{z})^2}.\label{eq:RicciNumerator}
\end{equation}

\paragraph{Computing the denominator.} Next, we work out the denominator. Using \eqref{eq:risodd} and \eqref{eq:rcangivewp}, the left bracket in the denominator of \eqref{eq:RicciDPCM} takes the form
\begin{equation}
	D_{\theta}+D_{-\theta}=\rho\,\frac{r^{\theta\prime}(-\ha{z})}{r^\theta(-\ha{z})}-\rho\,\frac{r^{\theta\prime}(\ha{z})}{r^\theta(\ha{z})}=\rho\,\frac{\wp'(\ha{z})}{r^\theta(\ha{z})\,r^{\theta}(-\ha{z})}.
\end{equation}
We can compute $D_\sigma+D_{-\sigma}$ in the same fashion. In the end, we obtain
\begin{equation}
	\{D_\theta+D_{-\theta}\}\{D_\sigma+D_{-\sigma}\}=\frac{\rho^2\,\wp'(\ha{z})^2}{r^\sigma(\ha{z})\,r^{\sigma}(-\ha{z})\,r^\theta(\ha{z})\,r^{\theta}(-\ha{z})}.
\end{equation}
Now we see that this exactly cancels the left fraction in \eqref{eq:RicciNumerator}. In the end, we have managed to rewrite the elements $R_\alpha^+$ as
\begin{equation}
	R_\alpha^+=\frac{4}{N}\sum_{\substack{\theta,\sigma\in\mathbb{A}\\\sigma-\theta=\alpha}}\sin^2\left(\pi\frac{\sigma\times\alpha}{N}\right)\frac{r^\sigma(2\ha{z})\,r^{-\theta}(2\ha{z})}{r^\alpha(\ha{z})^2}\label{eq:Rnexttolast}.
\end{equation}

\paragraph{Computing the sum.} This has greatly simplified the result. The last step is to expand the remaining product of $r^\sigma$ and $r^{-\theta}$ and perform the summation. One might wish to use \eqref{eq:risacomp} to get terms proportional to $r^{\alpha}(2\ha{z})$. However, note that we have to be careful with \eqref{eq:risacomp} if $z_1=z_2$, since the RHS contains $r(z_2-z_1)=r(0)$, which is divergent. We will thus consider displacing one of the $2\ha{z}$ by $\varepsilon$ and take the limit as $\varepsilon\ra 0$. We then have
\begin{equation}
	r^\sigma(2\ha{z}+\varepsilon)\,r^{-\theta}(2\ha{z})=r^\alpha(2\ha{z}+\varepsilon)\,r^{-\theta}(-\varepsilon)+r^\alpha(2\ha{z})\,r^\sigma(\varepsilon)\label{eq:rwithepsilon}.
\end{equation}
Next, we can use the expression \eqref{eq:defofralpha} of $r^{\alpha}$ to find its limiting behaviour for small values of the argument:
\begin{equation}
	r^{\alpha}(\varepsilon)=\frac{1}{\varepsilon}+\zeta(q_\alpha)-Q_\alpha+\mathcal{O}(\varepsilon).
\end{equation}
Inserting this into the expression above and taking the limit $\varepsilon\ra 0$ while using that $\zeta$ is odd, we find
\begin{equation}
	r^\sigma(2\ha{z})\,r^{-\theta}(2\ha{z})=-r^{\alpha\prime}(2\ha{z})+r^\alpha(2\ha{z})\big\{\zeta(q_\sigma)-Q_\sigma\big\} - r^\alpha(2\ha{z})\big\{\zeta(q_\theta)-Q_\theta\big\}.
\end{equation}
Importantly, this has separated the $\sigma$- and $\theta$-dependence. Indeed, notice that the first term in the expression above is independent of $\sigma$ and $\theta$: we are then free to perform the sum of the $\sin^2$-terms overall $\sigma\in\mathbb{A}$, which evaluates to $N^2/2$. On the other hand, the second and third terms depend only on $\sigma$ and $\theta$, respectively. Let us focus on the $\sigma$-dependent term, the third one being treated in the same way. We note that it is odd with respect to the reflection $\sigma\mapsto - \sigma$, \textit{i.e.} $\zeta(q_\sigma)-Q_\sigma=-\zeta(q_{-\sigma})+Q_{-\sigma}$. In contrast, the $\sin^2$ factor is even: since we are summing over all $\sigma\in\mathbb{A}$, we then find that the sum must vanish. Thus, the final result is
\begin{equation}\label{eq:RApp}
	R_\alpha^+=-2N\frac{r^{\alpha\prime}(2\ha{z})}{r^\alpha(\ha{z})^2}\,,
\end{equation} 
as claimed in the equation \eqref{eq:Rr} of the main text.

\paragraph{Rewriting the sum.} In order to use the result \eqref{eq:RApp} to extract the $\beta$-functions of the model, as we will do in the next subsection, it is useful to re-express $R_\alpha^+$ in terms of $\zeta$-functions. To achieve this, write the denominator as $-r^{\alpha}(\ha{z})\,r^{-\alpha}(-\ha{z})$. One can then consider an $\varepsilon$-expansion along the lines of \eqref{eq:rwithepsilon}, but this times with respect to the variable $\alpha$; in the end, we find
\begin{equation}
	R_\alpha^+=2N\frac{\zeta(2\ha{z}+q_\alpha)-\zeta(2\ha{z})-Q_\alpha}{\zeta(2\ha{z}+q_\alpha)-2\zeta(\ha{z})-\zeta(q_\alpha)}\,.
\end{equation}
Lastly, one can use the following addition theorem for $\zeta$-functions:
\begin{equation}
	\zeta(z_1+z_2)=\zeta(z_1)+\zeta(z_2)+\frac{1}{2}\frac{\wp'(z_1)-\wp'(z_2)}{\wp(z_1)-\wp(z_2)}\,.
\end{equation}
Using this twice, the numerator becomes
\begin{equation}
	\zeta(2\ha{z}+q_\alpha)-\zeta(2\ha{z})-Q_\alpha=-\frac{1}{2}\left[2\left\{Q_\alpha+\zeta-\zeta_\alpha\right\}-\frac{\wp'-\wp_\alpha'}{\wp-\wp_\alpha}+\frac{\wp''}{\wp'}\right]\,,
\end{equation}
while the denominator becomes
\begin{equation}
	\zeta(2\ha{z}+q_\alpha)-2\zeta(\ha{z})-\zeta(q_\alpha)=-\frac{\wp'}{\wp-\wp_\alpha}\,,
\end{equation}
where we have introduced the shorthand $f=f(\ha{z})$ and $f_\alpha=f(\ha{z}+q_\alpha)$ for $f=\zeta,\wp,\wp',\wp''$, which will simplify notations greatly in the next subsection.
In all, we have thus successfully simplified the torsionful Ricci tensor down to the expression
\begin{equation}
	R_\alpha^+=\frac{N}{\wp'}\left[2\Big\{\wp_\alpha-\wp\Big\}\Big\{Q_\alpha+\zeta-\zeta_\alpha+\frac{\wp''}{2\wp'}\Big\}-\Big\{\wp'_\alpha-\wp'\Big\}\right]\,.\label{eq:RElliptic}
\end{equation}
This is the equation \eqref{eq:RicciFlowofD} of the main text.

\subsection{Extracting the RG-flow and the \texorpdfstring{$\beta$-functions}{beta-functions}}\label{subsec:MatchingRG}

\paragraph{Variation of $\bm{D_\alpha}$ from running parameters.} We now wish to show that the Ricci-flow \eqref{eq:RGDalpha} of the coefficients $D_\alpha$ can be re-absorbed into a running of the parameters $(\ell_1,\ell_2,\rho,\ha{z})$ from which they are built. Instead of using the half-periods $\ell_1$ and $\ell_2$ directly, we will consider the Weierstrass invariants $g_2$ and $g_3$, defined by equation \eqref{eq:WeiInv} -- see Appendix \ref{app:Weierstrass} for more details. The data of $(g_2,g_3)$ is equivalent to that of $(\ell_1,\ell_2)$ but
the former are slightly easier to work with in the present context. It will also be convenient to use the modular discriminant, defined in equation \eqref{eq:ModDisc} as $\Delta=g_2^3-27g_3^2$. From now on, we thus see $D_\alpha$ as a function of $(\rho,\ha{z},g_2,g_3)$.

Our goal is to prove that the Ricci-flow \eqref{eq:RGDalpha} of $D_\alpha$ is induced by a variation of these parameters. To do so, we need to rewrite it as
\begin{equation}\label{eq:dDalphaParam}
    \frac{\dd D_\alpha}{\dd\mathfrak{t}}=\beta_{\rho}\frac{\partial D_\alpha}{\partial \rho}+\beta_{\hat{z}}\frac{\partial D_\alpha}{\partial \hat{z}}+\beta_{g_2}\frac{\partial D_\alpha}{\partial g_2}+\beta_{g_3}\frac{\p D_\alpha}{\p g_3}\,,
\end{equation}
for some coefficients $(\beta_\rho,\beta_{\hat{z}},\beta_{g_2},\beta_{g_3})$, which will then be interpreted as the $\beta$-functions of the parameters $(\rho,\ha{z},g_2,g_3)$. The first step in this endeavour, which will be the subject of this paragraph, is to compute the partial derivatives appearing on the right-hand side of this equation. Two of them are relatively simple:
\begin{equation}
	\frac{\partial D_\alpha}{\partial \rho}=Q_\alpha+\zeta-\zeta_\alpha\qquad \text{ and } \qquad \frac{\partial D_\alpha}{\partial \ha{z}}=-\rho\left\{\wp-\wp_\alpha \right\}\,.
\end{equation}
However, the $g_2$ and $g_3$ derivatives are substantially more complicated. For instance, we have
\begin{equation}
	\frac{\partial D_\alpha}{\partial g_2}=\rho\left[\frac{\p Q_\alpha}{\p g_2}+\frac{\p\zeta}{\p g_2}-\frac{\p\zeta_\alpha}{\p g_2}\right],
\end{equation}
with each derivative in turn being given by
\begin{equation*}
	\frac{\p Q_\alpha}{\p g_2}=\frac{2g^2_2Q_\alpha-3g_2g_3q_\alpha}{8\Delta}\,,\qquad \frac{\p \zeta}{\p g_2}=\frac{2(g_2^2+18g_3\wp)\zeta-g_2\left(3g_3+2g_2\wp\right)\ha{z}+18g_3\wp'}{8\Delta}\,,
\end{equation*}
\begin{equation}
    \frac{\p \zeta_\alpha}{\p g_2}=\frac{2(g_2^2+18g_3\wp_\alpha)\zeta_\alpha-g_2\left(3g_3+2g_2\wp_\alpha\right)\ha{z}+3g_3(6\wp'_\alpha-g_2q_\alpha-12\wp_\alpha Q_\alpha)}{8\Delta}\,,
\end{equation}
using various identities obeyed by the Weierstrass functions. Combining this together, we find
\begin{equation*}
	\frac{\partial D_\alpha}{\partial g_2}=	\frac{\rho}{8\Delta}\Big[2g_2^2\left\{Q_\alpha+\zeta-\zeta_\alpha\right\}-2g_2^2\ha{z}\left\{\wp-\wp_\alpha\right\}+ 18g_3\left\{2Q_\alpha\wp_\alpha+\wp'-\wp'_\alpha\right\}+36g_3\left\{\wp\zeta-\wp_\alpha\zeta_\alpha\right\}\Big].
\end{equation*}
Similarly, the $g_3$ derivatives read
\begin{equation*}
	\frac{\p Q_\alpha}{\p g_3}=\frac{g_2^2q_\alpha-18g_3Q_\alpha}{4\Delta}\comma \frac{\p \zeta}{\p g_3}=-\frac{6(3g_3+2g_2\wp)\zeta-(g_2^2+18g_3\wp)\ha{z}+6g_2\wp'}{4\Delta}\,,
\end{equation*}
\begin{equation}
	\frac{\p \zeta_\alpha}{\p g_3} =-\frac{6(3g_3+2g_2\wp_\alpha)\zeta_\alpha-(g_2^2+18g_3\wp_\alpha)\ha{z}+g_2(6\wp'_\alpha-g_2q_\alpha-12\wp_\alpha Q_\alpha)}{4\Delta} \,,
\end{equation}
allowing us to write
\begin{equation*}
	\frac{\partial D_\alpha}{\partial g_3}	=\frac{\rho}{8\Delta}\Big[-36g_3\left\{Q_\alpha+\zeta-\zeta_\alpha\right\}+36g_3\ha{z}\left\{\wp-\wp_\alpha\right\}-12g_2\left\{2Q_\alpha\wp_\alpha+\wp'-\wp'_\alpha\right\}-24g_2\left\{\wp\zeta-\wp_\alpha\zeta_\alpha\right\}\Big].
\end{equation*}
To simplify the final expression, we will use the $\beta$-function of the modular discriminant $\Delta$, given by
\begin{equation}\label{eq:BetaDelta}
	 \beta_{\Delta}=3g_2^2\beta_{g_2}-54g_3\beta_{g_3}\,.
\end{equation}
We can then rewrite the desired form \eqref{eq:dDalphaParam} of the flow of $D_\alpha$ as
\begin{align}
&	 \label{eq:DdtElliptic}\frac{\dd D_\alpha}{\dd \mathfrak{t}}=\Big[\frac{3\rho}{4\Delta}\left\{3g_3\beta_{g_2}-2g_2\beta_{g_3}\right\}\Big]\Big[2\wp_\alpha\{Q_\alpha-\zeta_\alpha\}-\wp_\alpha'\Big] + \Big[\rho\beta_{\hat{z}}+\frac{\rho\ha{z}\,\beta_\Delta}{12\Delta}\Big]\wp_\alpha + \Big[\beta_\rho+\frac{\rho\,\beta_\Delta}{12\Delta}\Big]\Big[Q_\alpha-\zeta_\alpha\Big] \notag \\
&\hspace{50pt}	+\Big[\beta_\rho+\frac{\rho\,\beta_{\Delta}}{12\Delta}\Big]\zeta - \Big[\rho\beta_{\hat{z}}+\frac{\rho\ha{z}\,\beta_{\Delta}}{12\Delta}\Big]\wp+\Big[\frac{3\rho}{4\Delta}\{3g_3\beta_{g_2}-2g_2\beta_{g_3}\}\Big]\Big[\wp'+2\wp\zeta\Big].
\end{align}

We finally note that if one knows $\beta_{g_2}$ and $\beta_{g_3}$, one can easily find the $\beta$-function of $\ell_i$, which reads
\begin{equation}
    \beta_{\ell_i} =\frac{18g_3\left[L_i\beta_{g_2}+\ell_i\beta_{g_3}\right]-g_2\left[g_2\ell_i\beta_{g_2}+12L_i\beta_{g_3}\right]}{4\Delta}\,.\label{eq:betaellfrombetag}
\end{equation}

\paragraph{Matching with the Ricci-flow.} We now want to match the two expressions \eqref{eq:RGDalpha} and \eqref{eq:DdtElliptic} of the flow of $D_\alpha$. To do so, we reexpand the Ricci tensor \eqref{eq:RElliptic} appearing in \eqref{eq:RGDalpha} in terms whose $\alpha$-dependence match those of \eqref{eq:DdtElliptic}:
\begin{equation*}
	R_\alpha^+=\frac{N}{\wp'}\Big[2\wp_\alpha\{Q_\alpha-\zeta_\alpha\}-\wp_\alpha'\Big]+\Big[\frac{2N}{\wp'}\Big\{\zeta+\frac{\wp''}{2\wp'}\Big\}\Big]\wp_\alpha-\frac{2N\wp}{\wp'}\Big[Q_\alpha-\zeta_\alpha\Big]-\frac{N}{\wp'}\Big[2\wp\Big\{\zeta+\frac{\wp''}{2\wp'}\Big\}+\wp'\Big].
\end{equation*}
One then finds that the flows \eqref{eq:RGDalpha} and \eqref{eq:DdtElliptic} perfectly match for all $\alpha\in\mathbb{A}$ if one enforces the following relations between the $\beta$-functions:
\begin{equation*}
	3g_3\beta_{g_2}-2g_2\beta_{g_3}=\frac{4\hbar\,\Delta N}{3\rho\,\wp'}\,,\qquad \beta_{\hat{z}}+\frac{\ha{z}\,\beta_\Delta}{12\Delta}=\frac{2\hbar N}{\rho\,\wp'}\left[\zeta+\frac{\wp''}{2\wp'}\right]\,,\qquad \beta_\rho+\frac{\rho\,\beta_{\Delta}}{12\Delta}=-\frac{2\hbar N\wp}{\wp'}\,.
\end{equation*}
These equations are underdetermined: indeed, we have 3 relations for 4 $\beta$-functions, recalling that $\beta_\Delta$ depends on $\beta_{g_2}$ and $\beta_{g_3}$ through equation \eqref{eq:BetaDelta}. This means that we will get an unconstrained parameter $b$ in the final expressions of $(\beta_\rho,\beta_{\hat{z}},\beta_{g_2},\beta_{g_3})$. Explicitly, we find
    \begin{equation}\label{eq:AppBeta1}
    \beta_\rho=\hbar N\left( -\frac{2\wp(\ha{z})}{\wp'(\ha{z})}+ b\, \rho \right) \,,\qquad \beta_{\hat{z}}=\hbar N\left(\frac{2\zeta(\ha{z})\,\wp'(\ha{z})+\wp''(\ha{z})}{\rho\,\wp'(\ha{z})^2}+b\,\ha{z} \right)\,,
\end{equation}
\begin{equation}
    \beta_{g_2}=-4\hbar N\left(\frac{3\,g_3}{\rho\, \wp'(\ha{z})} + b\,g_2 \right)\,,\qquad \beta_{g_3}=-6\hbar N\left(\frac{g_2^2}{9\rho\, \wp'(\ha{z})} + b\,g_3 \right)\,.
\end{equation}
Note that there is some freedom in the way the unconstrained coefficient $b$ enters these expressions, as one could for instance consider shifts or dilations of it by any functions of the other parameters.
The form considered here has been chosen to facilitate the identification of $b$ in the $\p\Psi$-conjecture and its interpretation in terms of dilations of the spectral parameter. We refer to the main text and in particular subsections \ref{subsec:FindingtheRG-Flow} and \ref{subsec:ConjDPCM} for more details.\\

The $\beta$-function of the modular discriminant $\Delta$ can be easily determined from that of the Weirstrass invariants $(g_2,g_3)$ and the relation \eqref{eq:BetaDelta}. One then finds the particularly simple expression
\begin{equation}
    \beta_{\Delta}=-12\hbar N\,b \Delta \,.
\end{equation}
Finally, one can determine the $\beta$-function of the half-period $\ell_i$, which is related to $(\beta_{g_2},\beta_{g_3})$ by the identity \eqref{eq:betaellfrombetag}. This yields
\begin{equation}\label{eq:AppBeta2}
    \beta_{\ell_i} = \hbar N\left( \frac{2 L_i}{\rho\,\wp'(\ha{z})} + b\,\ell_i \right)\,.
\end{equation}
Together, the equations \eqref{eq:AppBeta1} and \eqref{eq:AppBeta2} form the result \eqref{eq:betafunctions} announced in the main text.

\newpage

\bibliographystyle{JHEP}

\begin{thebibliography}{10}

\bibitem{delduc_rg_2021}
	F.~Delduc, S.~Lacroix, K. Sfetsos and K. Siampos, ``\emph{RG flows of integrable sigma-models and the twist function}'',   \href{https://doi.org/10.1007/JHEP02(2021)065}{JHEP \textbf{02} (2021) 065} [\href{https://arxiv.org/abs/2010.07879}{{\ttfamily arXiv:2010.07879}}].

\bibitem{derryberry_lax_2021}
R.~Derryberry, ``\emph{{Lax formulation for harmonic maps to a moduli of bundles}}'' (2021), [\href{https://arxiv.org/abs/2106.09781}{{\ttfamily arXiv:2106.09781}}].

\bibitem{levin_hitchin_2003}
A.~M.~Levin, M.~A.~Olshanetsky and A.~Zotov,
``\emph{Hitchin systems\textendash{}symplectic hecke correspondence and two-dimensional version}'',
  \href{http://dx.doi.org/10.1007/s00220-003-0801-0}{Commun. Math. Phys. \textbf{236} (2003) 93} 
  [\href{https://arxiv.org/abs/nlin/0110045}{{\tt arXiv:nlin/0110045}}].

\bibitem{feigin_quantization_2009}
B.~Feigin and E.~Frenkel, ``\emph{{Quantization of soliton systems and Langlands duality}}'', in \emph{Exploration of
New Structures and Natural Constructions in Mathematical Physics}, \href{https://doi.org/10.1142/e032}{Adv. Stud. Pure Math. 61 (2011) 185–274} [\href{https://arxiv.org/abs/0705.2486}{{\ttfamily arXiv:0705.2486}}].

\bibitem{vicedo_integrable_2019}
B.~Vicedo, ``\emph{{On integrable field theories as dihedral affine Gaudin  models}}'', \href{https://doi.org/10.1093/imrn/rny128}{Int. Math. Res. Not. {\bfseries 2020} (2020) 15} [\href{https://arxiv.org/abs/1701.04856}{{\ttfamily arXiv:1701.04856}}].

\bibitem{delduc_assembling_2019}
F.~Delduc, S.~Lacroix, M.~Magro and B.~Vicedo, ``\emph{{Assembling integrable $\sigma$-models as affine Gaudin models}}'',
  \href{https://doi.org/10.1007/JHEP06(2019)017}{JHEP {\bfseries 06}
  (2019) 017} [\href{https://arxiv.org/abs/1903.00368}{{\ttfamily
  arXiv:1903.00368}}].
  
\bibitem{costello_gauge_2019}
K.~Costello and M.~Yamazaki, ``\emph{{Gauge Theory And Integrability, III}}'' (2019), [\href{https://arxiv.org/abs/1908.02289}{{\ttfamily arXiv:1908.02289}}].

\bibitem{vicedo_4d_2021}
B.~Vicedo, ``\emph{{4D Chern–Simons theory and affine Gaudin models}}'', \href{https://doi.org/10.1007/s11005-021-01354-9}{Lett. Math. Phys. \textbf{111} (2021) 24}
  [\href{https://arxiv.org/abs/1908.07511}{{\ttfamily arXiv:1908.07511}}].

\bibitem{levin_2d_2022}
A.~Levin, M.~Olshanetsky and A.~Zotov, ``\emph{{2D Integrable systems, 4D Chern\textendash{}Simons theory and affine Higgs bundles}}'', \href{https://doi.org/10.1140/epjc/s10052-022-10553-0}{Eur. Phys. J. C \textbf{82} (2022) no~7, 635} [\href{https://arxiv.org/abs/2202.10106}{{\ttfamily arXiv:2202.10106}}].

\bibitem{lacroix_4-dimensional_2022}
S.~Lacroix, 
``\emph{4-dimensional Chern-Simons theory and integrable field theories}'',
\href{https://doi.org/10.1088/1751-8121/ac48ed}{J. Phys. A {\bf 55} (2022) 083001} [\href{https://arxiv.org/abs/2109.14278}{{\ttfamily arXiv:2109.14278}}].

\bibitem{Lacroix:2023gig}
S.~Lacroix, 
``\emph{Lectures on classical Affine Gaudin models}'', [\href{https://arxiv.org/abs/2312.13849}{{\ttfamily arXiv:2312.13849}}].

\bibitem{Fateev:1992tk}
V.~A.~Fateev, E.~Onofri and A.~B.~Zamolodchikov, 
``\emph{Integrable deformations of the $O(3)$ sigma model. The sausage model}'',
\href{https://doi.org/10.1016/0550-3213(93)90001-6}{Nucl. Phys. B \textbf{406} (1993), 521-565}.

\bibitem{Fateev:1996ea}
V.~A.~Fateev,  
``\emph{The sigma model (dual) representation for a two-parameter family of integrable quantum field theories}'',
\href{https://doi.org/10.1016/0550-3213(96)00256-8}{Nucl. Phys. B \textbf{473} (1996), 509-538}.

\bibitem{Lukyanov:2012zt}
S.~L.~Lukyanov,
``\emph{The integrable harmonic map problem versus Ricci flow}'',
\href{https://doi.org/10.1016/j.nuclphysb.2012.08.002}{Nucl. Phys. B \textbf{865} (2012), 308-329} [\href{https://arxiv.org/abs/1205.3201}{{\ttfamily arXiv:1205.3201}}].

\bibitem{hassler_rg_2021}
F.~Hassler, ``\emph{RG flow of integrable $\mathcal{E}$-models}'',   \href{https://doi.org/10.1016/j.physletb.2021.136367}{Phys. Lett. B \textbf{818} (2021) 136367} [\href{https://arxiv.org/abs/2012.10451}{{\ttfamily arXiv:2012.10451}}].

\bibitem{Hassler:2023xwn}
F.~Hassler, S.~Lacroix, and B.~Vicedo, ``\emph{The Magic Renormalisability of
  Affine Gaudin Models}'', \href{https://doi.org/10.1007/JHEP12(2023)005}{JHEP \textbf{12} (2023) 005} [\href{https://arxiv.org/abs/2310.16079}{{\ttfamily arXiv:2310.16079}}].
  
\bibitem{Lacroix:2023qlz}
S.~Lacroix and A.~Wallberg, ``\emph{An elliptic integrable deformation of the Principal Chiral Model}'', \href{https://doi.org/10.1007/JHEP05(2024)006}{{JHEP \textbf{05} (2024) 006}}  [\href{https://arxiv.org/abs/2311.09301}{\ttfamily arXiv:2311.09301}].

\bibitem{Ecker:1972bm}
G.~Ecker and J.~Honerkamp, ``\emph{Application of invariant renormalization to the nonlinear chiral invariant pion lagrangian in the one-loop approximation}'', \href{https://doi.org/10.1016/0550-3213(71)90468-8}{\emph{Nucl. Phys. \textbf{B35} (1971), 481-492}}.

\bibitem{Honerkamp:1971sh}
J.~Honerkamp, ``\emph{Chiral multiloops}'', \href{https://doi.org/10.1016/0550-3213(72)90299-4}{\emph{Nucl. Phys. \textbf{B36} (1972), 130-140}}.

\bibitem{Friedan:1980jf}
D.~H.~Friedan, ``\emph{Nonlinear Models in Two Epsilon Dimensions}'', \href{https://doi.org/10.1103/PhysRevLett.45.1057}{Phys. Rev. Lett. \textbf{45} (1980), 1057-1060} and \emph{Nonlinear Models in Two + Epsilon Dimensions}, \href{https://doi.org/10.1016/0003-4916(85)90384-7}{Annals Phys. \textbf{163} (1985), 318-419}.

\bibitem{Curtright:1984dz}
T.~L.~Curtright and C.~K.~Zachos, ``\emph{Geometry, Topology and Supersymmetry in Nonlinear Models}'', \href{https://doi.org/10.1103/PhysRevLett.53.1799}{Phys. Rev. Lett. \textbf{53} (1984), 1799-1801}.

\bibitem{Kotousov:2022azm}
G.~A.~Kotousov, S.~Lacroix and J.~Teschner, ``\emph{Integrable sigma models at RG fixed points: quantisation as affine Gaudin models}'',   \href{https://doi.org/10.1007/s00023-022-01243-4}{Ann. Henri Poincaré (2022)} [\href{https://arxiv.org/abs/2204.06554}{{\ttfamily arXiv:2204.06554}}].

\bibitem{maillet_kac-moody_1985}
J.-M. Maillet,
``\emph{Kac-Moody algebra and extended Yang-Baxter relations
  in the {O}({N}) non-linear $\sigma$-model}'',
\href{https://doi.org/10.1016/0370-2693(85)91075-5}{Phys. Lett. B {\bf 162} (1985) 137--142}.

\bibitem{maillet_new_1986}
J.-M. Maillet,
``\emph{New integrable canonical structures in two-dimensional
  models}'',
\href{https://doi.org/10.1016/0550-3213(86)90365-2}{Nucl. Phys. B {\bf 269} (1986) 54--76}.

\bibitem{Klimcik:1995ux}
  C.~Klim\v{c}\'{\i}k and P.~\v{S}evera,
  ``\emph{Dual non-Abelian duality and the Drinfeld double}'',
  \href{https://doi.org/10.1016/0370-2693(95)00451-P}{Phys. Lett. B \textbf{351} (1995) 455--462} [\href{https://arxiv.org/abs/hep-th/9502122}{\ttfamily arXiv:hep-th/9502122}].

\bibitem{Klimcik:1995dy}
  C.~Klim\v{c}\'{\i}k and P.~\v{S}evera,
  ``\emph{Poisson-Lie T duality and loop groups of Drinfeld doubles}'',
  \href{https://doi.org/10.1016/0370-2693(96)00025-1}{Phys. Lett. B \textbf{372} (1996) 65--71} [\href{https://arxiv.org/abs/hep-th/9512040}{\ttfamily arXiv:hep-th/9512040}].
  
\bibitem{delduc_unifying_2020}
F.~Delduc, S.~Lacroix, M.~Magro and B.~Vicedo, ``\emph{{A unifying 2d action for  integrable $\sigma$-models from 4d Chern-Simons theory}}'', \href{https://doi.org/10.1007/s11005-020-01268-y}{Lett. Math. Phys. \textbf{110} (2020) 1645–1687}
  [\href{https://arxiv.org/abs/1909.13824}{{\ttfamily arXiv:1909.13824}}].

\bibitem{lacroix_integrable_2021}
S.~Lacroix and B.~Vicedo, 
  ``\emph{Integrable $\mathcal{E}$-Models, 4d Chern-Simons Theory and Affine Gaudin Models. I.~Lagrangian Aspects}'',
  \href{https://doi.org/10.3842/SIGMA.2021.058}{SIGMA \textbf{17} (2021) 058} [\href{https://arxiv.org/abs/2011.13809}{\ttfamily arXiv:2011.13809}].

\bibitem{Liniado:2023uoo}
J.~Liniado and B.~Vicedo, 
  ``\emph{Integrable Degenerate $\mathcal {E}$-Models from 4d Chern\textendash{}Simons Theory}'',
  \href{https://doi.org/10.1007/s00023-023-01317-x}{Annales Henri Poincare \textbf{24} (2023) no.10, 3421-3459} [\href{https://arxiv.org/abs/2301.09583}{\ttfamily arXiv:2301.09583}].
  
\bibitem{belavin_solutions_1982}
A.~A. Belavin and V.~G. Drinfel'd,
``\emph{Solutions of the classical Yang-Baxter equation for simple Lie algebras}'',
\href{https://doi.org/10.1007/BF01081585}{Funct. Anal. Its Appl. {\bf 16} (1982) 159--180}.

\bibitem{maillet_hamiltonian_1986}
J.M.~Maillet, ``\emph{{Hamiltonian Structures for Integrable Classical Theories From Graded Kac-Moody Algebras}}'', \href{https://doi.org/10.1016/0370-2693(86)91289-X}{Phys. Lett. 167B (1986) 401}.

\bibitem{Reyman:1988sf}
A.~G.~Reyman and M.~A.~Semenov-Tian-Shansky, ``\emph{{Compatible Poisson structures for Lax equations: an R matrix approach}}'', \href{https://doi.org/10.1016/0375-9601(88)90707-4}{PPhys. Lett. A \textbf{130} (1988) 456-460}.

\bibitem{Vicedo:2010qd}
B.~Vicedo, ``\emph{{The classical R-matrix of AdS/CFT and its Lie dialgebra structure}}'', \href{https://doi.org/10.1007/s11005-010-0446-9}{Lett. Math. Phys. \textbf{95} (2011) 249-274} [\href{https://arxiv.org/abs/1003.1192}{{\ttfamily arXiv:1003.1192}}].

\bibitem{belavin_discrete_1981}
A.~A. Belavin, ``\emph{Discrete groups and the integrability of quantum systems}'',
  \href{https://doi.org/10.1007/BF01078301}{Funct. Anal. Its Appl. {\bf 14} (1981) no.~4,  260--267}.

\bibitem{ToAppear:Gaudin}
S.~Lacroix and A.~Wallberg, {\it {To appear}}.

\bibitem{Zorich:2006sur}
A.~Zorich, ``\emph{{Flat surfaces}}'', Frontiers in number theory, physics, and geometry. I, (2006) pp. 437-583, Springer Berlin [\href{https://arxiv.org/abs/math/0609392}{{\ttfamily arXiv:math/0609392}}].

\bibitem{Bainbridge:2016hor}
M.~Bainbridge, J.~Smillie and B.~Weiss, ``\emph{{Horocycle dynamics: new invariants and the eigenform loci in the stratum H(1, 1)}}'',  \href{https://doi.org/10.1090/memo/1384}{Mem. Amer. Math. Soc. (2022) \textbf{280} no. 1384} [\href{https://arxiv.org/abs/1603.00808}{{\ttfamily arXiv:1603.00808}}].

\bibitem{Winsor:2022rel}
K.~Winsor, ``\emph{{Dense real Rel flow orbits and absolute period leaves}}'' (2022),   [\href{https://arxiv.org/abs/2207.04628}{{\ttfamily arXiv:2207.04628}}].

\bibitem{Lukyanov:2013wra}
S.~L.~Lukyanov, 
``\emph{ODE/IM correspondence for the Fateev model}'', 
\href{https://doi.org/10.1007/JHEP12(2013)012}{JHEP \textbf{12} (2013), 012}
[\href{https://arxiv.org/abs/1303.2566}{\ttfamily arXiv:1303.2566}].

\bibitem{Bazhanov:2013cua} 
  V.~V.~Bazhanov and S.~L.~Lukyanov,
``\emph{Integrable structure of quantum field theory: Classical flat connections versus quantum stationary states}'', 
\href{https://doi.org/10.1007/JHEP09(2014)147}{JHEP \textbf{09} (2014) 147}
[\href{https://arxiv.org/abs/1310.4390}{\ttfamily arXiv:1310.4390}].

\bibitem{Bazhanov:2013oya}
V.~V.~Bazhanov and S.~L.~Lukyanov,
``\emph{From Fuchsian differential equations to integrable QFT}'', 
\href{https://doi.org/10.1088/1751-8113/47/46/462002}{J. Phys. A \textbf{47} (2014) no.46, 462002}
[\href{https://arxiv.org/abs/1310.8082}{\ttfamily arXiv:1310.8082}].

\bibitem{Lacroix:2018fhf}
S.~Lacroix, B.~Vicedo and C.~A.~S.~Young,
``\emph{Affine Gaudin models and hypergeometric functions on affine opers}'',
  \href{http://dx.doi.org/10.1016/j.aim.2019.04.032}{Adv. Math. {\bf 350} (2019) 486} 
  [\href{https://arxiv.org/abs/1804.01480}{{\tt arXiv:1804.01480}}].
 
\bibitem{Lacroix:2018itd}
S.~Lacroix, B.~Vicedo and C.~A.~S. Young, ``\emph{{Cubic hypergeometric integrals
  of motion in affine Gaudin models}}'', \href{http://dx.doi.org/10.4310/ATMP.2020.v24.n1.a5}{Adv. Theor. Math. Phys. {\bf 24} (2020) 155}  [\href{https://arxiv.org/abs/1804.06751}{{\tt arXiv:1804.06751}}].

\bibitem{Gaiotto:2020dhf}
D.~Gaiotto, J.~H.~Lee, B.~Vicedo and J.~Wu,
``\emph{Kondo line defects and affine Gaudin models}'',
\href{https://doi.org/10.1007/JHEP01(2022)175}{JHEP \textbf{01} (2022) 175}
[\href{https://arxiv.org/abs/2010.07325}{{\ttfamily arXiv:2010.07325}}].

\bibitem{Kotousov:2021vih}
G.~A.~Kotousov and S.~L.~Lukyanov,
``\emph{ODE/IQFT correspondence for the generalized affine $ \mathfrak{sl} $(2) Gaudin model}'',
\href{https://doi.org/10.1007/JHEP09(2021)201}{JHEP \textbf{09} (2021) 201}
[\href{https://arxiv.org/abs/2106.01238}{{\ttfamily arXiv:2106.01238}}].

\bibitem{Franzini:2022duf}
T.~Franzini and C.~A.~S.~Young,
``\emph{Quartic Hamiltonians, and higher Hamiltonians at next-to-leading order, for the affine Gaudin model}'',
\href{https://doi.org/10.1088/1751-8121/acbacf}{J. Phys. A \textbf{56} (2023) no.10, 105201}
[\href{https://arxiv.org/abs/2205.15815}{{\ttfamily arXiv:2205.15815}}].
  
\bibitem{Valent:2009nv}
G.~Valent, C.~Klimcik and R.~Squellari, ``\emph{One loop renormalizability of the Poisson-Lie sigma models}'', \href{https://doi.org/10.1016/j.physletb.2009.06.001}{Phys. Lett. B \textbf{678} (2009) 143-148}  [\href{https://arxiv.org/abs/0902.1459}{\ttfamily arXiv:0902.1459}].
  
\bibitem{sfetsos_renormalization_2010}
K.~Sfetsos, K.~Siampos and D.~C.~Thompson,
  ``\emph{Renormalization of Lorentz non-invariant actions and manifest T-duality}'', \href{https://doi.org/10.1016/j.nuclphysb.2009.11.001}{Nucl. Phys. B \textbf{827} (2010) 545-564}  [\href{https://arxiv.org/abs/0910.1345}{\ttfamily arXiv:0910.1345}].
  
\bibitem{Klimcik:1996np}
  C.~Klim\v{c}\'{\i}k and P.~\v{S}evera,
  ``\emph{Dressing cosets}'',
  \href{https://doi.org/10.1016/0370-2693(96)00669-7}{Phys. Lett. B \textbf{381} (1996) 56--61} [\href{https://arxiv.org/abs/hep-th/9602162}{\ttfamily hep-th/9602162}].

\bibitem{Sfetsos:1999zm}
  K.~Sfetsos,
  ``\emph{Duality invariant class of two-dimensional field theories}'',
  \href{https://doi.org/10.1016/S0550-3213(99)00485-X}{Nucl. Phys. B \textbf{561} (1999) 316--340} [\href{https://arxiv.org/abs/hep-th/9904188}{\ttfamily hep-th/9904188}].

\bibitem{Severa:2018pag}
P.~\v{S}evera and F.~Valach, 
``\emph{Courant Algebroids, Poisson\textendash{}Lie T-Duality, and Type II Supergravities}'',
  \href{https://doi.org/10.1007/s00220-020-03736-x}{Commun. Math. Phys. \textbf{375} (2020) no.1, 307-344} [\href{https://arxiv.org/abs/1810.07763}{\ttfamily 1810.07763}].

\bibitem{cherednik_relativistically_1981}
I.~V.~Cherednik, ``\emph{Relativistically Invariant Quasiclassical Limits of Integrable Two-dimensional Quantum Models}'',
  \href{https://doi.org/10.1007/BF01086395}{Theor. Math. Phys. \textbf{47} (1981) 422-425}.

\bibitem{Sfetsos:2014jfa}
K.~Sfetsos and K.~Siampos, ``\emph{Gauged WZW-type theories and the all-loop anisotropic non-Abelian Thirring model}'',   \href{https://doi.org/10.1016/j.nuclphysb.2014.06.012}{Nucl. Phys. B \textbf{885} (2014), 583-599} [\href{https://arxiv.org/abs/1405.7803}{{\ttfamily arXiv:1405.7803}}].

\bibitem{Cvetic:2001zx}
M.~Cvetic, G.~W.~Gibbons, H.~Lu and C.~N.~Pope,  ``\emph{Cohomogeneity one manifolds of spin(7) and G(2) holonomy}'',   \href{https://doi.org/10.1103/PhysRevD.65.106004}{Phys. Rev. D \textbf{65} (2002), 106004} [\href{https://arxiv.org/abs/hep-th/0108245}{{\ttfamily arXiv:hep-th/0108245}}].
  
\bibitem{levine_universal_2023}
N.~Levine, ``\emph{Universal 1-loop divergences for integrable sigma models}'',   \href{https://doi.org/10.1007/JHEP03(2023)003}{JHEP {\bf 2023} (2023) 3} [\href{https://arxiv.org/abs/2209.05502}{{\ttfamily arXiv:2209.05502}}].

\bibitem{Levine:2023wvt}
N.~Levine, ``\emph{Equivalence of 1-loop RG flows in 4d Chern-Simons and integrable 2d sigma-models}'' (2023), [\href{https://arxiv.org/abs/2309.16753}{{\ttfamily arXiv:2309.16753}}].

\bibitem{ToAppear:Universal}
S.~Lacroix, N.~Levine and A.~Wallberg, {\it {To appear}}.

\bibitem{Young:2005jv}
C.~A.~S.~Young,  ``\emph{Non-local charges, Z(m) gradings and coset space actions}'',   \href{https://doi.org/10.1016/j.physletb.2005.10.090}{Phys. Lett. B \textbf{632} (2006), 559-565} [\href{https://arxiv.org/abs/hep-th/0503008}{{\ttfamily arXiv:hep-th/0503008}}].

\bibitem{delduc_classical_2013}
F.~Delduc, M.~Magro and B.~Vicedo, ``\emph{{On classical $q$-deformations of
  integrable $\sigma$-models}}'',
  \href{https://doi.org/10.1007/JHEP11(2013)192}{JHEP {\bfseries 1311}
  (2013) 192} [\href{https://arxiv.org/abs/1308.3581}{{\ttfamily arXiv:1308.3581}}].

\bibitem{hoare_integrable_2022}
B.~Hoare, 
``\emph{Integrable deformations of sigma models}'',
\href{https://doi.org/10.1088/1751-8121/ac4a1e}{J. Phys. A {\bf 55} (2022) 093001} [\href{https://arxiv.org/abs/2109.14284}{{\ttfamily arXiv:2109.14284}}].

\bibitem{Arutyunov:2020sdo}
G.~Arutyunov, C.~Bassi and S.~Lacroix, ``\emph{{New integrable coset sigma models}}'', \href{http://dx.doi.org/10.1007/JHEP03(2021)062}{JHEP \textbf{03} (2021) 062}
  [\href{https://arxiv.org/abs/2010.05573}{{\tt arXiv:2010.05573}}].

\bibitem{schmidtt_symmetric_2021}
D.~M.~Schmidtt, ``\emph{{Symmetric space $\lambda$-model exchange algebra from 4d holomorphic Chern-Simons theory}}'', \href{https://doi.org/10.1007/JHEP12(2021)004}{JHEP \textbf{21} (2020) 004} [\href{https://arxiv.org/abs/2109.05637}{{\ttfamily arXiv:2109.05637}}].

\bibitem{Delduc:2015xdm}
F.~Delduc, S.~Lacroix, M.~Magro and B.~Vicedo, ``\emph{{On the Hamiltonian integrability of the bi-Yang-Baxter sigma-model}}'',
  \href{https://doi.org/10.1007/JHEP03(2016)104}{JHEP \textbf{03} (2016) 104} [\href{https://arxiv.org/abs/1512.02462}{\ttfamily arXiv:1512.02462}].

\bibitem{Hasenfratz:1990zz}
P.~Hasenfratz, M.~Maggiore and F.~Niedermayer, ``\emph{{The Exact mass gap of the O(3) and O(4) nonlinear sigma models in d = 2}}'', \href{https://doi.org/10.1016/0370-2693(90)90685-Y}{Phys. Lett. B \textbf{245} (1990), 522-528}.

\bibitem{Evans:1995dn}
J.~M.~Evans and T.~J.~Hollowood, ``\emph{{Exact results for integrable asymptotically - free field theories}}'', \href{https://doi.org/10.1016/0920-5632(95)00622-2}{Nucl. Phys. B Proc. Suppl. \textbf{45} (1996) no.1, 130-139} [\href{https://arxiv.org/abs/hep-th/9508141}{{\ttfamily arXiv:hep-th/9508141}}].

\bibitem{Appadu:2018ioy}
C.~Appadu, T.~J.~Hollowood, D.~Price and D.~C.~Thompson, ``\emph{{Quantum Anisotropic Sigma and Lambda Models as Spin Chains}}'', \href{https://doi.org/10.1088/1751-8121/aadc6d}{J. Phys. A \textbf{51} (2018) no.40, 405401} [\href{https://arxiv.org/abs/1802.06016}{{\ttfamily arXiv:1802.06016}}].

\bibitem{Fateev:2018yos}
V.~A.~Fateev and A.~V.~Litvinov, ``\emph{{Integrability, Duality and Sigma Models}}'', \href{https://doi.org/10.1007/JHEP11(2018)204}{JHEP \textbf{11} (2018), 204} [\href{https://arxiv.org/abs/1804.03399}{{\ttfamily arXiv:1804.03399}}].

\bibitem{Fateev:2019xuq}
V.~Fateev, ``\emph{{Classical and Quantum Integrable Sigma Models. Ricci Flow, \textquotedblleft{}Nice Duality\textquotedblright{} and Perturbed Rational Conformal Field Theories}}'', \href{https://doi.org/10.1134/S1063776119100042}{J. Exp. Theor. Phys. \textbf{129} (2019) no.4, 566-590} [\href{https://arxiv.org/abs/1902.02811}{{\ttfamily arXiv:1902.02811}}].

\bibitem{Hoare:2019ark}
B.~Hoare, N.~Levine and A.~A.~Tseytlin, ``\emph{Integrable 2d sigma models: quantum corrections to geometry from RG flow}'',
\href{https://doi.org/10.1016/j.nuclphysb.2019.114798}{Nucl. Phys. B \textbf{949} (2019), 114798} [\href{https://arxiv.org/abs/1907.04737}{\ttfamily arXiv:1907.04737}].

\bibitem{Georgiou:2019nbz}
G.~Georgiou, E.~Sagkrioti, K.~Sfetsos and K.~Siampos, ``\emph{An exact symmetry in $\lambda$-deformed CFTs}'',
\href{https://doi.org/10.1007/JHEP01(2020)083}{JHEP \textbf{01} (2020), 083} [\href{https://arxiv.org/abs/1911.02027}{\ttfamily arXiv:1911.02027}].

\bibitem{Levine:2021fof}
N.~Levine and A.~A.~Tseytlin, ``\emph{Integrability vs. RG flow in $G \times G$ and $G \times G /H$ sigma models}'',
\href{https://doi.org/10.1007/JHEP05(2021)076}{JHEP \textbf{05} (2021), 076} [\href{https://arxiv.org/abs/2103.10513}{\ttfamily arXiv:2103.10513}].

\bibitem{Alfimov:2021sir}
M.~Alfimov and A.~Litvinov, ``\emph{On loop corrections to integrable 2D sigma model backgrounds}'',
\href{https://doi.org/10.1007/JHEP01(2022)043}{JHEP \textbf{01} (2022), 043} [\href{https://arxiv.org/abs/2110.05418}{\ttfamily arXiv:2110.05418}].

\bibitem{Costello:2020twi}
 K. Costello, ``\emph{Topological strings, twistors and Skyrmions}'', \href{https://web.math.ucsb.edu/~drm/WHCGP/wh3.pdf}{The Western Hemisphere Colloquium on Geometry and Physics (2020)}.

\bibitem{Bittleston:2020hfv}
R.~Bittleston and D.~Skinner, ``\emph{Twistors, the ASD Yang-Mills equations and 4d Chern-Simons theory}'',
\href{https://doi.org/10.1007/JHEP02(2023)227}{JHEP \textbf{02} (2023), 227} [\href{https://arxiv.org/abs/2011.04638}{\ttfamily arXiv:2011.04638}].

\bibitem{Penna:2020uky}
R.~F.~Penna, ``\emph{Twistor Actions for Integrable Systems}'',
\href{https://doi.org/10.1007/JHEP09(2021)140}{JHEP \textbf{09} (2021), 140} [\href{https://arxiv.org/abs/2011.05831}{\ttfamily arXiv:2011.05831}].

\bibitem{He:2021xoo}
Y.~J.~He, J.~Tian and B.~Chen, ``\emph{Deformed integrable models from holomorphic Chern-Simons theory}'',
\href{https://doi.org/10.1007/s11433-022-1931-x}{Sci. China Phys. Mech. Astron. \textbf{65} (2022) no.10, 100413} [\href{https://arxiv.org/abs/2105.06826}{\ttfamily arXiv:2105.06826}].

\bibitem{Cole:2023umd}
L.~T.~Cole, R.~A.~Cullinan, B.~Hoare, J.~Liniado and D.~C.~Thompson, ``\emph{Integrable Deformations from Twistor Space}'', [\href{https://arxiv.org/abs/2311.17551}{\ttfamily arXiv:2311.17551}].

\bibitem{fay_theta_1973}
J.~D. Fay, ``\emph{Theta Functions on Riemann surfaces}'',
  \href{https://doi.org/10.1007/BFb0060090}{Springer-Verlag (1973)}, Berlin, Heidelberg.

\bibitem{benini_homotopical_2022}
M.~Benini, A.~Schenkel and B.~Vicedo, ``\emph{{Homotopical analysis of 4d Chern-Simons theory and integrable field theories}}'',
  \href{https://doi.org/10.1007/s00220-021-04304-7}{Commun. Math. Phys. (2022) 1417--1443} [\href{https://arxiv.org/abs/2008.01829}{{\ttfamily arXiv:2008.01829}}].

\bibitem{nekrassov_four_1996}
N.~Nekrasov, ``\emph{{Four Dimensional Holomorphic Theories}}'',  PhD thesis, Princeton University (1996). \href{http://media.scgp.stonybrook.edu/papers/prdiss96.pdf}{Available online}.

\bibitem{costello_supersymmetric_2013}
K.~Costello, ``\emph{{Supersymmetric gauge theory and the Yangian}}'' (2013),   [\href{https://arxiv.org/abs/1303.2632}{{\ttfamily arXiv:1303.2632}}].

\bibitem{costello_gauge_2018}
K.~Costello, E.~Witten and M.~Yamazaki, ``\emph{{Gauge Theory and Integrability,  I}}'', 
  \href{https://doi.org/10.4310/ICCM.2018.v6.n1.a6}{ICCM Not. \textbf{06}, no~1 (2018), 46-119} [\href{https://arxiv.org/abs/1709.09993}{{\ttfamily arXiv:1709.09993}}].

\bibitem{costello_gauge_2018-1}
K.~Costello, E.~Witten and M.~Yamazaki, ``\emph{{Gauge Theory and Integrability, II}}'', 
  \href{https://doi.org/10.4310/ICCM.2018.v6.n1.a7}{ICCM Not. \textbf{06}, no.1 (2018), 120-146} [\href{https://arxiv.org/abs/1802.01579}{{\ttfamily arXiv:1802.01579}}].

\end{thebibliography}

\begingroup\raggedright\endgroup

\end{document}